\documentclass{article}

     \PassOptionsToPackage{numbers, compress}{natbib}

     \usepackage[final]{neurips_2024}

\usepackage[utf8]{inputenc} %
\usepackage[T1]{fontenc}    %
\usepackage[hypertexnames=false]{hyperref}       %
\usepackage{url}            %
\usepackage{booktabs}       %
\usepackage{amsfonts}       %
\usepackage{nicefrac}       %
\usepackage{microtype}      %
\usepackage{xcolor}         %

\usepackage{comment}
\usepackage{amsmath} 
\usepackage{amsthm}
\newcommand{\ourtitle}{\color{black}            {NeuralClothSim}}
\usepackage{siunitx}
\usepackage{adjustbox}
\usepackage{xcolor}
\usepackage{titletoc}
\usepackage{graphicx}%
\usepackage{wrapfig}
\usepackage{enumitem}
\definecolor{myred}{rgb}{0.8,0,0}
\definecolor{mygreen}{rgb}{0,0.8,0}
\usepackage{pifont}%
\newcommand{\cmark}{{\color{mygreen}\ding{51}}}
\newcommand{\xmark}{{\color{myred}\ding{55}}}

\newtheorem{theorem}{Theorem}[section]
\newtheorem{lemma}[theorem]{Lemma}
\definecolor{darkgreen}{rgb}{0.0, 0.4, 0.0}
\definecolor{darkorange}{rgb}{1.0, 0.55, 0.0}
\definecolor{lightblue}{rgb}{0.2, 0.50, 0.9}
\definecolor{applegreen}{rgb}{0.55, 0.71, 0.0}
\definecolor{ao}{rgb}{0.0, 0.5, 0.0}
\definecolor{darkpastelgreen}{rgb}{0.01, 0.75, 0.24}
\definecolor{emerald}{rgb}{0.31, 0.78, 0.47}
\definecolor{lightseagreen}{rgb}{0.13, 0.7, 0.67}
\definecolor{mint}{rgb}{0.24, 0.71, 0.54}
\definecolor{teal}{rgb}{0.0, 0.5, 0.5}
\definecolor{violet}{rgb}{0.56, 0.0, 1.0}
\definecolor{violet(ryb)}{rgb}{0.53, 0.0, 0.69}
\definecolor{purple(html/css)}{rgb}{0.5, 0.0, 0.5}
\definecolor{purple(munsell)}{rgb}{0.62, 0.0, 0.77}

\usepackage{soul}

\usepackage{eccvabbrv}

\title{NeuralClothSim: Neural Deformation Fields \\Meet the Thin Shell Theory } %

\author{%
  Navami Kairanda ~~~ Marc Habermann ~~~ Christian Theobalt ~~~ Vladislav Golyanik \\
  \\
  Max Planck Institute for Informatics, Saarland Informatics Campus
}

\begin{document}

\maketitle

\begin{abstract}

Despite existing 3D cloth simulators producing realistic results, they predominantly operate on discrete surface representations (\eg,~points and meshes) with a fixed spatial resolution, which often leads to large memory consumption 
and resolution-dependent simulations. 
Moreover, back-propagating gradients through the existing solvers is difficult, and they hence cannot be easily integrated 
into modern neural architectures.
In response,  
this paper re-thinks 
physically accurate cloth simulation: We propose NeuralClothSim, \textit{i.e.,} a new quasistatic cloth simulator using thin shells, in which surface deformation is encoded in neural network weights in the form of a neural field. 
Our memory-efficient solver operates on a new continuous coordinate-based surface representation called neural deformation fields (NDFs); it supervises NDF equilibria with the laws of the non-linear Kirchhoff-Love shell theory with a non-linear anisotropic material model. 
NDFs are adaptive: They 1) allocate their capacity to the deformation details and 2) allow surface state queries at arbitrary spatial resolutions without re-training. 
We show how to train NeuralClothSim while imposing hard boundary conditions and demonstrate multiple applications, such as material interpolation and simulation editing. 
The experimental results highlight the effectiveness of our continuous neural formulation. 
See our project page: \url{https://4dqv.mpi-inf.mpg.de/NeuralClothSim/}.

\end{abstract}

\section{Introduction} \label{sec:intro}
    
Realistic cloth simulation is a central, long-standing and challenging problem in computer graphics.
It arises in game engines, computer animation, movie production, digital art, and garment digitisation, only to name a few areas. 
To date, it has been mostly addressed with physics-based simulators operating on explicit geometric representations, \textit{i.e.,} meshes and particle systems.
While recent simulators~\cite{guo2018material, wang2021gpu, li2022diffcloth, li2018implicit, li2020p, Liang2019}
can produce realistic 3D simulations that obey various types of boundary conditions and consider secondary effects, but their operational principle remains limited in several ways. 
First, they work on discrete surface representations such as meshes and points inherently assuming a pre-defined spatial resolution that cannot be easily changed once the simulation is accomplished.
Second, re-running with different meshing of the same initial template leads to different folds and wrinkles, which is often problematic for downstream applications. 
Third, explicit geometries require notoriously large amounts of storage for the detailed simulation: 
the memory size grows linearly with the number of points. %
Moreover, it is difficult to integrate simulators into learning frameworks and to edit
the output 3D state without re-running the simulation.
\begin{figure}
\centering
\includegraphics[width=\linewidth]{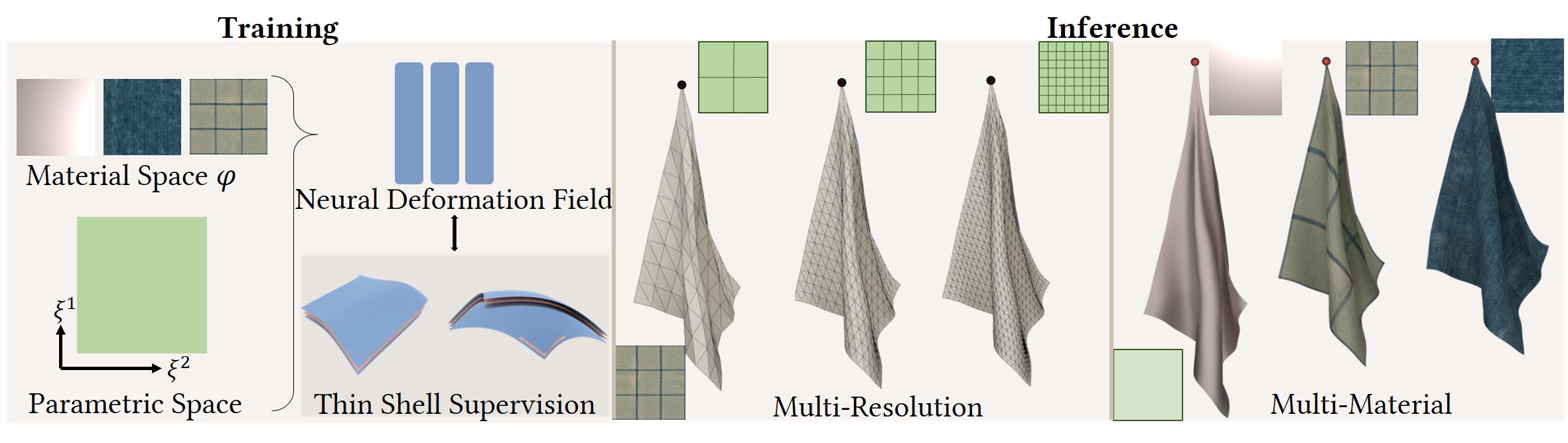} %
  \caption{
  NeuralClothSim is the first neural cloth simulator representing surface deformation as a neural field. %
  It is supervised for each target scenario with the laws of the Kirchhoff-Love thin shell theory with non-linear strain (left). 
  Once trained, the simulation can be queried \textit{continuously} and \textit{consistently} %
  enabling different spatial resolutions (center).
  NeuralClothSim can also incorporate learnt priors such as material properties that can be edited at test time (right). }
  \label{fig:teaser}  %
  \vspace{-5mm}
\end{figure}
\par 
The recent advances in physics-informed neural networks~\cite{raissi2019physics, hao2022physics} 
as well as the %
success of neural fields \cite{mildenhall2020nerf,xie2022neural, wang2021neus, yang2021geometry},
makes us question if continuous coordinate-based representations can alleviate these limitations. 
All these considerations motivate us to rethink the fundamentals of physically accurate cloth simulation and we introduce a new approach for cloth quasistatics, in which the surface deformation is encoded in neural network weights.
The proposed neural architecture is coordinate-based and has multiple advantages compared to previous simulators; see Fig.~\ref{fig:teaser} for an overview. 
Our neural fields are adaptive, \textit{i.e.,} the parameters are used to encode the deformations as they occur. 
As a matter of efficiency, we neither need to know the resolution in advance before the simulation nor do we require complex re-meshing schemes \cite{narain2012adaptive}. 
Realistic cloth simulation requires modelling geometric non-linearities and non-linear anisotropic elasticity. 
It involves large bending deformations and rigid transformations leading to non-linear point displacements. 
To efficiently model this, we rely on neural networks as they %
are good universal (non-linear) function approximators. 
\par
We model cloth simulation as a thin shell boundary-value problem with the deformation governed by the \textit{Kirchhoff-Love shell theory}.
In contrast to previous simulators using Kirchhoff-Love shell relying on isogeometric analysis \cite{lu2014dynamic} or subdivision surface algorithms \cite{guo2018material, grinspun2002charms}, 
we model thin shell deformations
as implicit
neural representations, \textit{i.e.,} 3D deformation fields encoding cloth quasistatics.
During training, our formulation supervises a neural deformation field (NDF), minimising the cloth's potential energy functional. 
In contrast to classical  simulators~\cite{Liang2019, li2022diffcloth} 
sensitive to the finite element discretisations of the initial surface, which could lead to inconsistent folds, we generate simulations with consistent drapes, folds, and wrinkles. %
This is important for downstream applications that might query (\textit{e.g.,} in the case of a renderer) or even modify (\textit{e.g.,} like inverse methods) 
the simulation with adaptive sampling. %
Next, our representation is memory-efficient, and the simulation states are generated directly in a compressed form.
In summary, our core technical contributions are as follows: 
\begin{itemize}[noitemsep,topsep=0pt]
    \item 
    A new continuous
    coordinate-based neural representation
    (Sec.~\ref{ssec:neural_deformation_field})---and a new neural solver for cloth quasistatics based on thin shell theory that
    accepts boundary conditions such as external forces or guiding motions (Sec.~\ref{ssec:initial_boundary_conditions}).
    \item Modelling of thin shell's deformation with non-linear Kirchhoff-Love theory 
    supervising the neural deformation fields (Sec.~\ref{subsec:optimisation}). %
    Upon convergence, the equilibrium state can be queried continuously and consistently. %
    \item Applications of the proposed neural simulator including %
    material interpolation 
    and fast editing of simulations according to updated simulation parameters (\cref{ssec:editing_simulation}).
\end{itemize} 

We want to point out that we do \textit{not} claim qualitative superiority over classical cloth simulation methods and \textit{completeness} of our
formulation (\eg, our method does not consider collisions). 
However, we believe that our new way of deeply integrating neural networks as a surface representation and solver into cloth simulation has the potential to stimulate future research in this direction, and we show 
that our formulation overcomes multiple fundamental limitations of existing discrete approaches. 

\section{Related Work} \label{sec:related}

\textit{Cloth Simulation}
is a well-studied problem \cite{BaraffWitkin1998, bridson2002robust, grinspun2003discrete, harmon2008robust, volino2000implementing, Liang2019, li2021codimensional, santesteban2022snug}, with the first methods dating back to the 1980s \cite{Barr1984, Terzopoulos1987}. %
The computational flow of the modern simulation approaches includes: Discretisation using the finite element method (FEM)~\cite{etzmuss2003fast, narain2012adaptive}, 
implicit time-integration~\cite{BaraffWitkin1998, li2020p}, %
frictional contact~\cite{li2018implicit, ly2020projective}, and collision handling~\cite{otaduy2009implicit, tang2018pscc, harmon2008robust}. %
Cloth simulators model real fabric behaviors~\cite{clyde2017modeling, wang2011data}, which is typically done by fitting constitutive material models.
\citet{Liang2019} and \citet{li2022diffcloth} introduced differentiable cloth simulators, %
which were subsequently shown to be also useful in 3D reconstruction as a physics-based prior \cite{kairanda2022f,li2023diffavatar}. 
Zhang \etal's approach \cite{zhang2022progressive, zhang2023progressive} %
enables interactive exploration of cloth parameters with progressively consistent quasistatics. 
Another category of methods constitutes \emph{neural cloth simulators}. 
\citet{pfaff2021learning} proposed to learn simulations using graph neural networks. %
\citet{bertiche2021pbns} is a neural simulator for static draping of garments on a virtual character. 
It is further extended with self-supervised approaches~\cite{santesteban2022snug, bertiche2022neural} to learning garment dynamics. 
They leverage physics-based loss terms and do not require simulated ground-truth data. %
However, these methods are application-oriented rather than approaches for general cloth simulation, as the garments are skinned to the human body and garment deformations are driven by body shape and poses. 
Several methods for cloth simulation rely on the Kirchhoff-Love shell theory~\cite{green2002subdivision,guo2018material}.  
The energy functionals in the theory require higher-order derivatives, which are not available for general unstructured triangle meshes. 
In their pioneering work, Cirak et al.~\cite{cirak2000subdivision} present Loop subdivision with control meshes that meet this additional $C^1$ interpolation requirement,
which is extended to %
dynamic cloth simulation with corotational strains~\cite{thomaszewski2006consistent}. %
NURBS isogeometry~\cite{lu2014dynamic} also enables continuity, whereas recent methods~\cite{clyde2017simulation, kopanicakova2019subdivision} rely on Catmull-Clark subdivision surfaces and model the geometric non-linearity of shells. 
All the aforementioned cloth simulators (traditional FEM, neural, and Kirchhoff-Love) use 
discrete surface representation (\ie, meshes) with several inherent limitations. %
The representation is not adaptive, and simulations suffer from coarse-to-fine inconsistency and are sensitive to initial discretisation.

\textit{Neural Fields.}
Recent approaches parameterising surfaces as neural fields~\cite{wang2021neus, sitzmann2020implicit, muller2022instant, mildenhall2020nerf, tretschk2021nonrigid} offer a promising alternative to meshes. 
As a common theme, these methods use
coordinate-based MLP for neural field parameterisation, which takes
coordinates in the spatio-temporal domain and returns the task-specific property, \textit{e.g.} occupancy or SDF values. 
For a detailed discussion, we refer to the survey of \citet{xie2022neural}.
However, none of the works focus on integrating such neural fields into the cloth simulation, which is the main goal of the proposed work.

\textit{Neural Networks for Solving PDEs/ODEs.} 
Several recent approaches \cite{rao2021physics, raissi2019physics, chen2018neural, zehnder2021ntopo, li2022pac}, also dubbed Physics-Informed Neural Networks (PINNs), 
leverage neural networks %
for solving tasks that are supervised by the laws of physics;
we refer to a recent survey from \citet{hao2022physics} for a detailed review. 
Chen \etal\cite{chen2022crom, chen2023implicit} use implicit neural representation to accelerate~\cite{chen2022crom} or replace~\cite{chen2023implicit} PDE solvers. 
However, they do not demonstrate thin-shell simulation.
While previous works such as Rao \etal~\cite{rao2021physics} and Zehnder \etal~\cite{zehnder2021ntopo} applied neural implicit representations for volumetric elastodynamic problems, our approach focuses on realistic thin-shell and cloth simulation. It addresses important simulation aspects such as geometric non-linearities and the integration of non-linear anisotropic models that are crucial for simulating large deformations and rotations. 
Another method~\cite{yang2021geometry} allows the processing of neural fields encoding geometric structures. 
Conceptually, the most closely related to ours is the work of \citet{bastek2023physics}, %
however, there are important differences to our work. 
First, they model linear small-strain regime for Naghdi shells, whereas we model the full non-linear stretching and bending behaviour of clothes. 
Second, we propose several architectural improvements ---periodic activation functions, periodic boundary conditions, data-driven orthotropic material model--- that are necessary for producing realistic wrinkles and folds, and demonstrate generalisation to point loads, different material and boundary values.
Next, we present a short background on Kirchhoff-Love theory that enables us to model a cloth deformation as a thin shell.

\section{Kirchhoff-Love Thin Shell Theory for Cloth Modeling}
\label{sec:background} 
Before we explain our method, we define our cloth representation.
We characterise cloth as a thin shell and model its behaviour with the Kirchhoff-Love theory \cite{love2013treatise, wempner2003mechanics}.
A thin shell is a 3D geometry with a high ratio of width to thickness. %
The shell continuum can be kinematically described by the \textit{midsurface} located in the middle of the thickness dimension and the \textit{director}, a unit vector directed along fibres in the shell that are initially perpendicular to the midsurface. 
The Kirchhoff hypothesis states the director remains straight and normal, and the shell thickness $h \in \mathbb{R}$ does not change with deformation (see inset).
We provide a detailed review of Kirchhoff-Love thin shell theory in App.~\ref{sec:background_suppl}. 

\textit{Notation.}
\label{ssec:notation}
Throughout the document, we use Greek letters for indexing quantities on the midsurface, \eg, $\mathbf{a}_{\alpha}, \alpha, \beta, ... = 1,2$, and Latin letters for indexing quantities on the shell, \eg, $\mathbf{g}_i, i, j, ... = 1,2,3$. %
Italic letters $a, A$ indicate scalars, lower case bold letters $\mathbf{a}$ indicate first-order tensors (vectors), and upper case bold letters $\mathbf{A}$ indicate second-order tensors. 
An index can appear as a superscript or subscript. 
\begin{wrapfigure}[5]{R}{0.32\textwidth} %
\vspace{-54pt}
\centering
	\includegraphics[width=\linewidth]{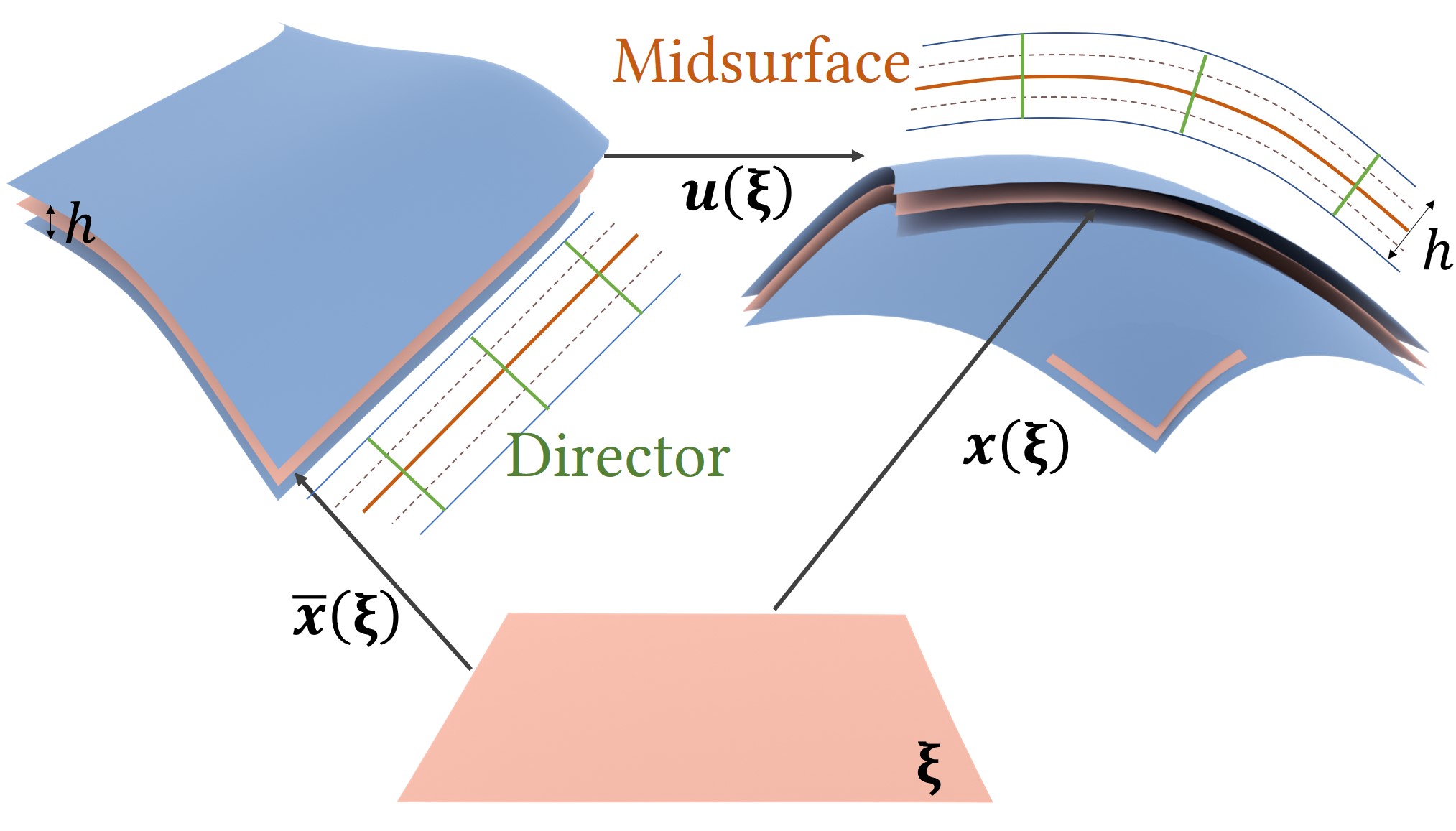} %
        \caption
	{Kirchhoff-Love shell}
	\label{fig:background}
\vspace{-34pt}
\end{wrapfigure}

Superscripts $(\cdot)^i$ refer to contravariant components of a tensor, which scale inversely with the change of basis, whereas subscripts $(\cdot)_i$ refer to covariant components that change in the same way as the basis transforms. 
Moreover, we use upper dot notation for time derivatives, lower comma notation for partial derivatives with respect to the curvilinear coordinates, $\xi^i$, and vertical bar for covariant derivatives, \eg, $\dot{\mathbf{u}} = \partial \mathbf{u} / \partial t $, $\mathbf{x,}_\alpha = \partial \mathbf{x} / \partial \xi^\alpha $,  and $u_\alpha |_\beta$, respectively. %
Geometric quantities with overbar notation $\bar{(\cdot)}$ refer to the reference configuration. %
Additionally, Einstein summation convention of repeated indices is used for tensorial operations, \eg, $\varphi_{\alpha \lambda} \varphi_\beta^\lambda = \varphi_{\alpha 1} \varphi_\beta^1 + \varphi_{\alpha 2} \varphi_\beta^2$. 
A detailed list of notations can be found in Tab.~\ref{tab:notation_supplemental} in Appendix ~\ref{sec:background_suppl}. 

\begin{figure}
	\includegraphics[width=\linewidth]{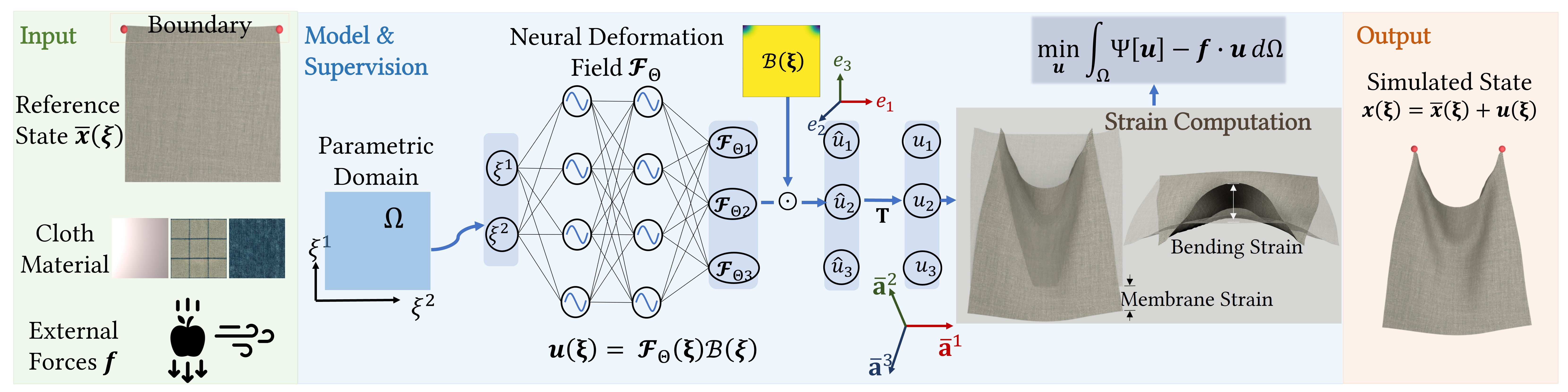}
	\caption
	{
 \textbf{NeuralClothSim} takes as input a thin shell in the reference state and its material properties, boundary motion and external forces. 
 It then learns an NDF, 
 \textit{i.e.,} a coordinate-based implicit 3D deformation field. 
 At inference, NDF can be \textit{continuously} queried for the deformed state of the surface at equilibrium using curvilinear coordinates from the parametric domain.  %
 We 
 use 
 the Kirchhoff-Love thin shell modelling to supervise the cloth quasistatics with the potential energy functional. 
	} 
	\label{fig:overview}
 \vspace{-5mm}
\end{figure}
 
\section{Method}\label{sec:method} 

We propose \ourtitle, \textit{i.e.,} a new approach for continuous and consistent quasistatic cloth simulation relying 
the thin shell theory. %
We seek to generate a complex simulation state at equilibria given a cloth geometry in a reference configuration, its material properties and external forces.
The physical basis for our cloth quasistatics is the nonlinear Kirchhoff-Love thin shell equations that model the stretching and bending 
of cloths in a unified manner. %
We parameterise the cloth states as a neural deformation field (NDF) %
defined over a continuous parametric domain %
(Sec.~\ref{ssec:neural_deformation_field}).
We explicitly account for positional and periodic boundary conditions, incorporated as 
hard constraints (Sec.~\ref{ssec:initial_boundary_conditions}).
NDF is optimised 
using 
a loss function based on the potential energy functional (Sec.~\ref{subsec:optimisation}).
Fig.~\ref{fig:overview} provides a method overview. 
\subsection{Neural Deformation Field (NDF)}\label{ssec:neural_deformation_field} 
At the core of our approach is a
\textit{neural deformation field} (NDF), a continuous representation of cloth quasistatics, entirely parameterised by a neural network.
Following Sec.~\ref{sec:background}, we model cloth geometry as a Kirchhoff-Love thin shell.
Given the rest state $\mathbf{\bar x}(\boldsymbol{\xi})$ of a cloth, we describe the equilibrium state $\mathbf{x}(\boldsymbol{\xi})$ of its midsurface under the action of external forces $\mathbf{f}(\boldsymbol{\xi})$ and boundary constraints $\mathcal{B}_d(\boldsymbol{\xi})$ using 
\begin{equation}
\small 
        \mathbf{x}(\boldsymbol{\xi}) = \mathbf{\bar{x}}(\boldsymbol{\xi}) + \mathbf{u}(\boldsymbol{\xi}), \text{ with } \boldsymbol{\xi} := (\xi^1, \xi^2) \in \Omega.%
        \label{eq:deformed_surface}
\end{equation}
The curvilinear coordinate space $(\xi^1, \xi^2)$ can (but does not need to) naturally correspond to the orthotropic
warp-weft structure of woven clothes.
As examples, the reference state associated with a flat square cloth of side $L$ in the $x$$y$-plane and that 
of a garment sleeve (radius $R$, length $L$) admitting a natural parameterisation with cylindrical coordinates are:
\begin{equation}
\small
\begin{split}
    \mathbf{\bar x}(\boldsymbol{\xi}) &= [\xi^1, \xi^2, 0]^\top, \quad \forall (\xi^1, \xi^2) \in [0,L]^2, \\
    \mathbf{\bar x}(\boldsymbol{\xi}) &= [R \cos\xi^1, \xi^2, R \sin\xi^1]^\top, \; \forall  \xi^1 \in [0,2\pi); \xi^2 \in [0,L].
    \label{eq:sleeve_napkin}
\end{split}
\end{equation}
Analytically defining surface parameterisations might not be feasible for reference geometries given as meshes. In such cases, we learn the reference parametrisation by fitting an MLP $\mathbf{\bar x}(\boldsymbol{\xi};\Upsilon)$ with parameters $\Upsilon$ to the reference mesh. 
Specifically, we learn $\mathbf{\bar x}$
by supervising it with the $\ell_2$-loss $\mathcal{L}(\Upsilon) = ||\mathbf{\bar x}(\boldsymbol{\hat \xi};\Upsilon) - \mathbf{\hat {\bar x}}||^2_2$,
where $\mathbf{\hat {\bar x}} \in \mathbb{R}^3, \boldsymbol{\hat \xi} \in \mathbb{R}^2$ are the vertices and texture coordinates of the given reference mesh. 
The advantage of this preprocessing over directly using the reference mesh is that we can continuously sample in the parametric domain by querying the MLP  
and compute all the geometric quantities %
at these points, similar to 
analytical access to the reference surface.
Our key idea is to regress the displacement field $\mathbf{u} (\boldsymbol{\xi})$ using an MLP $\mathcal{F}_\Theta: \Omega \to \mathbb{R}^3 $ and optimise its weights $\Theta$ to minimise the total potential energy of the thin-shell cloth.
Specifically, the NDF $\mathbf{u}$ is formulated as follows: 
\begin{equation}
\small
    \mathbf{u} (\boldsymbol{\xi};\Theta) = \mathcal{F}_\Theta (\mathcal{B}_p(\boldsymbol{\xi})) \mathcal{B}_d(\boldsymbol{\xi}), 
    \label{eq:deformation_field}
\end{equation}
where $\mathcal{B}_p(\boldsymbol{\xi})$ and $\mathcal{B}_d(\boldsymbol{\xi})$ are functions that respectively account for periodic and Dirichlet boundary conditions.
In Sec.~\ref{ssec:initial_boundary_conditions}, we elaborate on encoding such conditions as hard constraints. 

\par

Apart from being parameter-differentiable, \textit{i.e.,} the gradient $\nabla_\Theta\mathcal{F}_\Theta$ is defined everywhere,  $\mathcal{F}_\Theta$ needs to be input-differentiable, \textit{i.e.,} $\nabla^2_{\boldsymbol{\xi}}\mathcal{F}_\Theta$ 
must exist likewise, in order to compute the strains required for the Kirchhoff-Love energy functional. 
This restricts the activation function used in the network; only $C^2$-continuous non-linearities can be used.  
Therefore, we use periodic sine as the preferred activation function~\cite{sitzmann2020implicit} as it can represent high-frequency signals (needed for folds and wrinkles) while allowing for computing higher-order derivatives. 
Note that unlike NDF $\mathbf{u}(\boldsymbol{\xi};\Theta)$, %
we use GELU~\cite{hendrycks2016gaussian} activations for smoothly fitting the reference shape, $\mathbf{\bar x}(\boldsymbol{\xi};\Upsilon)$. 
Sec.~\ref{subsec:optimisation} describes the optimisation procedure to train the deformation field $\mathbf{u}(\boldsymbol{\xi};\Theta)$. 
\par
Once trained, $\mathcal{F}_\Theta$ provides \textit{continuous} access to the cloth quasistatics, where the network can be queried at any point in the spatial domain $\Omega$.
Based on the requirement for downstream applications, parametric input samples during inference can be \textit{different} and their number can be \textit{higher} than those during training, since it does not require the expensive computations of physical quantities; see Fig.~\ref{fig:teaser}. 
Thanks to our continuous formulation, at inference, different discretised meshing and texturing operations in the parametric domain $\Omega$ can be lifted from 2D to 3D using $\mathbf{u}(\boldsymbol{\xi};\Theta)$, which will lead to consistent result irrespective of the specific discretisation (see also Fig.~\ref{fig:initial_discrete_inconsistency_suppl}). 
\subsection{Boundary Conditions} 
\label{ssec:initial_boundary_conditions}
\begin{wrapfigure}[14]{R}{0.4\textwidth}
\centering
\vspace{-36pt}
        \includegraphics[width=0.4\textwidth]{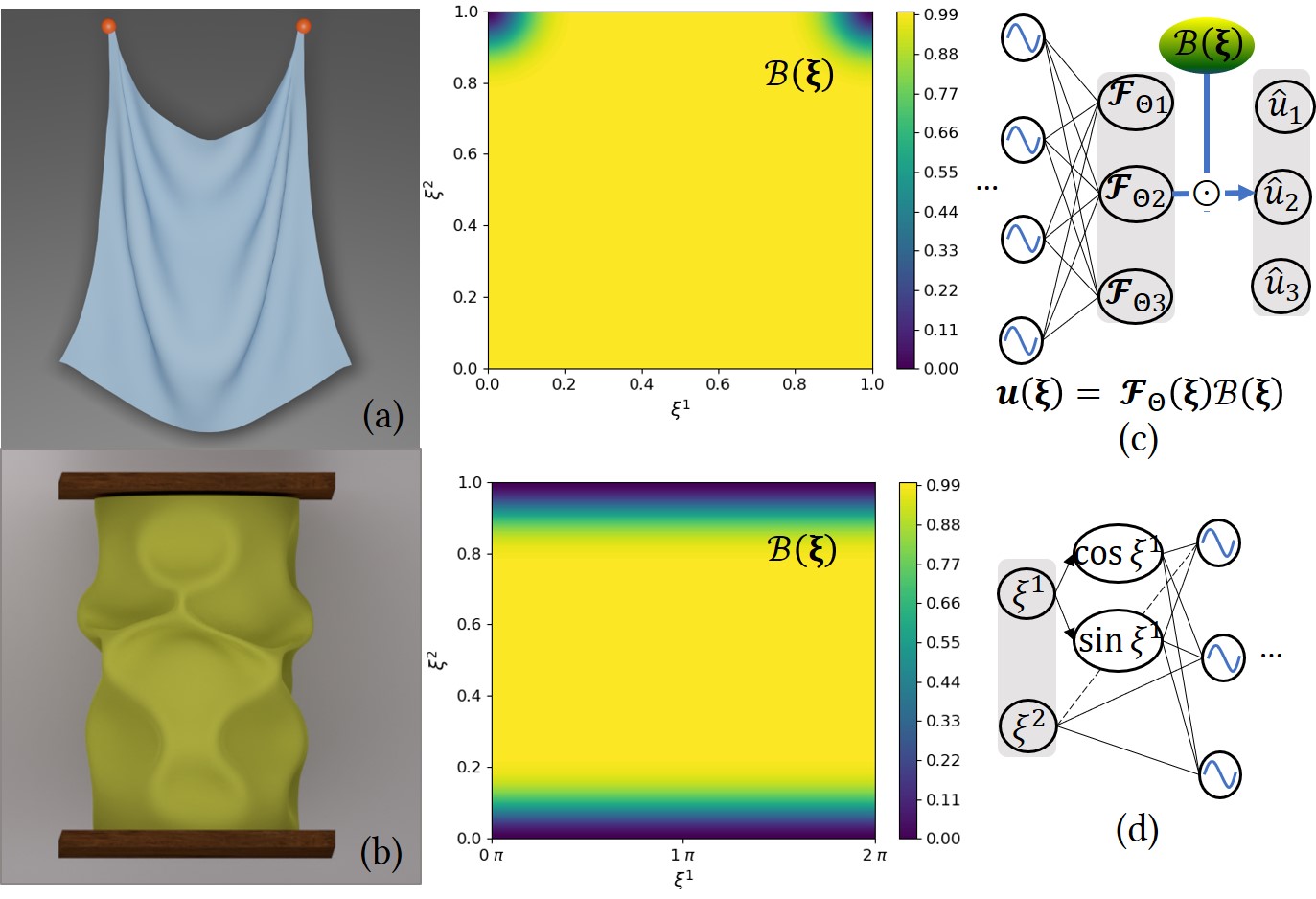} %
	\caption
        {
        \textbf{Boundary conditions.}
        In contrast to Dirichlet conditions that alter the network output (c), we impose periodic boundaries by remapping the network input to its sine and cosine values (d). %
	\label{fig:boundary_condition}
        }
 \vspace{-15pt}
\end{wrapfigure}

A practical cloth simulator allows for imposing conditions such as 
a user-specified corner motion; 
for %
most garments, the simulation needs to be continuous and consistent along the seams.
We seek to strictly enforce these conditions in our method. 
We achieve this by formulating boundary conditions as spatial distance functions, and seams as periodicity constraints along a curvilinear coordinate (such as the azimuthal angle of a cylindrically parameterised sleeve), and directly apply them to the NDF in Eq.~\eqref{eq:deformation_field}. 

\textit{Dirichlet Boundary Conditions.} 
To constrain boundary positions, we require $\mathbf{u}(\boldsymbol{\xi}_{\partial \Omega}) = \mathbf{0}$ for some specified list of parameter space points $\boldsymbol{\xi}_{\partial \Omega}$ along the boundary segment $\partial \Omega $. 
While we elaborate on the simpler case here, it is also possible to specify complex conditions \textit{i.e.,} $\mathbf{u}(\boldsymbol{\xi}_{\partial \Omega}) = \mathbf{b}(\boldsymbol{\xi}_{\partial \Omega})$, detailed in Appendix~\ref{sec:simulation_setup}. 
One solution is to sample points in the boundary segment and enforce the boundary conditions through separate loss terms. 
As shown in previous physics-informed neural networks~\cite{hao2022physics}, having competing objectives during training can lead to unbalanced gradients, which causes the network to often struggle with accurately learning the underlying solution. 
Further, there is no guarantee that the boundary conditions will always be enforced. 
Therefore, we propose to modify the NDF to embed \textit{essential} boundary conditions as hard constraints~\cite{lu2021physics}. 
Specifically, a distance function $\mathcal{B}_d(\boldsymbol{\xi})$ satisfying
    $\mathcal{B}_d(\boldsymbol{\xi}) = \begin{cases}
    0, \text{ if } \boldsymbol{\xi} \in \partial \Omega\\
    {>}0, \text{ otherwise if } \boldsymbol{\xi} \in \Omega
\end{cases}$
ensures that any instance of deformation field $\mathbf{u} (\boldsymbol{\xi}, t;\Theta)$ automatically satisfies the boundary conditions. 
We set
\begin{equation}
\begin{split}
 \mathcal{B}_d(\xi^1, \xi^2) := 1 - e ^{-((\xi^1 - \xi_{\partial \Omega}^1)^2 + (\xi^2 - \xi_{\partial \Omega}^2)^2)/\sigma} 
 \;\text{s.t.}\; (\xi_{\partial \Omega}^1, \xi_{\partial \Omega}^2) \in \partial \Omega, \quad
\forall (\xi^1, \xi^2) \in \Omega
\end{split}
\end{equation}
as a distance function with small support $\sigma=0.01$.
Fig.~\ref{fig:boundary_condition} provides an illustrative example.

The above formulation supports point and shape constraints in the cloth interior, \ie,~$\partial \Omega$ can likewise be a boundary segment inside the domain (\cref{fig:nonboundary_constraint}-appendix). 
Moreover, if the initial geometry is provided as a mesh (instead of an analytical definition), point constraints can be directly provided as mesh vertices, with $(\xi^1_{\partial \Omega}, \xi^2_{\partial \Omega})$ corresponding to texture coordinates of the vertex; see \cref{fig:initial_discrete_inconsistency_suppl}-(right). 

\textit{Periodic Boundary Conditions.}
In contrast to the positional or motion-dependent boundary conditions specified as per the user's desires, additional boundary conditions can arise from the geometric cloth parametrisation. 
Points along the panel seams of the garment share the world-space position and velocity, though they are mapped to different values in the parametric domain. 
We express continuity in geometry and simulation using periodic conditions. 
Consider any simulation involving a sleeve: 
Our method needs to guarantee the additional condition due to the parametrisation, \textit{i.e.,} $\mathbf{u}(\xi^1, \xi^2) = \mathbf{u}(\xi^1 \pm 2 n \pi, \xi^2)$. 
Whereas the Dirichlet condition is imposed by altering the network output, we \textit{strictly} impose periodic boundaries by modifying its input. 
Recall that any continuous periodic function can be written using its Fourier series. 
If $\mathbf{u}(\boldsymbol{\xi})$ is a periodic deformation field with period $P$ w.r.t.~the input coordinate $\xi^\lambda$, $\mathbf{u}(\boldsymbol{\xi})$ can be decomposed into a weighted sum 
$\{1, \sin ( 2n \pi \xi^\lambda/P ), \cos ( 2n \pi \xi^\lambda/P )\}, n \in \mathbb{N}$.
Due to the universal approximation power of MLP, only the first cosine and sine terms need to be considered, as the others can be expressed as the nonlinear continuous functions of %
$\cos(2\pi \xi^\lambda /P )$ and $ \sin(2\pi \xi^\lambda/P)$ \cite{lu2021physics}.
Hence, we map
$\xi^\lambda$ 
using $\xi^\lambda \mapsto \{\cos \xi^\lambda, \sin\xi^\lambda\}$ when feeding it to the MLP, enforcing periodicity of the predicted NDF along $\xi^\lambda$.
This completes the definition of boundary conditions %
applied during both training and inference.
\subsection{NDF Optimisation} \label{subsec:optimisation} 
We next %
explain optimisation in NDF learning. 
Note  
$\boldsymbol{\xi}$ for $\mathbf{u}(\boldsymbol{\xi})$ and derived quantities are dropped. 

\textit{Strain Computation.}
To compute the geometric strains due to the thin shell deformation, we evaluate the NDF on samples from the curvilinear coordinate space $\Omega$.
We generate $N_\Omega$ points using a stratified sampling approach. 
This ensures that the samples are random, yet well-distributed.
At each training iteration, we re-sample 
coordinates to learn an NDF that fully explores the continuous domain over the course of the optimisation. 
We evaluate NDF $\mathbf{u}(\boldsymbol{\xi})$ at all samples using Eq.~\ref{eq:deformation_field} and this prediction (\ie, $\hat{u}_i$) is assumed to be in the Cartesian coordinate system, 
\textit{i.e.,} $\mathbf{u} = \hat{u}_i \mathbf{e_i}$.
Our further strain computations (Eq.~\ref{eq:deformation_gradient}) require covariant deformation components in the reference contravariant basis, \textit{i.e.,} $\mathbf{u} = u_\alpha \mathbf{\bar a}^\alpha + u_3 \mathbf{\bar a}^3$, therefore we use the basis transformation matrix $ \mathbf{T} = [\mathbf{\bar a}^1\ \mathbf{\bar a}^2\ \mathbf{\bar a}^3 ]^{-1}$  for converting from Cartesian deformation coordinates to covariant coordinates (see \cref{sec:background_suppl} for detailed Kirchhoff-Love preliminaries).
While it is possible to predict in the local contravariant basis directly, the global basis is better suited for NDF training since the local basis vectors are not normalised, and the basis varies with the input position $\boldsymbol{\xi}$, especially noticeable for reference geometries such as sleeve
(\cref{fig:background_supplemental}-(b)-appendix).

Next, we describe the ingredients required to evaluate the internal strain energy $\Psi$.
Membrane strain $\boldsymbol \varepsilon = [ \varepsilon_{\alpha \beta}]$ and bending strain $\boldsymbol \kappa = [ \kappa_{\alpha \beta}]$ measure the in-plane stretching and the curvature change, respectively, and are defined as $\varepsilon_{\alpha \beta} := \frac{1}{2} (a_{\alpha \beta} - \bar{a}_{\alpha \beta}), $ and $\kappa_{\alpha \beta} := \bar{b}_{\alpha \beta} - b_{\alpha \beta}$ 
where $(\bar{a}_{\alpha \beta}, a_{\alpha \beta})$, and $(\bar{b}_{\alpha \beta}, b_{\alpha \beta})$ are the metric and curvature tensors of reference and deformed midsurface. 
With the assumptions of Kirchhoff-Love theory and following~\cite{basar2013mechanik}, we simplify these equations to directly operate on $\mathbf{u}$ and evaluate strains as 
\begin{equation}
\small 
    \begin{split}
    \varepsilon_{\alpha \beta} &= \frac{1}{2} %
    ({\color{darkorange}\varphi_{\alpha \beta} + \varphi_{\beta \alpha}}
    + {\color{teal}\varphi_{\alpha \lambda} \varphi_\beta^\lambda + \varphi_{\alpha 3} \varphi_{\beta 3}}), \\
    \kappa_{\alpha \beta} &= 
    {\color{darkorange}-\varphi_{\alpha 3}|_\beta - \bar{b}_\beta^\lambda \varphi_{\alpha \lambda}} 
    + {\color{teal}\varphi^\lambda_3 (\varphi_{\alpha \lambda} |_\beta + \frac{1}{2} \bar{b}_{\alpha \beta} \varphi_{\lambda 3} - \bar{b}_{\beta \lambda} \varphi_{\alpha 3})},  
    \end{split}
    \label{eq:strain}
\end{equation}

where the deformation gradients $\varphi_{\alpha \lambda}, \varphi_{\alpha 3}$ are the components of $\mathbf{u}_{,\alpha} $ such that
\begin{equation}
\small
    \begin{split}
    \mathbf{u}_{,\alpha} = \varphi_{\alpha \lambda} \mathbf{\bar a}^\lambda + \varphi_{\alpha 3} \mathbf{\bar a}^3, 
    \varphi_{\alpha \lambda} := u_\lambda |_\alpha - \bar{b}_{\alpha \lambda} u_3, %
     \;\text{and}\;  \varphi_{\alpha 3} := u_{3,\alpha} + \bar{b}_\alpha^\lambda u_\lambda.
    \end{split}
    \label{eq:deformation_gradient}
\end{equation}
We do not linearise the strain.
{\color{darkorange}Orange} and {\color{teal}teal} correspond to the linear and the non-linear components, respectively. 
To evaluate the derivatives of geometric quantities based on NDF $\mathbf{u}$ w.r.t.~inputs $\boldsymbol{\xi}$ (as part of strain computation), we use automatic differentiation 
of machine learning frameworks
\cite{paszke2019pytorch}. 

\textit{Cloth Material Model.}
A thin shell develops an internal potential energy due to deformation and the material's hyperelasticity. 
As in the cloth simulation literature~\cite{li2021codimensional, zhang2022progressive}, we write the internal hyperelastic energy density as a function of the stretching and bending strains, $\Psi(\boldsymbol{\varepsilon}, \boldsymbol{\kappa}, \xi^3; z(\mathbf{\bar x}), \boldsymbol{\Phi}, h)$. 
Here, $\boldsymbol{\Phi}$ are the cloth's material parameters and $\xi^3 \in [-\frac{h}{2}, \frac{h}{2}]$ is the thickness coordinate, and $z(\mathbf{\bar x})$ are geometric quantities derived from the reference midsurace $\mathbf{\bar x}$.
Our neural field-based cloth simulation is orthogonal to the research on material modelling and can, thus, be formulated with many different elastic models, as long as the elasticity can be represented as an energy density function.
For example, a linear isotropic~\cite{simo1989stress} stress-strain relationship leads to strain energy of the form 
$\Psi = \frac{1}{2} (D H^{\alpha \beta \lambda \delta} \varepsilon_{\alpha \beta} \varepsilon_{\lambda \delta} + B H^{\alpha \beta \lambda \delta} \kappa_{\alpha \beta} \kappa_{\lambda \delta}),$
where $D$ is the in-plane stiffness and $B$ is the bending stiffness computed as $D := \frac{Eh}{1 - \nu^2} \;\text{and}\; B := \frac{Eh^3}{12 (1 - \nu^2)},$
with Young's modulus $E$, Poisson's ratio $\nu$, and
$H^{\alpha \beta \lambda \delta} := \nu \bar{a}^{\alpha \beta} \bar{a}^{\lambda \delta} + \frac{1}{2} (1-\nu) (\bar{a}^{\alpha \lambda} \bar{a}^{\beta \delta} + \bar{a}^{\alpha \delta} \bar{a}^{\beta \lambda})$
with $\bar{a}^{\alpha \beta}$ being the contravariant metric tensors. 
Alternatively, we support the data-driven non-linear anisotropic material model of Clyde \etal \cite{clyde2017modeling} that has been carefully constructed to fit measured woven fabrics. 
We refer to \cref{sec:material_model} for the mathematical details of the non-linear model.

\textit{Energy Optimisation.}
A thin shell's stable equilibrium is characterised by the principle of minimum potential energy, i.e.~the sum of external potential energy owing to forces $\mathbf{f}$ and internal potential energy $\Psi$ due to material elasticity. 
The total potential energy $\mathcal{E}$ reads as
$\mathcal{E} [\mathbf{u}] = \int_{\Omega} \Psi \,d\Omega - \int_{\Omega} \mathbf{f} \cdot \mathbf{u} \,d\Omega, $
and the stable equilibrium deformation $\mathbf{u^*}$ can be found by minimising the energy functional subject to boundary constraints $\mathbf{u}(\xi^1, \xi^2) = \mathbf{b}(\xi^1, \xi^2) \text{ on } \partial \Omega$.
We take advantage of the variational structure of $\mathcal{E} [\mathbf{u}]$  and minimise it directly with gradient descent. 
All operations of our energy computation are naturally differentiable, and we estimate the integral %
as a sum
over continuous parametric domain. 
For linear isotropic materials, we arrive at the 
following loss function to optimise the MLP weights for a physically-principled cloth simulation encoded as $\mathbf{u}^* (\boldsymbol{\xi};\Theta)$: 
\begin{equation}
\begin{aligned}
\resizebox{0.94\textwidth}{!}{$
    \mathcal{L}(\Theta) = \frac{|\Omega|}{N_\Omega}\sum_{i=1}^{N_\Omega} \Big( \frac{1}{2} D \boldsymbol{\varepsilon}^\top (\boldsymbol{\xi}_i;\Theta) \mathbf{H} (\boldsymbol{\xi}_i) \boldsymbol{\varepsilon}(\boldsymbol{\xi}_i;\Theta)  
    + \frac{1}{2} B \boldsymbol{\kappa}^\top (\boldsymbol{\xi}_i;\Theta) \mathbf{H} (\boldsymbol{\xi}_i) \boldsymbol{\kappa} (\boldsymbol{\xi}_i;\Theta) 
    - \mathbf{f}^\top (\boldsymbol{\xi}_i) \mathbf{u} (\boldsymbol{\xi}_i;\Theta) \Big)\sqrt{\bar{a}(\boldsymbol{\xi}_i)},
\label{eq:main_loss}
$}
\end{aligned}
\end{equation} 
where $\boldsymbol \varepsilon(\boldsymbol\xi;\Theta) \in \mathbb{R}^4$, $\boldsymbol \kappa(\boldsymbol\xi;\Theta) \in \mathbb{R}^4$ are vectorised strains computed using (\ref{eq:strain}); $|\Omega| =  \int_{\Omega} \,d\xi^1 d\xi^2 $ is the area of the parametric domain; $\mathbf{H}(\boldsymbol\xi) \in \mathbb{R}^{4 \times 4}$ depends only on the reference surface.
For data-driven materials~\cite{clyde2017modeling}, the strain energy is additionally a function of thickness coordinate $\xi^3$. 
Hence, we integrate 
$\mathcal{E}$ along the thickness with the Simpson's 3-point rule (similar to \cite{clyde2017simulation}) \ie,
$\mathcal{E} [\mathbf{u}] = \int_{\Omega} \int_{-\frac{h}{2}}^{\frac{h}{2}} \Psi \,d\xi^3 \,d\Omega - \int_{\Omega} \int_{-\frac{h}{2}}^{\frac{h}{2}} \mathbf{f} \cdot \mathbf{\tilde{u}} \,d\xi^3 \,d\Omega, $ where $\mathbf{\tilde{u}} = \mathbf{u} + \xi^3 \mathbf{w}$ is the deformation for a point on the shell continuum and $\mathbf{w}$ quantifies the change in the midsurface orientation (see \cref{ssec:shell_kinematics}).

\section{Experimental Evaluation} \label{sec:results}
\begin{wrapfigure}[8]{R}{0.5\textwidth} %
\vspace{-48pt}
\centering
	\includegraphics[width=\linewidth]{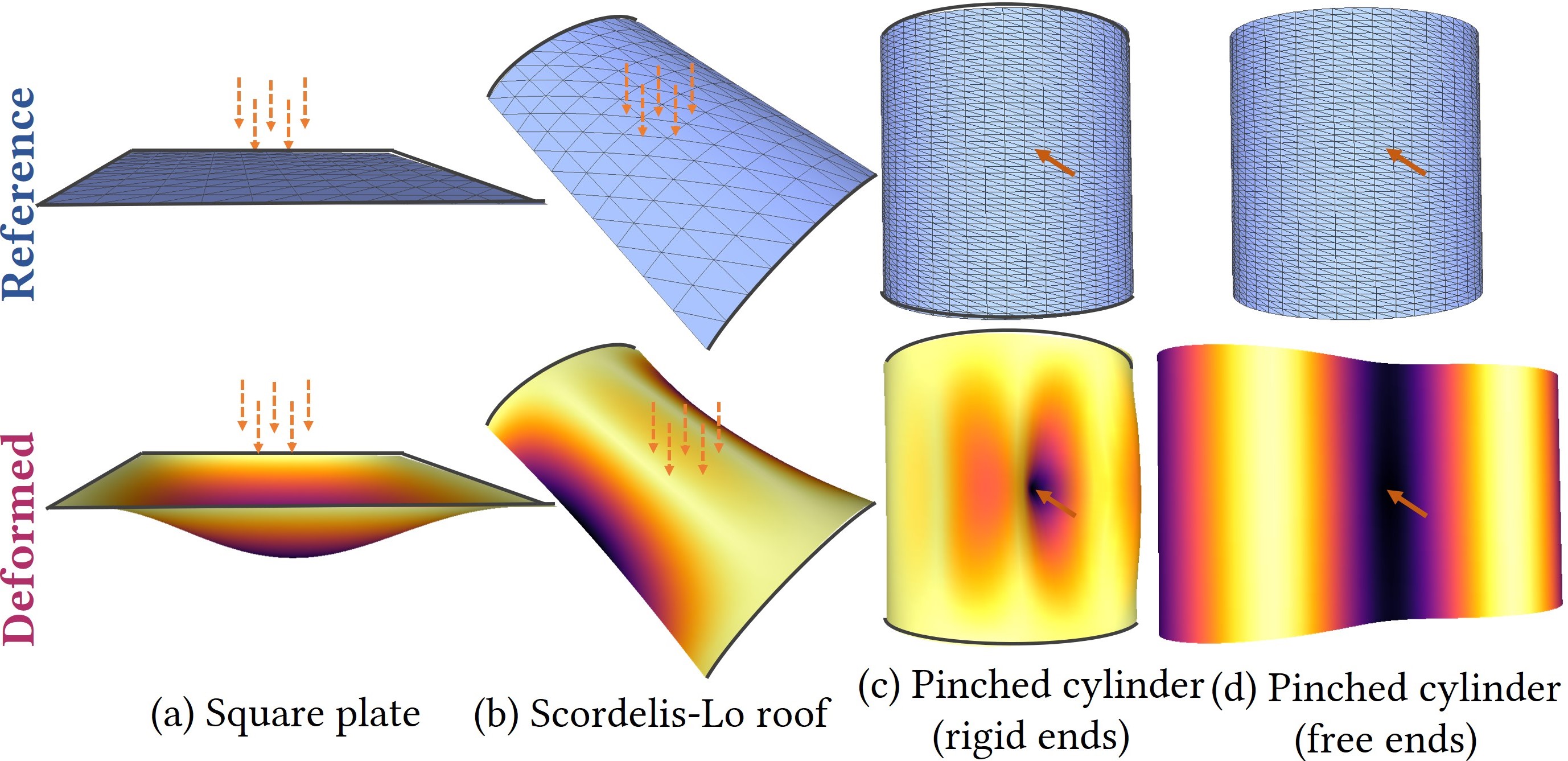} %
	\caption
	{
        \textbf{Belytschko obstacle course} 
        for which we generate accurate displacements (rescaled for better visualisation). %
	}
	\label{fig:belytschko}
        \vspace{-56pt}
\end{wrapfigure}

We next present 
the qualitative and empirical results highlighting the new characteristics of our continuous neural fields, including validation (\cref{sec:obstacle_course}), simulation results (\cref{sec:qualitative}), comparison to prior works (\cref{sec:comparison}), and applications (\cref{sec:ablation}).

\subsection{Obstacle Course} 
\label{sec:obstacle_course} 
\begin{table}
\small{ 
\caption
{
\textbf{Quantitative evaluation.}
We validate the displacements obtained with our method on the Belytschko obstacle course with analytical solutions from~\cite{belytschko1985stress, timoshenko1959theory}.
\citet{guo2021deep} use different material and match the corresponding reference result. %
Below, we show the ablation. 
We highlight that our method outperforms prior works and baselines by a large margin.
}
\begin{center}
	
\begin{tabular}{ m{8em}  m{6em}  m{7.5em} m{7.8em}  m{7.5em}  }%
    \toprule
    \textbf{Method}  & \textbf{Square plate}  & \textbf{Scordelis-Lo roof}  & \textbf{Rigid-end cylinder} & \textbf{Free-end cylinder} \\ %
    \midrule
    Analytical	                    &   0.487            &  0.3024        &     \num{1.825e-5}      &  \num{4.52e-4}   \\ %
    Guo \textit{et al.}            &   2.566*             &   n/a         &   n/a        &   n/a    \\ %
    Bastek \textit{et al.}            &   n/a             &   0.297         &   n/a        &   n/a    \\ %
    Ours, full	                &   0.487                &  0.3018             &   \num{1.81e-5}    &   \num{4.58e-4}     \\
    \midrule
    Ours, no periodicity   &   n/a                &  n/a              &     \num{3.6e-9}       &      \num{3.13e-6}  \\ %
    Ours, GELU             &   0.496              &  0.288              &   \num{1.74e-5}        &   \num{5.7e-4}    \\
    \bottomrule
    \end{tabular}
    \end{center}
    \label{tab:belytschko}
     \vspace{-3mm}
}
\end{table}

A scrupulously modelled thin shell, and consequently cloth, must be able to handle inextensional bending modes, complex membrane states of stress, and rigid body motion without straining. 
Therefore, for validation, we use the engineering obstacle course of benchmark problems from~\citet{belytschko1985stress}, for which the analytical solutions are known for linearised functionals. 
Such problems were previously used in computer graphics \cite{grinspun2006computing} for testing the performance 
of finite mesh elements. 
Specifically, we test 
our method on the square plate~\cite{timoshenko1959theory}, the Scordelis-Lo roof, and the pinched cylinder with rigid diaphragms and free ends examples, for which the original and our deformed shells are shown in Fig.~\ref{fig:belytschko}. 
See Tab.~\ref{tab:belytschko} for converged numerical results. 
The results, which show that our method outperforms prior works by a significant margin, demonstrate our method's excellent modelling ability.
We further present details of the experiments with the square plate and pinched cylinder, including the experimental setup and visualisations of the full displacement fields, in App.~\ref{sec:obstacle_course_appendix}. 

\textit{Scordelis-Lo Roof}
is a non-flat reference shape subject to complex membrane strains, 
i.e.~an open cylindrical shell with radius $R = \qty{25}{\metre}$, length $ L = \qty{50}{\metre}$ and subtends an angle of $\qty{80}{\degree}$. 
It is supported with two rigid diaphragms at the ends and loaded by gravity $ \mathbf{f} = [0, -90, 0]^\top$. 
The shell's material is given as 
$E = \qty{4.32e8}{\pascal}, \nu = 0$ and 
thickness $h = \qty{0.25}{\metre}$. 
We obtain the maximum vertical displacement $u_2$ at the centre of the edge (averaged over the two sides) as $0.3018$, closely approximating the analytical $u_2 = 0.3024$~\cite{belytschko1985stress}. 

\subsection{Qualitative Results} 
\label{sec:qualitative}
We next present our simulation results.
The experiments are performed with the values $E = \qty{5000}{\pascal}, \nu = 0.25, h = \qty{0.0012}{\metre}$ for the linear isotropic material,
and with parameters from \citet{clyde2017modeling} for the nonlinear orthotropic material. 
For the supplemental video, we extend the method to visualise the deformation trajectory (\ie, transition 
from the reference to the equilibrium state).
Details on boundary conditions, external forces, and time-stepping %
can be found in Appendix~\ref{sec:simulation_setup}.

\textit{Napkin.}
We first consider 
a square napkin of length $L = \qty{1}{\metre}$, falling freely under the effect of gravitational force. 
The napkin has a flat reference state in the \textit{xy}-plane given by (\ref{eq:sleeve_napkin}), and the gravitational force field is applied along the negative \textit{y}-axis, \textit{i.e.,} external force density $\mathbf{f} = [0, -9.8 \rho, 0]^\top$. %
We specify a fixed boundary condition at the top left corner to constrain the napkin movement. 
The meshes extracted from the trained NDF are visualised in Fig.~\ref{fig:teaser}-(center).
Note that apart from its realism, 
one can also query the simulation at arbitrary resolution in the case of \ourtitle.  
Next, we perform another experiment with a napkin subject to gravity and dynamic boundary condition, \textit{i.e.,} in which the corners move inwards.
This leads to fold formation at the top, as visualised in 
Fig.~\ref{fig:boundary_condition}-(a) and in~\cref{fig:material_model} for varying fabrics such as cotton and silk. 

\textit{Sleeve.}
We also consider a cylindrical shell and perform sleeve compression and twisting. 
In both cases, we consider the reference state (\ref{eq:sleeve_napkin}) with $L = \qty{1}{\metre}$ and $R = \qty{0.25}{\metre}$. %
See Figs.~\ref{fig:boundary_condition}-(b) and \ref{fig:pinn_comparison}-(b) for visualisations. %
In the first case, we apply torsional motion on the sleeve, \textit{i.e.,} 
a total rotation of $\frac{3\pi}{4}$ around the \textit{y-}axis to both the top and bottom rims. 
The optimised 
NDF forms wrinkles at the centre as expected~\cite{guo2018material}. %

In the second case, we compress the sleeve to produce the characteristic buckling effect. 
There are no external forces here and the compression is entirely specified by boundary conditions. 
We achieve a total displacement of $\qty{0.2}{\metre}$ due to 
compression with the inward motion of the top and bottom rims along the cylinder axis; %
see Fig.~\ref{fig:boundary_condition}-(right). 
The demonstrated simulation is a representative example of strain localisation, with noticeable diamond patterns of shell buckling.

\subsection{Comparisons to Previous Methods}\label{sec:comparison} 
\begin{figure}
        \includegraphics[width=\linewidth]{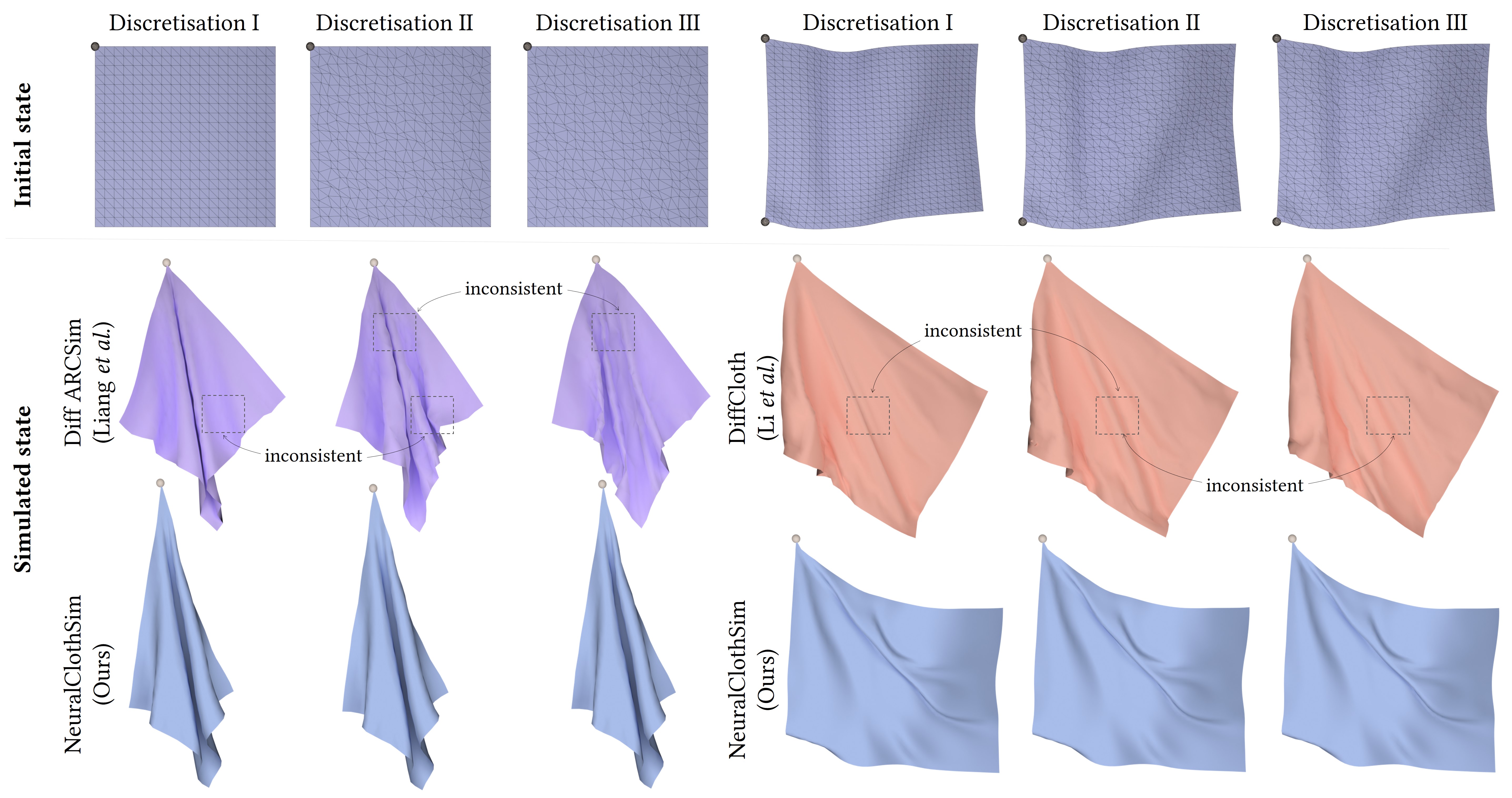}
	\caption
	{
        \textbf{Simulation consistency.}
        At different initial state discretisations, FEM-based simulators %
        lead to inconsisies with 
        often 
        differences in the folds or wrinkles. 
        In contrast, ours 
        overfits an MLP to the reference mesh and encodes the surface evolution using another MLP 
        (\textit{continuous} neural fields).
        lead to inconsistent results with often occurring differences in the folds or wrinkles. 
	}
	\label{fig:initial_discrete_inconsistency_suppl}
        \vspace{-5mm}
\end{figure}

In this section, we compare NeuralClothSim to state-of-the-art FEM cloth %
simulators and physics-informed neural networks for shell structures. 
We do not compare to other neural simulators \cite{bertiche2022neural, santesteban2022snug}, as they do not support simulating non-garment cloths, whereas ours is a general neural cloth simulator.

\textit{Cloth Simulators.}
Next, we validate the consistency of cloth simulations at different discretisations of the reference state. 
We consider two 
scenarios: 
1) A napkin with a fixed corner under gravity simulated with our approach and DiffARCSim~\cite{Liang2019} (Fig.~\ref{fig:initial_discrete_inconsistency_suppl}-left), and 
2) a flag  with two fixed corners deforming under wind and gravity simulated with our approach and DiffCloth~\cite{li2022diffcloth} %
(Fig.~\ref{fig:initial_discrete_inconsistency_suppl}-right). 
In both scenarios, we simulated ours and compared methods thrice, starting with a marginally perturbed meshing of the same initial geometry resulting in different mesh discretisations. 
\begin{wrapfigure}[15]{R}{0.4\textwidth}
\vspace{-12pt}
\centering
	\includegraphics[width=\linewidth]{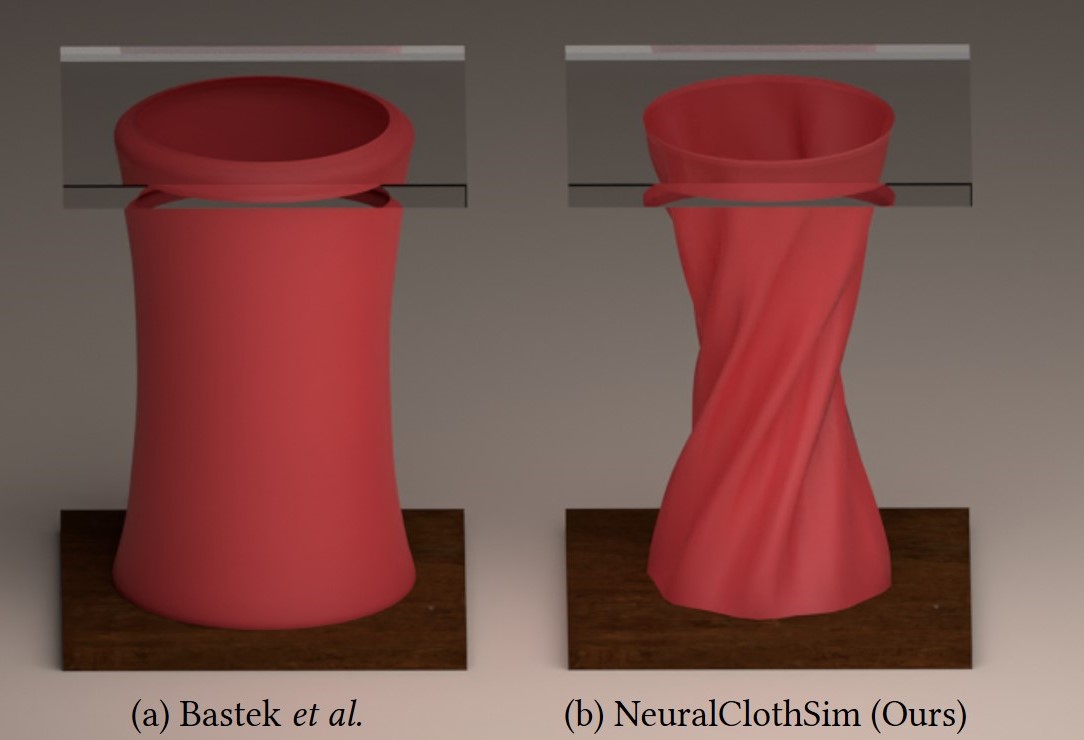}
	\caption
	{
        \textbf{Comparison to Bastek \etal}
        on 
        sleeve twisting. %
        While the cylinder %
        in (a) 
        twists without 
        wrinkles, our result (b) is correctly wrinkled, similar to \cite{guo2018material}. 
	}
	\label{fig:pinn_comparison}
\end{wrapfigure}

In the case of NeuralClothSim, we learn the reference parameterisation by fitting an MLP $\mathbf{\bar x}(\boldsymbol{\xi};\Upsilon)$ for each initial discretisation followed by NDF optimisation. %
We find that the simulated meshes extracted from NDF %
are consistent for all discretisations. 
In contrast, for competing FEM-based methods, simulation is sensitive to the discretisation; %
while multiple simulations with the same initial mesh produce identical results, slightly different meshing generates inconsistent simulations.
Theoretically, a well-defined FEM-based cloth solver should lead to consistent simulation results under different discretisation at high mesh resolutions. 
To investigate this, we perform an additional experiment where we increase the resolution of the ARCSim simulation ($10k$ vertices) so that the computation time roughly matches ours. %
However, the results still contain noticeable inconsistencies (\cref{fig:arcsim_inconsistency}-appendix), possibly due to several operations that are highly discretisation-dependent \cite{wang2011data} (such as the bending model relying on the dihedral angles).
In contrast, our method leads to consistent results already at much coarser discretisations (400 vertices, \cref{fig:initial_discrete_inconsistency_suppl}).%
We next evaluate 
the memory efficiency %
for simulations generated by NeuralClothSim, DiffARCSim and DiffCloth. %
The simulations are chosen to be of similar complexity, and qualitative results are visualised in Fig.~\ref{fig:resolution_inconsistency}-appendix.
In \cref{fig:memory_resolution}-appendix, we then plot the memory requirement as a function of spatial resolution. 
Memory is recorded for the simulated mesh states for the compared methods and weights of the NDF network for ours. %
While the memory requirement of finite-element-based methods grows linearly as the function of the number of vertices in the simulated cloth, our approach requires a constant and comparably small memory volume to store the quasistatic simulations. 
For better memory efficiency, existing simulators offer adaptive refinement (such as ARCSim~\cite{narain2012adaptive}) by re-meshing at each time step (coarse triangulation is used at smooth regions, and fine meshes are used for wrinkles). 
However, this requires additional computation and loss of important characteristics, such as differentiability. 
In contrast, our approach is adaptive without the overhead and without losing correspondence and differentiability due to re-meshing. 

\textit{PINNs for Shells.}
Bastek \textit{et al.}~\cite{bastek2023physics} focus on engineering scenarios, and 
we compare the solutions 
to the Scordelis-Lo roof and a square plate in Tab.~\ref{tab:belytschko}; both ours and theirs closely match the reference solutions.
Bastek \textit{et al.} note instabilities during training a neural network trained on a point load, therefore, define the Gaussian force kernel in their pinched hemisphere example. 
In contrast to theirs, we propose a new loss function for point loads (see App.~\ref{sec:obstacle_course_appendix}) addressing the pinched cylinder obstacle course. 
While Bastek \textit{et al.} show converged results on engineering examples, their method cannot capture the high-frequency signal (folds and wrinkles) required for cloth simulation; see \cref{fig:pinn_comparison} for an illustrative example.
The main reasons for their failure are 1) the linear strain and 2) that their activation function (GELU) can capture only smooth signals.

\subsection{Ablation and Applications} \label{sec:ablation} 
\textit{Ablation.}
We evaluate the following ablated versions of our approach: 1) Contravariant coordinate system for NDF components, 2) Using a linear approximation of the strains instead of our model, 3) Variants of boundary constraint imposition and 4) Choice of activation functions for NDF. 
For the latter two, we present the numerical results in Tab.~\ref{tab:belytschko}. 
See \cref{sec:appendix_ablations} for further details.

\textit{Material-conditioning.}
\label{ssec:material_conditioning}
NeuralClothSim can incorporate learnt priors: Our NDF 
can be directly extended by making it dependent on the material properties, \textit{i.e.,} it can accept the material parameters as an extra input. 
This is possible since the material parameter space is typically low-dimensional. 
Once such NeuralClothSim modification is trained, we can edit the simulated geometry at test time, as shown in Fig.~\ref{fig:teaser}-(bottom right). 
We provide implementation details of conditioning in \cref{sec:applications_suppl}.

\textit{Simulation editing.} 
\label{ssec:editing_simulation} 
For high-dimensional scene parameters such as reference pose and external forces, 
we can edit simulations: 
The user can interrupt the training of NDF at any point, change the parameters and continue the training. 
Moreover, 
editing can also be done after full convergence (aka pre-training) and then fine-tuned with gradually modified design parameters. 
Editing an NDF has multiple advantages over NDF training 
from scratch: It is computationally and memory efficient, and provides access to interpolated simulations.
We provide further details 
and results in \cref{sec:applications_suppl}.

\section{Discussion and Conclusion} \label{sec:conclusion}

NeuralClothSim closely matches reference values in challenging 
cloth 
deformation scenarios (e.g.~the Belytschko course), thanks to compact NDF representation governed by the non-linear Kirchhoff-Love shell theory with (non-)linear orthotropic material. 
An extended NDF allows test-time interpolation of material properties and simulation editing. 
In contrast to the previous mesh-based simulators, NeuralClothSim enables
querying continuous and consistent equilibrium
cloth states. 
The shown results are physically plausible in different scenarios 
under time-varying external forces and boundary motions. 
We also 
see multiple avenues for future research, such as adding dynamic effects, \ie~inertia and damping. 
Our simulator currently does not support  contacts and friction
necessary for many 
applications beyond what is demonstrated 
here 
(\textit{cf.}~Appendix~\ref{sec:limitations} on this standalone research problem). 

In conclusion, we see NeuralClothSim as an exciting 
step towards neural-field-based continuous and differentiable cloth simulation. 
Inverse problems in vision 
could benefit from its multi-resolution consistency. 
While there is a long way 
until 
other functionalities such as collision handling are unlocked, we believe it can pave the way towards 
a new generation of physics simulation engines.

{
    \small
    \bibliographystyle{ieeenat_fullname}
    \bibliography{main}

\begin{thebibliography}{68}
\providecommand{\natexlab}[1]{#1}
\providecommand{\url}[1]{\texttt{#1}}
\expandafter\ifx\csname urlstyle\endcsname\relax
  \providecommand{\doi}[1]{doi: #1}\else
  \providecommand{\doi}{doi: \begingroup \urlstyle{rm}\Url}\fi

\bibitem[Baraff and Witkin(1998)]{BaraffWitkin1998}
David Baraff and Andrew Witkin.
\newblock Large steps in cloth simulation.
\newblock In \emph{Annual Conference on Computer Graphics and Interactive
  Techniques}, 1998.

\bibitem[Barr(1984)]{Barr1984}
Alan~H. Barr.
\newblock Global and local deformations of solid primitives.
\newblock In \emph{Annual Conference on Computer Graphics and Interactive
  Techniques}, 1984.

\bibitem[Basar and Kr{\"a}tzig(2013)]{basar2013mechanik}
Yavuz Basar and Wilfried~B Kr{\"a}tzig.
\newblock \emph{Mechanik der Fl{\"a}chentragwerke: theorie,
  berechnungsmethoden, anwendungsbeispiele}.
\newblock Springer-Verlag, 2013.

\bibitem[Ba{\c{s}}ar et~al.(2000)Ba{\c{s}}ar, Itskov, and
  Eckstein]{bacsar2000composite}
Y Ba{\c{s}}ar, M Itskov, and A Eckstein.
\newblock Composite laminates: nonlinear interlaminar stress analysis by
  multi-layer shell elements.
\newblock \emph{Computer Methods in Applied Mechanics and Engineering}, 2000.

\bibitem[Bastek and Kochmann(2023)]{bastek2023physics}
Jan-Hendrik Bastek and Dennis~M Kochmann.
\newblock Physics-informed neural networks for shell structures.
\newblock \emph{European Journal of Mechanics-A/Solids}, 2023.

\bibitem[Belytschko et~al.(1985)Belytschko, Stolarski, Liu, Carpenter, and
  Ong]{belytschko1985stress}
Ted Belytschko, Henryk Stolarski, Wing~Kam Liu, Nicholas Carpenter, and Jame~SJ
  Ong.
\newblock Stress projection for membrane and shear locking in shell finite
  elements.
\newblock \emph{Computer Methods in Applied Mechanics and Engineering}, 1985.

\bibitem[Bertiche et~al.(2021)Bertiche, Madadi, and Escalera]{bertiche2021pbns}
Hugo Bertiche, Meysam Madadi, and Sergio Escalera.
\newblock Pbns: physically based neural simulation for unsupervised garment
  pose space deformation.
\newblock \emph{ACM Transactions on Graphics (TOG)}, 2021.

\bibitem[Bertiche et~al.(2022)Bertiche, Madadi, and
  Escalera]{bertiche2022neural}
Hugo Bertiche, Meysam Madadi, and Sergio Escalera.
\newblock Neural cloth simulation.
\newblock \emph{ACM Transactions on Graphics (TOG)}, 2022.

\bibitem[Bridson et~al.(2002)Bridson, Fedkiw, and Anderson]{bridson2002robust}
Robert Bridson, Ronald Fedkiw, and John Anderson.
\newblock Robust treatment of collisions, contact and friction for cloth
  animation.
\newblock In \emph{{ACM} Transactions on Graphics}, 2002.

\bibitem[Chen et~al.(2023)Chen, Wu, Grinspun, Zheng, and
  Chen]{chen2023implicit}
Honglin Chen, Rundi Wu, Eitan Grinspun, Changxi Zheng, and Peter~Yichen Chen.
\newblock Implicit neural spatial representations for time-dependent pdes.
\newblock In \emph{International Conference on Machine Learning (ICML)}, 2023.

\bibitem[Chen et~al.(2022)Chen, Xiang, Cho, Chang, Pershing, Maia, Chiaramonte,
  Carlberg, and Grinspun]{chen2022crom}
Peter~Yichen Chen, Jinxu Xiang, Dong~Heon Cho, Yue Chang, GA Pershing,
  Henrique~Teles Maia, Maurizio Chiaramonte, Kevin Carlberg, and Eitan
  Grinspun.
\newblock Crom: Continuous reduced-order modeling of pdes using implicit neural
  representations.
\newblock \emph{arXiv preprint arXiv:2206.02607}, 2022.

\bibitem[Chen et~al.(2018)Chen, Rubanova, Bettencourt, and
  Duvenaud]{chen2018neural}
Ricky T.~Q. Chen, Yulia Rubanova, Jesse Bettencourt, and David Duvenaud.
\newblock Neural ordinary differential equations.
\newblock \emph{Advances in Neural Information Processing Systems (NeurIPS)},
  2018.

\bibitem[Choi and Ko(2005)]{choi2005stable}
Kwang-Jin Choi and Hyeong-Seok Ko.
\newblock Stable but responsive cloth.
\newblock In \emph{ACM SIGGRAPH 2005 Courses}. 2005.

\bibitem[Cirak et~al.(2000)Cirak, Ortiz, and
  Schr{\"o}der]{cirak2000subdivision}
Fehmi Cirak, Michael Ortiz, and Peter Schr{\"o}der.
\newblock Subdivision surfaces: a new paradigm for thin-shell finite-element
  analysis.
\newblock \emph{Internat. J. Numer. Methods Engrg.}, 2000.

\bibitem[Clyde et~al.(2017{\natexlab{a}})Clyde, Teran, and
  Tamstorf]{clyde2017modeling}
David Clyde, Joseph Teran, and Rasmus Tamstorf.
\newblock Modeling and data-driven parameter estimation for woven fabrics.
\newblock In \emph{Proc. ACM SIGGRAPH / Eurographics Symposium on Computer
  Animation (SCA)}, 2017{\natexlab{a}}.

\bibitem[Clyde et~al.(2017{\natexlab{b}})Clyde, Teran, and
  Tamstorf]{clyde2017simulation}
David Clyde, Joseph Teran, and Rasmus Tamstorf.
\newblock Simulation of nonlinear kirchhoff-love thin shells using subdivision
  finite elements.
\newblock In \emph{Proc. ACM SIGGRAPH / Eurographics Symposium on Computer
  Animation (SCA)}, 2017{\natexlab{b}}.

\bibitem[Clyde(2017)]{clyde2017numerical}
David~Corwin Clyde.
\newblock \emph{Numerical Subdivision Surfaces for Simulation and Data Driven
  Modeling of Woven Cloth}.
\newblock University of California, Los Angeles, 2017.

\bibitem[Etzmu{\ss} et~al.(2003)Etzmu{\ss}, Keckeisen, and
  Stra{\ss}er]{etzmuss2003fast}
Olaf Etzmu{\ss}, Michael Keckeisen, and Wolfgang Stra{\ss}er.
\newblock A fast finite element solution for cloth modelling.
\newblock In \emph{Proc. of The Pacific Conference on Computer Graphics and
  Applications}, 2003.

\bibitem[Green et~al.(2002)Green, Turkiyyah, and Storti]{green2002subdivision}
Seth Green, George Turkiyyah, and Duane Storti.
\newblock Subdivision-based multilevel methods for large scale engineering
  simulation of thin shells.
\newblock In \emph{Proceedings of the seventh ACM symposium on Solid modeling
  and applications}, 2002.

\bibitem[Grinspun et~al.(2002)Grinspun, Krysl, and
  Schr{\"o}der]{grinspun2002charms}
Eitan Grinspun, Petr Krysl, and Peter Schr{\"o}der.
\newblock Charms: A simple framework for adaptive simulation.
\newblock \emph{{ACM} Transactions on Graphics}, 2002.

\bibitem[Grinspun et~al.(2003)Grinspun, Hirani, Desbrun, and
  Schr{\"o}der]{grinspun2003discrete}
Eitan Grinspun, Anil~N Hirani, Mathieu Desbrun, and Peter Schr{\"o}der.
\newblock Discrete shells.
\newblock In \emph{Proc. ACM SIGGRAPH / Eurographics Symposium on Computer
  Animation (SCA)}, 2003.

\bibitem[Grinspun et~al.(2006)Grinspun, Gingold, Reisman, and
  Zorin]{grinspun2006computing}
Eitan Grinspun, Yotam Gingold, Jason Reisman, and Denis Zorin.
\newblock Computing discrete shape operators on general meshes.
\newblock In \emph{Computer Graphics Forum}, 2006.

\bibitem[Guo et~al.(2021)Guo, Zhuang, and Rabczuk]{guo2021deep}
Hongwei Guo, Xiaoying Zhuang, and Timon Rabczuk.
\newblock A deep collocation method for the bending analysis of kirchhoff
  plate.
\newblock \emph{arXiv preprint arXiv:2102.02617}, 2021.

\bibitem[Guo et~al.(2018)Guo, Han, Fu, Gast, Tamstorf, and
  Teran]{guo2018material}
Qi Guo, Xuchen Han, Chuyuan Fu, Theodore Gast, Rasmus Tamstorf, and Joseph
  Teran.
\newblock A material point method for thin shells with frictional contact.
\newblock \emph{{ACM} Transactions on Graphics}, 2018.

\bibitem[Hao et~al.(2022)Hao, Liu, Zhang, Ying, Feng, Su, and
  Zhu]{hao2022physics}
Zhongkai Hao, Songming Liu, Yichi Zhang, Chengyang Ying, Yao Feng, Hang Su, and
  Jun Zhu.
\newblock Physics-informed machine learning: A survey on problems, methods and
  applications.
\newblock \emph{arXiv preprint arXiv:2211.08064}, 2022.

\bibitem[Harmon et~al.(2008)Harmon, Vouga, Tamstorf, and
  Grinspun]{harmon2008robust}
David Harmon, Etienne Vouga, Rasmus Tamstorf, and Eitan Grinspun.
\newblock Robust treatment of simultaneous collisions.
\newblock \emph{{ACM} Transactions on Graphics}, 2008.

\bibitem[Hendrycks and Gimpel(2016)]{hendrycks2016gaussian}
Dan Hendrycks and Kevin Gimpel.
\newblock Gaussian error linear units (gelus).
\newblock \emph{arXiv preprint arXiv:1606.08415}, 2016.

\bibitem[Kairanda et~al.(2022)Kairanda, Tretschk, Elgharib, Theobalt, and
  Golyanik]{kairanda2022f}
Navami Kairanda, Edith Tretschk, Mohamed Elgharib, Christian Theobalt, and
  Vladislav Golyanik.
\newblock f-sft: Shape-from-template with a physics-based deformation model.
\newblock In \emph{Computer Vision and Pattern Recognition (CVPR)}, 2022.

\bibitem[Kiendl et~al.(2015)Kiendl, Hsu, Wu, and Reali]{kiendl2015isogeometric}
Josef Kiendl, Ming-Chen Hsu, Michael~CH Wu, and Alessandro Reali.
\newblock Isogeometric kirchhoff--love shell formulations for general
  hyperelastic materials.
\newblock \emph{Computer Methods in Applied Mechanics and Engineering}, 2015.

\bibitem[Kingma and Ba(2014)]{kingma2014adam}
Diederik~P Kingma and Jimmy Ba.
\newblock Adam: A method for stochastic optimization.
\newblock \emph{arXiv preprint arXiv:1412.6980}, 2014.

\bibitem[Kopanicakova et~al.(2019)Kopanicakova, Krause, and
  Tamstorf]{kopanicakova2019subdivision}
Alena Kopanicakova, Rolf Krause, and Rasmus Tamstorf.
\newblock Subdivision-based nonlinear multiscale cloth simulation.
\newblock \emph{SIAM Journal on Scientific Computing}, 2019.

\bibitem[Li et~al.(2020)Li, Tang, Tong, Cai, Zhao, and Manocha]{li2020p}
Cheng Li, Min Tang, Ruofeng Tong, Ming Cai, Jieyi Zhao, and Dinesh Manocha.
\newblock P-cloth: interactive complex cloth simulation on multi-gpu systems
  using dynamic matrix assembly and pipelined implicit integrators.
\newblock \emph{{ACM} Transactions on Graphics}, 2020.

\bibitem[Li et~al.(2018)Li, Daviet, Narain, Bertails-Descoubes, Overby, Brown,
  and Boissieux]{li2018implicit}
Jie Li, Gilles Daviet, Rahul Narain, Florence Bertails-Descoubes, Matthew
  Overby, George~E Brown, and Laurence Boissieux.
\newblock An implicit frictional contact solver for adaptive cloth simulation.
\newblock \emph{{ACM} Transactions on Graphics}, 2018.

\bibitem[Li et~al.(2021)Li, Kaufman, and Jiang]{li2021codimensional}
Minchen Li, Danny~M Kaufman, and Chenfanfu Jiang.
\newblock Codimensional incremental potential contact.
\newblock \emph{{ACM} Transactions on Graphics}, 2021.

\bibitem[Li et~al.(2022{\natexlab{a}})Li, Qiao, Chen, Jatavallabhula, Lin,
  Jiang, and Gan]{li2022pac}
Xuan Li, Yi-Ling Qiao, Peter~Yichen Chen, Krishna~Murthy Jatavallabhula, Ming
  Lin, Chenfanfu Jiang, and Chuang Gan.
\newblock Pac-nerf: Physics augmented continuum neural radiance fields for
  geometry-agnostic system identification.
\newblock In \emph{International Conference on Learning Representations
  (ICLR)}, 2022{\natexlab{a}}.

\bibitem[Li et~al.(2022{\natexlab{b}})Li, Du, Wu, Xu, and
  Matusik]{li2022diffcloth}
Yifei Li, Tao Du, Kui Wu, Jie Xu, and Wojciech Matusik.
\newblock Diffcloth: Differentiable cloth simulation with dry frictional
  contact.
\newblock \emph{{ACM} Transactions on Graphics}, 2022{\natexlab{b}}.

\bibitem[Li et~al.(2023)Li, Chen, Larionov, Sarafianos, Matusik, and
  Stuyck]{li2023diffavatar}
Yifei Li, Hsiao-yu Chen, Egor Larionov, Nikolaos Sarafianos, Wojciech Matusik,
  and Tuur Stuyck.
\newblock Diffavatar: Simulation-ready garment optimization with differentiable
  simulation.
\newblock \emph{arXiv preprint arXiv:2311.12194}, 2023.

\bibitem[Liang et~al.(2019)Liang, Lin, and Koltun]{Liang2019}
Junbang Liang, Ming Lin, and Vladlen Koltun.
\newblock Differentiable cloth simulation for inverse problems.
\newblock In \emph{Advances in Neural Information Processing Systems
  (NeurIPS)}, 2019.

\bibitem[Love(2013)]{love2013treatise}
Augustus Edward~Hough Love.
\newblock \emph{A treatise on the mathematical theory of elasticity}.
\newblock Cambridge university press, 2013.

\bibitem[Lu and Zheng(2014)]{lu2014dynamic}
Jia Lu and Chao Zheng.
\newblock Dynamic cloth simulation by isogeometric analysis.
\newblock \emph{Computer Methods in Applied Mechanics and Engineering}, 2014.

\bibitem[Lu et~al.(2021)Lu, Pestourie, Yao, Wang, Verdugo, and
  Johnson]{lu2021physics}
Lu Lu, Raphael Pestourie, Wenjie Yao, Zhicheng Wang, Francesc Verdugo, and
  Steven~G Johnson.
\newblock Physics-informed neural networks with hard constraints for inverse
  design.
\newblock \emph{SIAM Journal on Scientific Computing}, 2021.

\bibitem[Ly et~al.(2020)Ly, Jouve, Boissieux, and
  Bertails-Descoubes]{ly2020projective}
Micka{\"e}l Ly, Jean Jouve, Laurence Boissieux, and Florence
  Bertails-Descoubes.
\newblock Projective dynamics with dry frictional contact.
\newblock \emph{ACM Transactions on Graphics (TOG)}, 2020.

\bibitem[Mildenhall et~al.(2020)Mildenhall, Srinivasan, Tancik, Barron,
  Ramamoorthi, and Ng]{mildenhall2020nerf}
Ben Mildenhall, Pratul~P Srinivasan, Matthew Tancik, Jonathan~T Barron, Ravi
  Ramamoorthi, and Ren Ng.
\newblock Nerf: Representing scenes as neural radiance fields for view
  synthesis.
\newblock In \emph{European Conference on Computer Vision (ECCV)}, 2020.

\bibitem[M{\"u}ller et~al.(2022)M{\"u}ller, Evans, Schied, and
  Keller]{muller2022instant}
Thomas M{\"u}ller, Alex Evans, Christoph Schied, and Alexander Keller.
\newblock Instant neural graphics primitives with a multiresolution hash
  encoding.
\newblock \emph{{ACM} Transactions on Graphics}, 2022.

\bibitem[Narain et~al.(2012)Narain, Samii, and O'brien]{narain2012adaptive}
Rahul Narain, Armin Samii, and James~F O'brien.
\newblock Adaptive anisotropic remeshing for cloth simulation.
\newblock \emph{{ACM} Transactions on Graphics}, 2012.

\bibitem[Otaduy et~al.(2009)Otaduy, Tamstorf, Steinemann, and
  Gross]{otaduy2009implicit}
Miguel~A Otaduy, Rasmus Tamstorf, Denis Steinemann, and Markus Gross.
\newblock Implicit contact handling for deformable objects.
\newblock In \emph{Compututer Graphics Forum}, 2009.

\bibitem[Paszke et~al.(2019)Paszke, Gross, Massa, Lerer, Bradbury, Chanan,
  Killeen, Lin, Gimelshein, Antiga, et~al.]{paszke2019pytorch}
Adam Paszke, Sam Gross, Francisco Massa, Adam Lerer, James Bradbury, Gregory
  Chanan, Trevor Killeen, Zeming Lin, Natalia Gimelshein, Luca Antiga, et~al.
\newblock Pytorch: An imperative style, high-performance deep learning library.
\newblock \emph{Advances in Neural Information Processing Systems (NeurIPS)},
  2019.

\bibitem[Pfaff et~al.(2021)Pfaff, Fortunato, Sanchez-Gonzalez, and
  Battaglia]{pfaff2021learning}
Tobias Pfaff, Meire Fortunato, Alvaro Sanchez-Gonzalez, and Peter Battaglia.
\newblock Learning mesh-based simulation with graph networks.
\newblock In \emph{International Conference on Learning Representations
  (ICLR)}, 2021.

\bibitem[Raissi et~al.(2019)Raissi, Perdikaris, and
  Karniadakis]{raissi2019physics}
Maziar Raissi, Paris Perdikaris, and George~E Karniadakis.
\newblock Physics-informed neural networks: A deep learning framework for
  solving forward and inverse problems involving nonlinear partial differential
  equations.
\newblock \emph{Journal of Computational Physics}, 2019.

\bibitem[Rao et~al.(2021)Rao, Sun, and Liu]{rao2021physics}
Chengping Rao, Hao Sun, and Yang Liu.
\newblock Physics-informed deep learning for computational elastodynamics
  without labeled data.
\newblock \emph{Journal of Engineering Mechanics}, 2021.

\bibitem[Santesteban et~al.(2022)Santesteban, Otaduy, and
  Casas]{santesteban2022snug}
Igor Santesteban, Miguel~A Otaduy, and Dan Casas.
\newblock {SNUG}: {S}elf-{S}upervised {N}eural {D}ynamic {G}arments.
\newblock \emph{Computer Vision and Pattern Recognition (CVPR)}, 2022.

\bibitem[Simo and Fox(1989)]{simo1989stress}
Juan~C Simo and David~D Fox.
\newblock On a stress resultant geometrically exact shell model. part i:
  Formulation and optimal parametrization.
\newblock \emph{Computer Methods in Applied Mechanics and Engineering}, 1989.

\bibitem[Sitzmann et~al.(2020)Sitzmann, Martel, Bergman, Lindell, and
  Wetzstein]{sitzmann2020implicit}
Vincent Sitzmann, Julien Martel, Alexander Bergman, David Lindell, and Gordon
  Wetzstein.
\newblock Implicit neural representations with periodic activation functions.
\newblock In \emph{Advances in Neural Information Processing Systems
  (NeurIPS)}, 2020.

\bibitem[Tang et~al.(2018)Tang, Liu, Tong, and Manocha]{tang2018pscc}
Min Tang, Zhongyuan Liu, Ruofeng Tong, and Dinesh Manocha.
\newblock Pscc: Parallel self-collision culling with spatial hashing on gpus.
\newblock \emph{Proceedings of the ACM on Computer Graphics and Interactive
  Techniques}, 2018.

\bibitem[Terzopoulos et~al.(1987)Terzopoulos, Platt, Barr, and
  Fleischer]{Terzopoulos1987}
Demetri Terzopoulos, John Platt, Alan Barr, and Kurt Fleischer.
\newblock Elastically deformable models.
\newblock \emph{SIGGRAPH Comput. Graph.}, 1987.

\bibitem[Thomaszewski et~al.(2006)Thomaszewski, Wacker, and
  Stra{\ss}er]{thomaszewski2006consistent}
Bernhard Thomaszewski, Markus Wacker, and Wolfgang Stra{\ss}er.
\newblock A consistent bending model for cloth simulation with corotational
  subdivision finite elements.
\newblock In \emph{Proc. ACM SIGGRAPH / Eurographics Symposium on Computer
  Animation (SCA)}, 2006.

\bibitem[Timoshenko et~al.(1959)Timoshenko, Woinowsky-Krieger,
  et~al.]{timoshenko1959theory}
Stephen Timoshenko, Sergius Woinowsky-Krieger, et~al.
\newblock \emph{Theory of plates and shells}.
\newblock McGraw-hill New York, 1959.

\bibitem[Tretschk et~al.(2021)Tretschk, Tewari, Golyanik, Zollh\"{o}fer,
  Lassner, and Theobalt]{tretschk2021nonrigid}
Edgar Tretschk, Ayush Tewari, Vladislav Golyanik, Michael Zollh\"{o}fer,
  Christoph Lassner, and Christian Theobalt.
\newblock Non-rigid neural radiance fields: Reconstruction and novel view
  synthesis of a dynamic scene from monocular video.
\newblock In \emph{{IEEE} International Conference on Computer Vision
  ({ICCV})}. {IEEE}, 2021.

\bibitem[Volino and Thalmann(2000)]{volino2000implementing}
Pascal Volino and N~Magnenat Thalmann.
\newblock Implementing fast cloth simulation with collision response.
\newblock In \emph{Proceedings Computer Graphics International 2000}, 2000.

\bibitem[Wang(2021)]{wang2021gpu}
Huamin Wang.
\newblock Gpu-based simulation of cloth wrinkles at submillimeter levels.
\newblock \emph{{ACM} Transactions on Graphics}, 2021.

\bibitem[Wang et~al.(2011)Wang, O'Brien, and Ramamoorthi]{wang2011data}
Huamin Wang, James~F O'Brien, and Ravi Ramamoorthi.
\newblock Data-driven elastic models for cloth: modeling and measurement.
\newblock \emph{{ACM} Transactions on Graphics}, 2011.

\bibitem[Wang et~al.(2021)Wang, Liu, Liu, Theobalt, Komura, and
  Wang]{wang2021neus}
Peng Wang, Lingjie Liu, Yuan Liu, Christian Theobalt, Taku Komura, and Wenping
  Wang.
\newblock Neus: Learning neural implicit surfaces by volume rendering for
  multi-view reconstruction.
\newblock \emph{NeurIPS}, 2021.

\bibitem[Wempner and Talaslidis(2003)]{wempner2003mechanics}
Gerald Wempner and Demosthenes Talaslidis.
\newblock Mechanics of solids and shells.
\newblock \emph{CRC, Boca Raton}, 2003.

\bibitem[Xie et~al.(2022)Xie, Takikawa, Saito, Litany, Yan, Khan, Tombari,
  Tompkin, Sitzmann, and Sridhar]{xie2022neural}
Yiheng Xie, Towaki Takikawa, Shunsuke Saito, Or Litany, Shiqin Yan, Numair
  Khan, Federico Tombari, James Tompkin, Vincent Sitzmann, and Srinath Sridhar.
\newblock Neural fields in visual computing and beyond.
\newblock In \emph{Computer Graphics Forum (Eurographics State of the Art
  Reports)}, 2022.

\bibitem[Yang et~al.(2021)Yang, Belongie, Hariharan, and
  Koltun]{yang2021geometry}
Guandao Yang, Serge Belongie, Bharath Hariharan, and Vladlen Koltun.
\newblock Geometry processing with neural fields.
\newblock \emph{Advances in Neural Information Processing Systems (NeurIPS)},
  2021.

\bibitem[Zehnder et~al.(2021)Zehnder, Li, Coros, and
  Thomaszewski]{zehnder2021ntopo}
Jonas Zehnder, Yue Li, Stelian Coros, and Bernhard Thomaszewski.
\newblock Ntopo: Mesh-free topology optimization using implicit neural
  representations.
\newblock \emph{Advances in Neural Information Processing Systems}, 2021.

\bibitem[Zhang et~al.(2022)Zhang, Dumas, Fei, Jacobson, James, and
  Kaufman]{zhang2022progressive}
Jiayi~Eris Zhang, J{\'e}r{\'e}mie Dumas, Yun Fei, Alec Jacobson, Doug~L James,
  and Danny~M Kaufman.
\newblock Progressive simulation for cloth quasistatics.
\newblock \emph{{ACM} Transactions on Graphics}, 2022.

\bibitem[Zhang et~al.(2023)Zhang, Dumas, Fei, Jacobson, James, and
  Kaufman]{zhang2023progressive}
Jiayi~Eris Zhang, J{\'e}r{\'e}mie Dumas, Yun Fei, Alec Jacobson, Doug~L James,
  and Danny~M Kaufman.
\newblock Progressive shell qasistatics for unstructured meshes.
\newblock \emph{ACM Transactions on Graphics (TOG)}, 2023.

\end{thebibliography}
}

\clearpage
\appendix
\section*{\centering NeuralClothSim: Neural Deformation Fields Meet the Thin Shell Theory ---Appendices---}
\begin{center}
    \textbf{Navami Kairanda ~~~ Marc Habermann ~~~ Christian Theobalt ~~~ Vladislav Golyanik} \\ 
    \smallskip 
    Max Planck Institute for Informatics, Saarland Informatics Campus
\end{center} 
\setcounter{figure}{0}
\renewcommand*\thetable{\Roman{table}}
\renewcommand*\thefigure{\Roman{figure}}

\renewcommand{\contentsname}{}
\vspace{5pt}
\hspace{12pt}{\large Table of Contents}\vspace{-37pt}

\vspace{15mm}
\startcontents[sections]
\printcontents[sections]{l}{1}{\setcounter{tocdepth}{2}}
\vspace{5mm}

Sections referenced with numbers refer to the main matter. 
All referenced figures and equations are per default from this document, unless they are followed by the ``(main matter)'' mark. %
\section{Implementation Details}\label{sec:implementation_details}

We implement NeuralClothSim in PyTorch~\cite{paszke2019pytorch} and compute the geometric quantities on the reference shape and on the NDF using its tensor operations; 
the first and second-order derivatives are calculated using automatic differentiation. 
Our network architecture for NDF is an MLP with sine activations (SIREN)~\cite{sitzmann2020implicit} with five hidden layers and $512$ units in each layer. 
We empirically set SIREN's frequency parameter to $\omega_0=30$ for all experiments (we observed that choosing $\omega_0=1$ does not permit folds). 
Although we sample from $(\xi^1, \xi^2) \in \Omega,  t \in [0,T], T=1$, we normalise samples to $(\xi^1, \xi^2, t) \in [0,1]^3$ when feeding the input to MLP as per the initialisation principle of SIREN. 
Note that all physical quantities are computed in the original domains $\Omega, [0,T]$ and the gradients are tracked in their scaled versions. 
For training, we use $N_\Omega = 20{\times}20$ and $N_t = 20$. At test time, we sample much higher for visualisation, usually with $N_\Omega = 100{\times}100$ and $N_t = 30$. 
For material conditioning, we use a single random material sample per training iteration. 
NeuralClothSim's training time amounts to ${\sim}10-30$ minutes for most experiments, and the number of training iterations equals ${\sim}2000-5000$. 
We use ADAM~\cite{kingma2014adam} optimiser with a learning rate of $10^{-4}$ and run our simulator on a single NVIDIA Quadro RTX 8000 GPU with 48 GB of global memory. 

\section{Kirchhoff-Love Thin Shell Theory} \label{sec:background_suppl} 

\begin{table}
 \caption
    {
    \textbf{Notations.}
    We omit separation of quantities in undeformed (overbar, \eg $\bar{\mathbf{x}}$) and deformed configurations.
    Moreover, we list the tensors with their covariant components but omit the contravariant and mixed variant versions. Instead of subscripts (covariant components), they are represented with superscripts or a mix of superscripts and subscripts.}
\begin{center}
\resizebox{\columnwidth}{!}{
\begin{tabular}{llll}
    
    \toprule
    \textbf{Symbol}  & \textbf{Description} & \textbf{Symbol}  & \textbf{Description}  \\
    
    \midrule
    $  \xi^\alpha \in \Omega, \boldsymbol\xi = (\xi^1, \xi^2)    $      & Curvilinear coordinates    & $t\in [0,T]$           &         Time   \\              
    $\xi^3   \in [-\frac{h}{2}, \frac{h}{2}], h \in \mathbb{R}     $ &  Thickness coordinate   & $\dot{\mathbf{u}}    : \Omega \times [0,T] \to \mathbb{R}^3 $ &  Velocity      \\
    
    $\mathbf{x}  : \Omega \to \mathbb{R}^3   $           &      Midsurface  representation  & $\mathcal{I} : [0,T] \to  \mathbb{R}$                &            Initial distance function                                          \\ %
    $\mathbf{a}_\alpha  : \Omega \to \mathbb{R}^3 $           &      Midsurface tangent vectors  & $\mathcal{B} : \Omega  \to  \mathbb{R}$                &          Boundary condition              \\ %
    $\mathbf{a}_3 : \Omega \to \mathbb{R}^3 $                &    Unit normal to midsurface    &  $\mathcal{F}_\Theta : \Omega  \to  \mathbb{R}^3$                &         Neural deformation field (NDF)                       \\ 
    $a_{\alpha \beta} : \Omega \to \mathbb{R} $                &       Metric tensor on midsurface       &$\rho \in  \mathbb{R}$                &    Mass density                                        \\ %
    $b_{\alpha \beta} : \Omega \to \mathbb{R} $                &       Curvature tensor on midsurface    & $E \in  \mathbb{R}$                &    Young's modulus                    \\ %

    $\mathbf{r}	    : \Omega \times [-\frac{h}{2}, \frac{h}{2}] \to \mathbb{R}^3 $ &      Shell  representation  & $\nu \in  \mathbb{R}$                &    Poisson's ratio                        \\ 
    $\mathbf{g}_i   : \Omega \times [-\frac{h}{2}, \frac{h}{2}] \to \mathbb{R}^3  $           &     Tangent base vectors on shell    & $k_{11}, k_{12}, k_{22}, G_{12} $ & Infinitesimal strain parameters                     \\
    $g_{i j} : \Omega \times [-\frac{h}{2}, \frac{h}{2}] \to \mathbb{R} $                &          Metric tensor on shell   & $\mu_{ji}, \alpha_{ji}, d_j$   & Nonlinear material response             \\
    $\mathbf{f}   : \Omega \to  \mathbb{R}^3$   &        External force     & $\mathbf{d}_1, \mathbf{d}_2 $&     Warp/weft material directions             \\ 
    $\mathbf{u} : \Omega \to \mathbb{R}^3     $       &      Midsurface deformation & $\tilde{E}_{\alpha \beta} :  \Omega \times [-\frac{h}{2}, \frac{h}{2}] \to \mathbb{R} $                &     Orthotropic strain            \\ 
    $\mathbf{\tilde{u}} : \Omega \times [-\frac{h}{2}, \frac{h}{2}] \to \mathbb{R}^3     $       &      Shell deformation    &$v_{\alpha}|_\beta :  \Omega \to \mathbb{R} $                &         Covariant derivative          \\ 
    $\mathbf{w} : \Omega \to \mathbb{R}^3     $       &      Midsurface orientation change    & $\Gamma_{\alpha \beta}^\lambda :  \Omega \to \mathbb{R} $                &    Christoffel symbol         \\ 
    $ \varphi_{\alpha \beta}:  \Omega \to  \mathbb{R}$ &  Deformation gradients  & $ \mathcal{E}:\Omega \to  \mathbb{R}$                &       Potential energy    \\
    $E_{i j} :  \Omega \times [-\frac{h}{2}, \frac{h}{2}] \to \mathbb{R} $                &     Green-Lagrange strain  & $\Psi :  \Omega \to \mathbb{R} $                &          Hyperelastic strain energy    \\
    $\varepsilon_{\alpha \beta} : \Omega \to \mathbb{R}   $   & Membrane strain & $\kappa_{\alpha \beta} : \Omega \to \mathbb{R} $                &      Bending strain           \\ %

    \bottomrule
    \end{tabular}
}
    \end{center}
   
    \label{tab:notation_supplemental}
     \vspace{-1mm}
\end{table}

In this section, we briefly review the Kirchhoff-Love thin shell theory following \cite{cirak2000subdivision, kiendl2015isogeometric}; detailed treatment of the subject can be found in \cite{basar2013mechanik}.
We already introduced physical and mathematical notations in Sec.~\ref{ssec:notation}-(main matter). 
A detailed list of notations can be found in Tab.~\ref{tab:notation_supplemental}.
We next present concepts from the differential geometry of surfaces to explain the midsurface and director (\cref{subsec:preliminaries}).
We then follow with the shell parameterisation and computation of strain measures on and off the midsurface (\cref{ssec:shell_kinematics}). 
Further, we present the hyperelastic material models that relate the strains to the internal stress (\cref{sec:material_model}) and finally review the energy principles for equilibrium deformation (\cref{ssec:equilibrium_deformation}). 
Moreover, we provide a proof of the simplified strain formulation in \cref{sec:strain_computation} as well as additional results from tensor algebra that are relevant for the computations (\cref{sec:tensor_algebra}).
\subsection{Geometric Preliminaries} \label{subsec:preliminaries}
In Kirchhoff-Love shell theory, the shell midsurface completely determines the strain components throughout the thickness. 
Therefore, we review those aspects of the differential geometry of surfaces that are essential for understanding the shell theory. 
\par 
Let us represent the midsurface as a 2D manifold in the 3D space, as shown in Fig.~\ref{fig:background_supplemental}. 
It can be described by a smooth map, $\mathbf{x}:\Omega \subset \mathbb{R}^2 \to \mathbb{R}^3$ on the parametric domain $ \Omega $. 
Any position $\mathbf{x}(\xi^1, \xi^2)$ on the surface is uniquely identified using the convective curvilinear coordinates $(\xi^1, \xi^2) \in \Omega$.
As positions can be specified using Cartesian coordinates $\mathbf{x} = x_i \mathbf{e}_i$, it follows that the invertible maps $x_i = x_i(\xi^1, \xi^2)$ and $\xi^ \alpha = \xi^\alpha(x_1, x_2, x_3)$ exist.
We define a local covariant basis to conveniently express local quantities on the surface.
Such a basis is constructed using $\mathbf{a}_\alpha$, the set of two vectors tangential to the curvilinear coordinate lines $\xi^\alpha$: 
\begin{align}\label{eq:covariant_basis} 
    \mathbf{a}_\alpha := \mathbf{x,}_\alpha. 
\end{align} 
\par 
To measure the distortion of length and angles, we compute the covariant components of the symmetric metric tensor (also known as the first fundamental form): 
\begin{align}
a_{\alpha \beta} = a_{\beta \alpha} := \mathbf{a}_\alpha \cdot \mathbf{a}_\beta. 
\end{align}
The corresponding contravariant components of the surface metric tensors denoted by $a^{\alpha \lambda}$ can be obtained using the following identity: 
\begin{align}
a^{\alpha \lambda} a_{\lambda \beta} = \delta_{\alpha \beta}, 
\end{align} 
where $\delta_{\alpha \beta}$ stands for the Kronecker delta. 
$a^{\alpha \lambda}$ can be used to compute the contravariant basis 
defined as $\mathbf{a}^\alpha \cdot \mathbf{a}_\beta = \delta_{\alpha \beta}$, as follows: 
\begin{align}
\mathbf{a}^\alpha = a^{\alpha \lambda} \mathbf{a}_\lambda. 
\end{align}
While the covariant base vector $\mathbf{a}_\alpha$ is tangent to the $\xi^\alpha$ line, the contravariant base vector $\mathbf{a}^\alpha$ is normal to $\mathbf{a}_\beta$ when $\alpha \neq \beta$. 
Generally, $\mathbf{a}_\alpha$ and $\mathbf{a}^\alpha$ need not be unit vectors. 
\par
The shell director coincides with $\mathbf{a}_3$, the unit normal to the midsurface, and, therefore, computed as the cross product of the tangent base vectors: 
\begin{align}
\mathbf{a}_3 := \frac{\mathbf{a}_1 \times \mathbf{a}_2}{|\mathbf{a}_1 \times \mathbf{a}_2 |},\;\;\mathbf{a}^3 = \mathbf{a}_3.  
\end{align} 
The second fundamental form---which measures the curvature of the midsurface---can be defined with $\mathbf{a}_3$ as: 
\begin{align}\label{eq:second_ff} 
b_{\alpha \beta} &:= -\mathbf{a}_\alpha \cdot \mathbf{a}_{3,\beta} = -\mathbf{a}_\beta \cdot \mathbf{a}_{3,\alpha} =  \mathbf{a}_{\alpha, \beta} \cdot \mathbf{a}_3. 
\end{align}
Finally, the surface area differential $\,d\Omega$ relates to the reference coordinates via the determinant of the metric tensor: 
\begin{align}
    \,d\Omega = \sqrt{a} \,d \xi^1 \,d \xi^2,\;\text{where}\;\sqrt{a} := |\mathbf{a}_1 \times \mathbf{a}_2 |.
\end{align}
\subsection{Kirchhoff-Love Shell Kinematics} \label{ssec:shell_kinematics}
\begin{figure}
\centering
	\includegraphics[width=\linewidth]{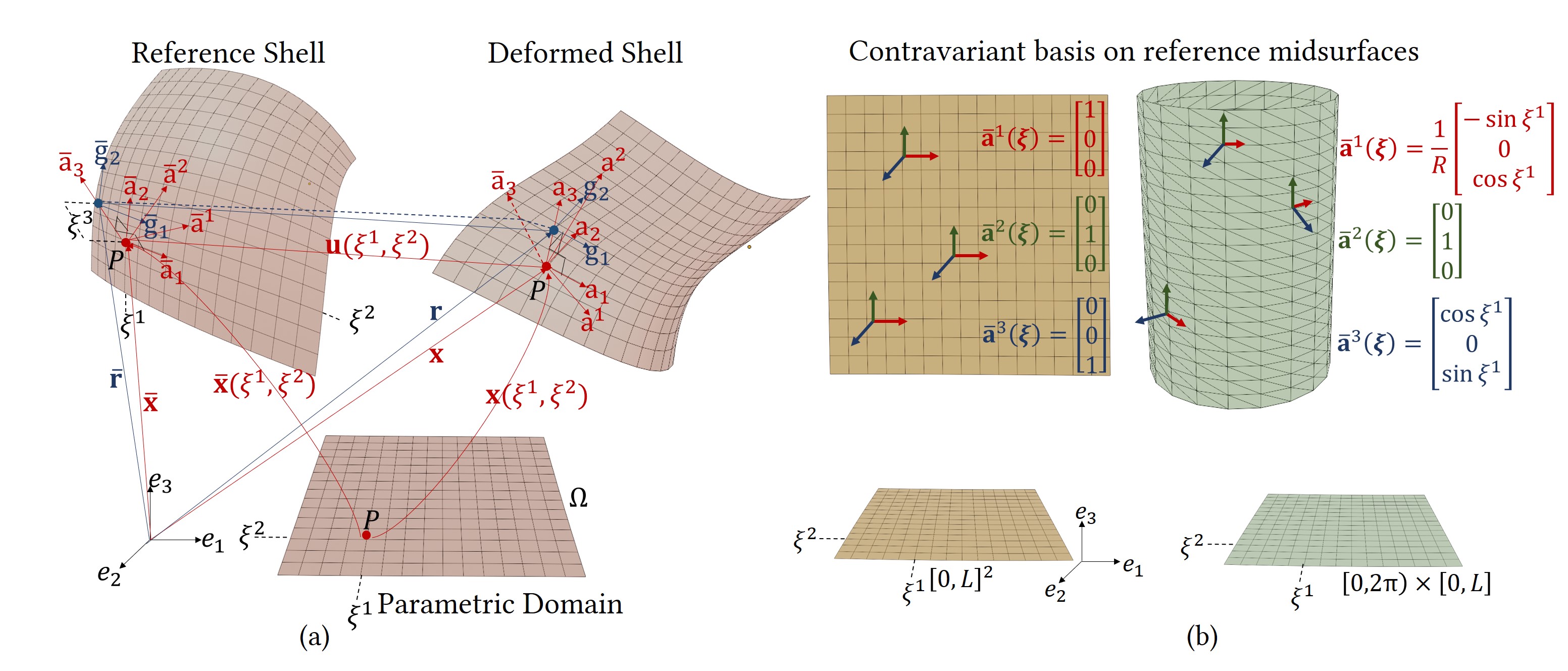}
	\caption
	{
        \textbf{(a) Kirchhoff-Love thin shell. }%
        A thin shell can be kinematically described by the midsurface 
        (here: reference and deformed midsurfaces) %
        and the
        director (here, $\mathbf{\bar{a}}_3$). %
        Any material point $P$ on the midsurface is then parameterised with curvilinear coordinates $(\xi^1, \xi^2)$, whereas a point on the shell continuum requires an additional thickness coordinate $\xi^3$.
        Geometric quantities on the midsurface (off the midsurface or on the shell continuum) are coloured red (blue). 
        \textbf{(b) Contravariant basis for midsurfaces in the reference configuration.} 
        While a local contravariant basis coincides with the global Cartesian coordinate system for a planar reference shell, such a basis varies in magnitude and direction across any circular section of the cylinder.
        Local basis relies on the surface parameterisation, therefore the derived basis vectors need not be normalised (notice how $\bar{\mathbf{a}}^1(\boldsymbol{\xi})$ scales inversely with the radius).
	}
	\label{fig:background_supplemental}
        \vspace{-1mm}
\end{figure}

The Kirchhoff-Love model proposes a reduced kinematic parameterisation of a thin shell characterised by a 2D midsurface and shell director. 
It relies on the Kirchhoff hypothesis, \textit{i.e.,} the director initially perpendicular to the midsurface remains straight and normal, and the shell thickness $h \in \mathbb{R}$ does not change with deformation; see \cref{fig:background_supplemental}. %
\par 
The position vector $\mathbf{\bar{r}}$ of a material point in the reference configuration of the shell continuum can be parametrised with curvilinear coordinates $(\xi^1, \xi^2)$ and thickness coordinate $\xi^3$ as: 
\begin{align}
   \mathbf{\bar{r}}(\xi^1, \xi^2, \xi^3) = \mathbf{\bar{x}}(\xi^1, \xi^2) + \xi^3 \mathbf{\bar{a}_3}(\xi^1, \xi^2),\;\text{s.t.}\;& -\frac{h}{2} \leq  \xi^3 \leq \frac{h}{2},  
\end{align}
where $\mathbf{\bar{x}}(\xi^1, \xi^2)$ represents the midsurface.
\par 
The shell adopts a deformed configuration under the action of applied forces $\mathbf{f}$.
Analogously, the deformed position vector $\mathbf{r}$ is represented as
\begin{align}
    \mathbf{r}(\xi^1, \xi^2, \xi^3) = \mathbf{x}(\xi^1, \xi^2) + \xi^3 \mathbf{a}_3(\xi^1, \xi^2),\;\text{s.t.}\;& -\frac{h}{2} \leq  \xi^3 \leq \frac{h}{2},  
\end{align}
where the deformed director $\mathbf{a}_3$ coincides with the unit normal.
\par
As a consequence, the overall deformation of the Kirchhoff-Love shell is fully described by the displacement field $\mathbf{u}(\xi^1, \xi^2)$ of the midsurface, \textit{i.e.,} 
\begin{align}
    \mathbf{x}(\xi^1, \xi^2) = \mathbf{\bar{x}}(\xi^1, \xi^2) + \mathbf{u}(\xi^1, \xi^2). 
\end{align}
Analogous to the deformation field $\mathbf{u} := \mathbf{x} - \bar{\mathbf{x}}$ of the midsurface, we define $\mathbf{w}$ as the difference vector of unit normals to the midsurface, \textit{i.e.,}
\begin{align}
\mathbf{w} := \mathbf{a}_3 - \bar{\mathbf{a}}_3 = w_\lambda \mathbf{\bar a}^\lambda + w_3 \mathbf{\bar a}^3 = w^\lambda \mathbf{\bar a}_\lambda + w^3 \mathbf{\bar a}_3.    
\label{eq:difference_vector}
\end{align}
Using this formulation, deformations on the shell continuum can be described by the field $\mathbf{\tilde{u}}(\xi^1, \xi^2, \xi^3) = \mathbf{u}(\xi^1, \xi^2) + \xi^3 \mathbf{w}(\xi^1, \xi^2)$. 
The difference vector $\mathbf{w}$ describes the change in the orientation of the midsurface, enabling us to quantify bending. 
A simplified way to compute the components $w_i$ of $\mathbf{w}$ is provided in \eqref{eq:difference_vector_components}.

The tangent base vectors at a point on the shell continuum are denoted by $\mathbf{g}_i := \mathbf{r,}_i$ and expressed by those of the midsurface as: 
\begin{align}
\begin{split}
    \mathbf{g}_\alpha &= \mathbf{a}_\alpha + \xi^3  \mathbf{a}_{3,\alpha}, \\
    \mathbf{g}_3 &= \mathbf{a}_3. 
    \label{eq:g_basis_vector}
\end{split}
\end{align}
The corresponding covariant components of the metric tensor are then obtained using 
\begin{align}
g_{ij} := \mathbf{g}_i \cdot \mathbf{g}_j. 
\label{eq:g_metric_tensor}
\end{align} 
To measure strain, we use the symmetric Green-Lagrange strain tensor $\mathbf{E} = E_{i j} \bar{\mathbf{g}}^i \otimes \bar{\mathbf{g}}^j$, since it discards the rotational degrees of freedom from tangent base vector $\mathbf{g}_i$ while retaining the stretch and shear information. 
It is defined as the difference between the metric tensors on the deformed and undeformed configurations of the shell, \textit{i.e.,} 
\begin{align}
    E_{i j} := \frac{1}{2} (g_{i j} - \bar{g}_{i j}).
    \label{eq:green_lagrange_strain}
\end{align}
Using (\ref{eq:g_basis_vector}) and (\ref{eq:g_metric_tensor}), note that transverse shear strain measuring the shearing of the director vanishes ($E_{\alpha 3} = 0$) and the stretching of the director is identity, \textit{i.e.,} $E_{3 3} = 1$; hence, the strain simplifies to 
\begin{align}
    E_{\alpha \beta} =  \varepsilon_{\alpha \beta} + \xi^3 \kappa_{\alpha \beta}, 
\end{align}
with membrane strain measuring the in-plane stretching defined as
\begin{align}
    \varepsilon_{\alpha \beta} := \frac{1}{2} (a_{\alpha \beta} - \bar{a}_{\alpha \beta}), 
    \label{eq:membrane_strain}
\end{align}
and bending strain measuring the change in curvature defined as 
\begin{align}
    \kappa_{\alpha \beta} := \bar{b}_{\alpha \beta} - b_{\alpha \beta}. 
    \label{eq:bending_strain}
\end{align}
\subsection{Material Elasticity Model}
\label{sec:material_model}
\begin{figure}
\centering
\includegraphics[width=\linewidth]{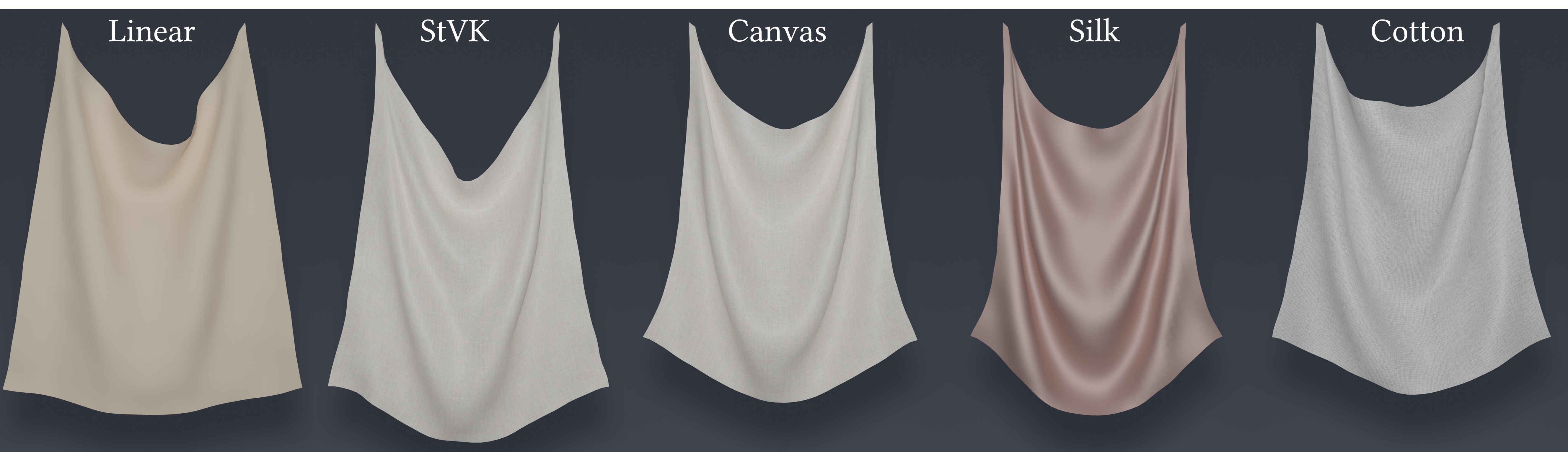} %
  \caption{
  \textbf{Material model.}
  Simulation of stable equilibria of $\qty{1}{\metre}\times\qty{1}{\metre}$ napkin with corners held $\qty{60}{\cm}$ apart. %
  From left to right, we visualise linear isotropic, linear anisotropic St.Venant-Kirchhoff (canvas), and non-linear anisotropic canvas, silk and cotton materials from \citet{clyde2017modeling}.
  }
  \label{fig:material_model}
\end{figure}
Our NeuralClothSim is orthogonal to the research on material modelling and can, thus, be formulated with many different elastic behaviours.
We demonstrate results with a simple linear isotropic model~\cite{cirak2000subdivision}, and the data-driven anisotropic non-linear model from Clyde \etal~ \cite{clyde2017modeling}, as well as the St Venant-Kirchhoff variant of the Clyde's model. 
\par
\textit{Linear Isotropic Material.}
Given the material Young's modulus $E$, Poisson's ratio $\nu$, a linear isotropic stress-strain relationship leads to hyperelastic strain energy density \cite{simo1989stress} of the form 
\begin{align}
    \Psi = \frac{1}{2} (D H^{\alpha \beta \lambda \delta} \varepsilon_{\alpha \beta} \varepsilon_{\lambda \delta} + B H^{\alpha \beta \lambda \delta} \kappa_{\alpha \beta} \kappa_{\lambda \delta}),
\end{align}
where $D$ is the in-plane stiffness and $B$ is the bending stiffness computed as 
\begin{align}
    D := \frac{Eh}{1 - \nu^2} \;\text{and}\; B := \frac{Eh^3}{12 (1 - \nu^2)}, 
\end{align}
and
\begin{align}
    H^{\alpha \beta \lambda \delta} := \nu \bar{a}^{\alpha \beta} \bar{a}^{\lambda \delta} + \frac{1}{2} (1-\nu) (\bar{a}^{\alpha \lambda} \bar{a}^{\beta \delta} + \bar{a}^{\alpha \delta} \bar{a}^{\beta \lambda}).
    \label{eq:H}
\end{align}
Here, $\Psi$ is the sum of the membrane strain energy density (the first term) and the bending strain energy density (the second term). 
\par
\textit{Non-linear Orthotropic Material.}
While the linear isotropic model is simple and  sufficient to demonstrate our NeuralClothSim formulation, 
a data-driven model with estimated fabric material parameters can generate highly realistic cloth simulations. 
Therefore, we additionally demonstrate our method with the non-linear anisotropic material model from Clyde~\etal~\cite{clyde2017modeling}, and its simplification to St. Venant-Kirchhoff model~\cite{bacsar2000composite}.
We show the simulation results with the material model with varying materials, such as cotton and silk in \cref{fig:material_model}, and describe the model next.
\par
Clyde \etal present an orthotropic constitutive model that accurately represents the anisotropy introduced by the warp and weft structure of woven cloth.
More concretely, they write the hyperelastic strain energy density as $\Psi (\mathbf{E}, \mathbf{D}, \boldsymbol{\Phi})$ 
where $\mathbf{E}$ is the Green-Lagrange strain~\eqref{eq:green_lagrange_strain}, 
$\mathbf{D}=[\mathbf{d}_1, \mathbf{d}_2, \mathbf{d}_3]$ is the reference configuration warp/weft orthotropy ($\mathbf{d}_1, \mathbf{d}_2$) and normal ($\mathbf{d}_3$) directions, 
and $\boldsymbol{\Phi}$ being the fabric parameters. 
We follow the technique of \cite{bacsar2000composite} to determine the material directions $\mathbf{D}$. 
Orthotropy directions are computed as tangents to the midsurface with the warp direction $\mathbf{d}_1$ coinciding with the normalised covariant base vector, \ie, 
\begin{equation}
   \mathbf{d}_1 =\frac{\mathbf{\bar{a}}_1}{\left\|\mathbf{\bar{a}}_1\right\|}, \mathbf{d}_3 = \mathbf{\bar{a}}_3, \text{ and } \mathbf{d}_2 = \mathbf{d}_3 \times \mathbf{d}_1.
\end{equation}
\par
Next, the orthotropic components $\tilde{E}_{ij}$ of the strain are obtained by expressing $\mathbf{E}$ in the material basis, with
$\tilde{\mathbf{E}} = \mathbf{D}^\top \mathbf{E} \mathbf{D}$. 
Due to the Kirchhoff-Love kinematic assumptions, any stretches and shears in the out-of-plane direction $\mathbf{d}_3$ vanish, \ie $\tilde{E}_{i 3} = \tilde{E}_{3 i} = 0$.
Finally, the Clyde model's strain energy density intuitively separates the distinct deformation modes and is defined as,
\begin{align}
    \Psi = \frac{k_{11}}{2} \eta_1(\tilde{E}_{11}^2) + k_{12} \eta_2(\tilde{E}_{11}\tilde{E}_{22}) + \frac{k_{22}}{2} \eta_3(\tilde{E}_{22}^2) + G_{12}\eta_4(\tilde{E}_{12}^2),
\end{align}

where $\{k_{11},k_{12},k_{22}, G_{12} \}$ describe the cloth's infinitesimal (linear) strain behaviour, whereas the function $\eta_j$ describes the nonlinear response to larger strains with,
\begin{align}
    \eta_j(x) = \sum_{i=1}^{d_j} \frac{\mu_{ji}}{\alpha_{ji}} ((x+1)^{\alpha_{ji}}-1).
    \label{eq:non_linear_strain_response}
\end{align}
We obtain the values for material parameters $\boldsymbol{\Phi} = \{k_{11},k_{12},k_{22}, G_{12}, \mu_{ji}, \alpha_{ji}, d_j \}_{i=1, j \in [1,...,4]}^{d_j} $ from \cite{clyde2017numerical} and model the silk, canvas and cotton fabrics.
Additionally, we can arrive at the orthotropic (linear) St. Venant-Kirchhoff model \cite{bacsar2000composite} by simply choosing $\eta_j(x) = x $ for all $j$ (see ~\cref{fig:material_model}).
\par
When optimising the NDF, the non-linear response \eqref{eq:non_linear_strain_response} gives unpredictable results for strains outside the fitting dataset. 
For a reasonable strain extrapolation, we use quadratic Taylor expansion around the closest valid strain, as proposed in \cite{clyde2017numerical} (see \cite{li2021codimensional}-supplement for the derivatives).  
Towards this, we leverage the strain cutoffs $\tilde{E}_{\alpha \beta}^{\mathrm{min}}$, and $\tilde{E}_{\alpha \beta}^{\mathrm{max}}$ provided as part of the material dataset. 

\subsection{Equilibrium Deformation} \label{ssec:equilibrium_deformation}
Under the action of external forces and boundary conditions, a thin shell deforms and achieves an equilibrium configuration. 
Its stable equilibrium state is characterised by the principle of minimum potential energy, which is the sum of external potential energy owing to applied forces and internal potential energy due to material elasticity.

While all the geometric quantities in (\ref{eq:covariant_basis})--(\ref{eq:H}) are defined at each material point $(\xi^1, \xi^2) \in \Omega$, the energy is integrated over the parametric domain $\Omega$. 
Considering the total potential energy of the shell $\mathcal{E} [\mathbf{u}]$ is given by the sum of elastic potential energy $\Psi$ and the potential energy due to the external force density $\mathbf{f}$, we obtain: 
\begin{align}
    \mathcal{E} [\mathbf{u}] = \int_{\Omega} \Psi \,d\Omega - \int_{\Omega} \mathbf{f} \cdot \mathbf{u} \,d\Omega. 
    \label{eq:potential_energy}
\end{align}
\par
Next, the stable equilibrium deformation of the shell can be found by minimising the potential energy functional subject to boundary constraints: 
\begin{align}
\begin{split}
&\mathbf{u^*} = \arg \min_\mathbf{u} \mathcal{E} [\mathbf{u}], \text{ subject to } \\
& \mathbf{u}(\xi^1, \xi^2) = \mathbf{b}(\xi^1, \xi^2) \text{ on } \partial \Omega.
\end{split}
\end{align}
The above definition of hyperelastic energy of the shell requires the displacement field $\mathbf{u} \in H^2(\Omega \mapsto \mathbb{R}^3) $
that must necessarily have square-integrable first and second derivatives. 

\subsection{Tensor Algebra}\label{sec:tensor_algebra}

We provide additional results from tensor algebra~\cite{wempner2003mechanics}) that are relevant for strain computations on shells (see Sec.~\ref{sec:strain_computation}). 
\par
Based on the coordinate system for the tensor components, a tensor can be covariant (\eg, $A_{\alpha \beta}$), contravariant (\eg, $A^{\alpha \beta}$) and may even have mixed character, \ie, partly contravariant and partly covariant in different indices (\eg, $A_\alpha^\beta$). 
For computing $\varphi_\beta^\lambda$, and $\varphi^\lambda_3$ in \eqref{eq:strain}-(main matter), we use the following rule from shell theory that 
transforms a covariant tensor to a mixed 
one: 
\begin{align} 
\begin{split} 
A^\alpha_\beta &= A_{\beta \lambda} \bar{a}^{\lambda \alpha}, \\ 
A^\alpha_3 &= A_{\lambda 3} \bar{a}^{\lambda \alpha}. 
\end{split} 
\end{align} 
\par
A tensor of $n$-th order has $n$ indices. For example, $v_\alpha$ is first-order, and $H^{\beta \alpha \lambda \delta}$ is fourth. 
For computing the covariant derivatives of the first-order tensor $u_\rho |_\alpha$ and the second-order tensor $\varphi_{\alpha \lambda} |_\beta$ in \eqref{eq:strain}, 
we use the following rules: 
\begin{align}
\begin{split}
& v_\alpha |_\beta = v_{\alpha,\beta}  - v_\lambda \Gamma_{\alpha \beta}^\lambda, \text{ and } \\
& A_{\alpha \beta}|_\gamma = A_{\alpha \beta, \gamma} - A_{\lambda \beta} \Gamma_{\alpha \gamma}^\lambda - A_{\alpha \lambda } \Gamma_{\beta \gamma}^\lambda,  
\end{split}
\end{align}
where $\Gamma_{\alpha \beta}^\lambda$ is the Christoffel symbol given by
\begin{align}
\Gamma_{\alpha \beta}^\lambda := \mathbf{\bar{a}}^\lambda \cdot \mathbf{\bar{a}}_{\alpha, \beta}. 
\end{align}
\par
Some tensors arising in the kinematic description of Kirchhoff-Love thin shells are symmetric with respect to indices $\alpha$ and $\beta$, \textit{i.e.,}  $A_{\alpha \beta}=A_{\beta \alpha}$. 
We exploit the symmetry for efficient computations of the following tensors: 
$a_{\alpha \beta}$, $b_{\alpha \beta}$, $\varepsilon_{\alpha \beta}$,  $\kappa_{\alpha \beta}$, and $\Gamma_{\alpha \beta}^\lambda$. 
\par
In the case of linear elastic material, we also exploit the symmetry of fourth-order symmetric tensor $\mathbf{H}$: 
\[
    H^{\alpha \beta \lambda \delta} = H^{\beta \alpha \lambda \delta} = H^{\beta \alpha \delta \lambda} = H^{\alpha \beta \delta \lambda} = H^{\lambda \delta \alpha \beta}. 
\]
This property means that only six independent components (after applying symmetry) need to be computed (\textit{i.e.,} $H^{1111}$, $H^{1112}$, $H^{1122}$, $H^{1212}$, $H^{1222}$, and $H^{2222}$).

\subsection{Proof of Strain Computation} 
\label{sec:strain_computation}

According to the Kirchhoff-Love theory, the Green-Lagrange strain associated with the deformation of a thin shell is decomposed into the stretching and bending strains of the midsurface.
One could compute them using Eqs.~\eqref{eq:membrane_strain} and \eqref{eq:bending_strain}, written in terms of the reference state $\bar{\mathbf{x}}$ and the deformed state $\mathbf{x}$ of the midsurface. 
As an easier alternative, we directly evaluate strains with the NDF $\mathbf{u}$ of the midsurface using \eqref{eq:strain}-(main matter).
Next, we prove that the two formulations are identical following~\cite{basar2013mechanik}). 
\begin{lemma}[Deformation gradient]
\label{lemma:deformation_gradient}
Deformation gradient $\mathbf{u}_{,\alpha} $ can be written as 
$\mathbf{u}_{,\alpha} = \varphi_{\alpha \lambda} \mathbf{\bar a}^\lambda + \varphi_{\alpha 3} \mathbf{\bar a}^3$ where the components of the gradients $\varphi_{\alpha \lambda}, \varphi_{\alpha 3}$ are defined as
\begin{align}
    \begin{split}
    \varphi_{\alpha \lambda} &:= u_\lambda |_\alpha - \bar{b}_{\alpha \lambda} u_3, \text{ and } \\
    \varphi_{\alpha 3} &:= u_{3,\alpha} + \bar{b}_\alpha^\lambda u_\lambda.
    \end{split}
\end{align}
\end{lemma}
\begin{proof}
Given deformation field $\mathbf{u}$ of the midsurface described in contravariant basis as $\mathbf{u} = u_\lambda \mathbf{\bar a}^\lambda + u_3 \mathbf{\bar a}^3$ , 
we compute the deformation gradient as follows: 
\begin{align}
    \begin{split}
    \mathbf{u}_{,\alpha} = \mathbf{u}|_\alpha &= u_\lambda|_\alpha \mathbf{\bar a}^\lambda + u_\lambda \mathbf{\bar a}^\lambda|_\alpha + u_3|_\alpha \mathbf{\bar a}^3 + u_3 \mathbf{\bar a}^3|_\alpha \\
     &= u_\lambda|_\alpha \mathbf{\bar a}^\lambda + u_\lambda \bar{b}_\alpha^\lambda \mathbf{\bar a}^3 + u_{3,\alpha} \mathbf{\bar a}^3 - u_3 \bar{b}_{\alpha \lambda} \mathbf{\bar a}^\lambda, 
    \end{split}
\end{align} 
where we use the following identities from the shell theory~\cite{basar2013mechanik} to arrive at the bottom part of the previous equation: 
\begin{align} 
    \begin{split}
    & \mathbf{\bar a}^3|_\alpha = \mathbf{\bar a}^3_{,\alpha} = -\bar{b}_{\alpha \lambda} \mathbf{\bar a}^\lambda = -\bar{b}_\alpha^\lambda \bar{\mathbf{a}}_\lambda, \\
    & \mathbf{\bar a}^\lambda|_\alpha = \bar{b}_\alpha^\lambda \mathbf{\bar a}^3. 
    \end{split}
    \label{eq:identities_derivatives}
\end{align} 
Finally, we rewrite them as 
\begin{align}  
    \mathbf{u}_{,\alpha} &= \varphi_{\alpha \lambda} \mathbf{\bar a}^\lambda + \varphi_{\alpha 3} \mathbf{\bar a}^3. 
    \label{eq:deformation_derivative}
\end{align}
\end{proof}

\begin{theorem}[Membrane strain]
Membrane strain (quantifying/measuring in-plane stretching) can be written as a function of the deformation gradient in the following form: 
\begin{align}
    \varepsilon_{\alpha \beta} &= \frac{1}{2} %
    (\varphi_{\alpha \beta} + \varphi_{\beta \alpha}
    + \varphi_{\alpha \lambda} \varphi_\beta^\lambda + \varphi_{\alpha 3} \varphi_{\beta 3} ). 
\end{align}
\end{theorem}
\par
\begin{proof}
We start with membrane strain given as the difference of metric tensors (first fundamental form) \eqref{eq:membrane_strain}: 
\begin{align}
    \begin{split}
    \varepsilon_{\alpha \beta} &:= \frac{1}{2} (a_{\alpha \beta} - \bar{a}_{\alpha \beta}),  \\
    \varepsilon_{\alpha \beta} &= \frac{1}{2} (\mathbf{a}_\alpha \cdot \mathbf{a}_\beta - \bar{\mathbf{a}}_\alpha \cdot \bar{\mathbf{a}}_\beta). 
    \end{split}
\end{align}
Substituting the tangent basis vectors $\mathbf{a}_\alpha$ 
\begin{align}
 \begin{split}
\mathbf{a}_\alpha = \mathbf{x},_\alpha = \bar{\mathbf{x}},_\alpha +  \mathbf{u},_\alpha 
 = \bar{\mathbf{a}}_\alpha + \mathbf{u},_\alpha
 \end{split}
 \label{eq:deformed_tangent}
\end{align}
gives us updated strain in terms of deformation $\mathbf{u}$: 
\begin{align}
    \varepsilon_{\alpha \beta} = \frac{1}{2} (\bar{\mathbf{a}}_\alpha \cdot \mathbf{u,_\beta} + \bar{\mathbf{a}}_\beta \cdot \mathbf{u,_\alpha} + \mathbf{u},_\alpha \cdot \mathbf{u},_\beta). 
\end{align}

Assuming the following identities from Kirchhoff-Love shell hypothesis~\cite{basar2013mechanik}: 
\begin{align}
    \begin{split}
    & \bar{\mathbf{a}}_\alpha \cdot \bar{\mathbf{a}}^\beta = \delta^\beta_\alpha, 
     \bar{\mathbf{a}}^3 = \bar{\mathbf{a}}_3,
    \bar{\mathbf{a}}_\alpha \cdot \bar{\mathbf{a}}^3 = \bar{\mathbf{a}}^\alpha \cdot \bar{\mathbf{a}}_3 = 0,
    \bar{\mathbf{a}}^3 \cdot \bar{\mathbf{a}}^3 = 1, \text{ and } \\
    & \bar{a}^{\alpha \beta} = \bar{\mathbf{a}}^\alpha \cdot \bar{\mathbf{a}}^\beta, 
    A^\alpha_\beta = A_{\beta \lambda} \bar{a}^{\lambda \alpha},  
    \end{split}
    \label{eq:kl_identities}
\end{align}
and considering the above lemma for the deformation gradient $\mathbf{u}_{,\alpha}$, 
we finally obtain 
the 
target 
formulation for strain: 
\begin{align}
    \varepsilon_{\alpha \beta} &= \frac{1}{2} %
    (\varphi_{\alpha \beta} + \varphi_{\beta \alpha}
    + \varphi_{\alpha \lambda} \varphi_\beta^\lambda + \varphi_{\alpha 3} \varphi_{\beta 3} ). 
\end{align}
\end{proof}
\begin{theorem}[Bending strain]
Bending strain (measuring the change in curvature) can be written as a function of the deformation gradient in the following form: 
\begin{align}
    \kappa_{\alpha \beta} \approx 
    -\varphi_{\alpha 3}|_\beta - \bar{b}_\beta^\lambda \varphi_{\alpha \lambda} 
    + \varphi^\lambda_3 (\varphi_{\alpha \lambda} |_\beta + \frac{1}{2} \bar{b}_{\alpha \beta} \varphi_{\lambda 3} - \bar{b}_{\beta \lambda} \varphi_{\alpha 3}). 
\end{align}
\end{theorem}
\begin{proof}

The bending strain of the midsurface is defined as the difference of curvature tensors (second fundamental form) in the reference and deformed configurations: 
\begin{align}
    \begin{split}
   \kappa_{\alpha \beta} &:= \bar{b}_{\alpha \beta} - b_{\alpha \beta} \\
   \kappa_{\alpha \beta} &= \mathbf{a}_\alpha \cdot \mathbf{a}_{3,\beta} - \bar{\mathbf{a}}_\alpha \cdot \bar{\mathbf{a}}_{3,\beta}
    \end{split}
\end{align}
Using \eqref{eq:deformed_tangent} and \eqref{eq:difference_vector}, we rewrite strain using deformation gradients as:%
\begin{align}
    \kappa_{\alpha \beta} &= (\bar{\mathbf{a}}_\alpha + \mathbf{u,_\alpha}) \cdot (\bar{\mathbf{a}}_{3,\beta} + \mathbf{w,_\beta}) - \bar{\mathbf{a}}_\alpha \cdot \bar{\mathbf{a}}_{3,\beta}
\end{align}
Further simplification and applying identity \eqref{eq:identities_derivatives} leads to: 
 \begin{align}
    \kappa_{\alpha \beta} &= \bar{\mathbf{a}}_\alpha \cdot \mathbf{w,_\beta} + \mathbf{u,_\alpha} \cdot \mathbf{w,_\beta} - \bar{b}_\beta^\lambda \mathbf{u,_\alpha} \cdot \bar{\mathbf{a}}_\lambda
\end{align}
Using Lemma~\ref{lemma:deformation_gradient} for deformation gradients
$\mathbf{u}_{,\alpha}$ and $\mathbf{w}_{,\beta}$, \textit{i.e,}
$\mathbf{w}_{,\beta} = (w_\lambda |_\beta - \bar{b}_{\lambda \beta} w_3)\mathbf{\bar a}^\lambda + (w_{3,\beta} + \bar{b}_\beta^\lambda w_\lambda) \mathbf{\bar a}^3$ 
and 
with the shell identities of \eqref{eq:kl_identities}, we arrive at:
    
\begin{align}
    \begin{split}
    \kappa_{\alpha \beta} = 
   w_\alpha |_\beta - \bar{b}_{\alpha \beta} w_3 - \bar{b}_\beta^\lambda \varphi_{\alpha \lambda} 
   + \varphi_\alpha^\lambda (w_\lambda|_\beta - \bar{b}_{\lambda \beta} w_3) \\ 
   + \varphi_{\alpha 3} (w_{3,\beta} + \bar{b}_\beta^\lambda w_\lambda)
    \end{split}
\end{align}
With the Kirchhoff-Love normal hypothesis and neglecting cubic terms, we can approximate the components of $\mathbf{w}$ as the following: 
\begin{align}
    \begin{split}
    w_3 \approx - \frac{1}{2} w_\lambda w^\lambda = - \frac{1}{2} \varphi_{\alpha 3} \varphi_3^\alpha , \\
    w_\alpha \approx - \varphi_{\alpha 3} + \varphi_{\alpha}^\lambda \varphi_{\lambda 3}.
    \label{eq:difference_vector_components}
    \end{split}
\end{align}
First eliminating the component $w_3$ and subsequently $w_\alpha$, we arrive at the target strain formulation:
\begin{align}
    \begin{split}   
    \kappa_{\alpha \beta} \approx 
   w_\alpha |_\beta -  \bar{b}_\beta^\lambda \varphi_{\alpha \lambda} + \frac{1}{2} \bar{b}_{\alpha \beta} w_\lambda w^\lambda + w^\lambda|_\beta \varphi _{\alpha \lambda} + \bar{b}_\beta^\lambda w_\lambda \varphi_{\alpha 3} \\
    \kappa_{\alpha \beta} \approx 
    -\varphi_{\alpha 3}|_\beta - \bar{b}_\beta^\lambda \varphi_{\alpha \lambda} 
    + \varphi^\lambda_3 (\varphi_{\alpha \lambda} |_\beta + \frac{1}{2} \bar{b}_{\alpha \beta} \varphi_{\lambda 3} - \bar{b}_{\beta \lambda} \varphi_{\alpha 3}).
     \end{split}
\end{align}
\end{proof}

\begin{figure*}[t!]
	\includegraphics[width=\linewidth]{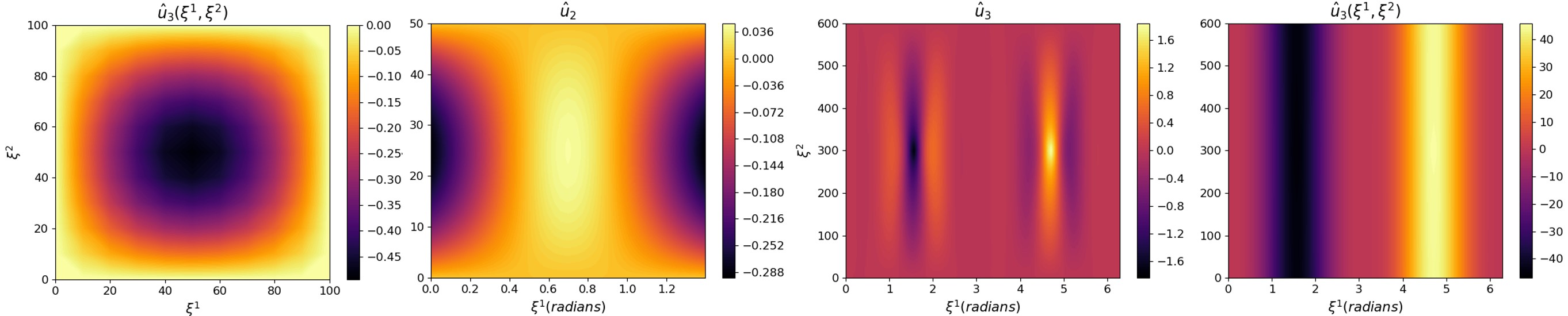}
	\caption
	{
        \textbf{Belytschko obstacle course: 
        Visualisation of the deformation fields 
        predicted by our NeuralClothSim.} 
        From the left to the right: Square plate, Scordelis-Lo roof, pinched cylinder 
        with fixed boundary conditions and pinched cylinder with free boundary conditions. 
        For the pinched cylinder, the results are re-scaled versions with $E=\qty{30}{}$. %
	}
	\label{fig:belytschko_displacement_field}
\end{figure*}

\section{Belytschko Obstacle Course}
\label{sec:obstacle_course_appendix}

In the following, we provide detailed information for reproducing the Belytschko obstacle course experiments from Sec.~\ref{sec:obstacle_course}-(main matter). 
We visualise the NDF along the direction of applied load in Fig.~\ref{fig:belytschko_displacement_field} that closely matches the reference solutions~\cite{belytschko1985stress}. %

\subsection{Square Plate}
In the first test case, we consider a simple bending problem of a flat square shell~\cite{cirak2000subdivision}. 
It is simply supported at all edges and is subject to a uniform load.
The plate has a side length of $L = \qty{100}{\metre}$ and a thickness $h = \qty{1}{\metre}$ and, therefore, falls under the scope of Kirchhoff-Love thin shell theory. 
The material parameters are given as $E = \qty{1e7}{\pascal}$ and $\nu = 0$. 
We represent the reference geometry with \eqref{eq:sleeve_napkin}-(main matter) and impose Dirichlet boundary constraints by constructing a distance function to the plate edges.   
This is followed by training an NDF to solve for quasi-static displacement minimising the total potential energy \eqref{eq:potential_energy} subject to uniformly distributed external load $\mathbf{f} = [0, 0, -1]^\top$.
With the simply supported constraints along the boundary defined by $\partial \Omega = \{(\xi^1, 0), (0, \xi^2), (\xi^1,L), (L,\xi^2) \}$, we define NDF as follows: 
\begin{align}
\begin{split}
    &\mathbf{u}(\boldsymbol\xi; \Theta)  = \mathcal{F}_\Theta(\boldsymbol\xi) \mathcal{B}(\boldsymbol\xi), \\
    \text{ s.t. } & \mathcal{B}(\boldsymbol\xi) := \xi^1 \xi^2 (L - \xi^1) (L - \xi^2). 
\end{split}
\end{align}
 
We train with the loss \eqref{eq:main_loss}-(main matter) for 2500 iterations and illustrate the solution in Fig.~\ref{fig:belytschko}-(main matter) where the displacement is scaled up by a factor of $50$. 
The maximum displacement $u_3$ at the centre of the plate is found to be 0.487 after convergence and exactly matches the reference solution~\cite{timoshenko1959theory}. 
Fig.~\ref{fig:belytschko_displacement_field}-(left) shows the obtained NDF along the $z$-axis for the square plate.

\subsection{Scordelis-Lo Roof}

The reference geometry of the Scordelis-Lo roof is given by the following parametric expression: 
\begin{align}
\begin{split}
    & \mathbf{\bar x}(\boldsymbol{\xi}) = [R \cos(\xi^1 + \qty{50}{\degree}), R \sin(\xi^1 + \qty{50}{\degree}), \xi^2]^\top, \\
    &\forall  \xi^1 \in [0,\qty{80}{\degree}); \xi^2 \in [0,L], \text{ with }
    R = \qty{25}{\metre}, L = \qty{50}{\metre}, \text{ and } h = \qty{0.25}{\metre}.
    \label{eq:scordelis}
\end{split}
\end{align}
Concerning the boundary conditions, the structure is supported with a rigid diaphragm along the edges, \textit{i.e.,} 
$\partial \Omega = \{(\xi^1, 0),(\xi^1, L)\}$. 
The material properties are set as $E = \qty{4.32e8}{\pascal}, \nu = 0$ 
and a uniformly distributed load $\mathbf{f} = [0, -90, 0]^\top$ is applied to it.

We optimise NDF under boundary conditions as follows: 
\begin{align} 
\begin{split} 
    & \mathbf{u}(\boldsymbol\xi; \Theta) = [\mathcal{F}_{\Theta 1} \mathcal{B}(\boldsymbol\xi), \mathcal{F}_{\Theta 2} \mathcal{B}(\boldsymbol\xi), \mathcal{F}_{\Theta 3} ]^\top, \\
    \text{ s.t. } & \mathcal{B}(\boldsymbol\xi) := \xi^2 (L - \xi^2). 
\end{split} 
\end{align} 
Fig.~\ref{fig:belytschko_displacement_field}-(second on the left) visualises the computed NDF along the $y$-axis. 

\subsection{Pinched Cylinder}
\textit{Pinched Cylinder.}
Finally, we consider the pinched cylinder problem, \textit{i.e.,} 
one of the most severe tests for both inextensional bending modes and complex membrane states. 
As shown in~Fig.~\ref{fig:belytschko}-(main matter), a cylindrical shell is pinched with two diametrically opposite unit loads applied at the middle of the shell.
We consider two cases: First, a shell with ends supported by rigid diaphragms~\cite{belytschko1985stress} (similar to Scordelis-Lo roof), and second, a cylinder with free ends~\cite{timoshenko1959theory}. 
We define the cylinder geometry with \eqref{eq:sleeve_napkin}-(main matter), where $R = \qty{300}{\metre}, L = \qty{600}{\metre}$ and the thickness is set to $h = \qty{3}{\metre}$;
the material properties are given as $E = \qty{3e6}{\pascal}, \nu = 0.3$. 
In contrast to the previous test geometries, which required specifying only the Dirichlet boundary conditions, we additionally account for the periodicity constraint along the circular cross-sections. 
To model this, we define NDF as $\mathcal{F}_\Theta (\cos \xi^1, \sin \xi^1, \xi^2)$, instead of the default case $\mathcal{F}_\Theta (\xi^1, \xi^2)$.
A crucial challenge of pinched cylinder test case is 
modelling load at singular points in the sample space. 
To achieve this, we adapt the potential energy functional \eqref{eq:potential_energy}-(main matter)---described previously for uniformly distributed forces~---to the point load setting, rewriting it as $\mathcal{E}^\mathrm{pot} [\mathbf{u}] = \int_{\Omega} \Psi \,d\Omega - \textstyle \sum_{\Omega_0} \mathbf{f} \cdot \mathbf{u},$

where $\Omega_0$
is the set of points of the load application.
We apply point loads 
$\mathbf{f} \in \{[0, 0, 1]^\top, [0, 0, -1]^\top\}$
at diametrically opposite points 
$\Omega_0 = \{(\qty{90}{\degree}, 300),(\qty{270}{\degree}, 300)\}$. %
In the case of distributed load, we previously proposed computing external and hyperelastic strain energy at an identical set of stratified samples in the parametric domain. 
We depart from this setting for point loads: At each training iteration, we sample all points from $\Omega_0$ for external energy, whereas random stratified samples are used for computing strain energy.
To speed up the convergence, we set $E = \qty{30}{\pascal}$ instead of the original value $E = \qty{3e6}{\pascal}$; this simply scales the displacement field in the linear setting as shown in~\cite{bastek2023physics}. 
As mentioned, the constrained cylinder is supported with a rigid diaphragm along the edges, \textit{i.e.,} $\partial \Omega = \{(\xi^1, 0),(\xi^1, L)\}$, 
therefore,
we optimise NDF using 
\begin{align} 
\begin{split} 
    & \mathbf{u}(\boldsymbol\xi; \Theta) = [\mathcal{F}_{\Theta 1} \mathcal{B}(\boldsymbol\xi), \mathcal{F}_{\Theta 2}, \mathcal{F}_{\Theta 3} \mathcal{B}(\boldsymbol\xi)]^\top, \\
     \text{ s.t. } & \mathcal{B}(\boldsymbol\xi) := \xi^2 (L - \xi^2). 
\end{split} 
\end{align} 
Next, we consider 
a pinched cylinder with free ends, \ie $\partial \Omega = \emptyset$.
Without any boundary constraints, the cylinder can move rigidly due to the applied force, and such rigid body motion should be factored out. 
Therefore, to suppress it, we restrict the displacement of the point under the load in directions other than the direction of the force vector. 
We achieve this by enforcing $\hat{u}_1=0, \hat{u}_2=0$ at load points. 
The NDF parametrisation factoring out the rigid motion reads as: 
\begin{align}
\begin{split}
    \mathbf{u}(\boldsymbol\xi; \Theta) &= [\mathcal{F}_{\Theta 1} \mathcal{B}_1(\boldsymbol\xi)\mathcal{B}_2(\boldsymbol\xi), \mathcal{F}_{\Theta 2}  \mathcal{B}_1(\boldsymbol\xi)\mathcal{B}_2(\boldsymbol\xi), \mathcal{F}_{\Theta 3}]^\top, \\
   \text{ s.t. } \mathcal{B}_1(\boldsymbol\xi) &:= 1 - e ^{-((\xi^1-\qty{90}{\degree})^2 + (\xi^2-300)^2)/\sigma}, \text{ and } \\
    \mathcal{B}_2(\boldsymbol\xi) &:= 1 - e ^{-((\xi^1-\qty{270}{\degree})^2 + (\xi^2-300)^2)/\sigma}.
    \end{split}
\end{align}
In both examples with the pinched cylinder, we monitor the displacements under the loading point. 
As shown in Fig.~\ref{fig:belytschko}-(main matter) and Tab.~\ref{tab:belytschko}-(main matter), it qualitatively and quantitatively converges to the reference solution. 
Fig.~\ref{fig:belytschko_displacement_field}-(second from the right, and the rightmost) shows the obtained NDFs along the $z$-axis for the two cases of the pinched cylinder.

\begin{wrapfigure}[8]{R}{0.5\textwidth}
\centering
        \vspace{-36pt}
	\includegraphics[width=\linewidth]{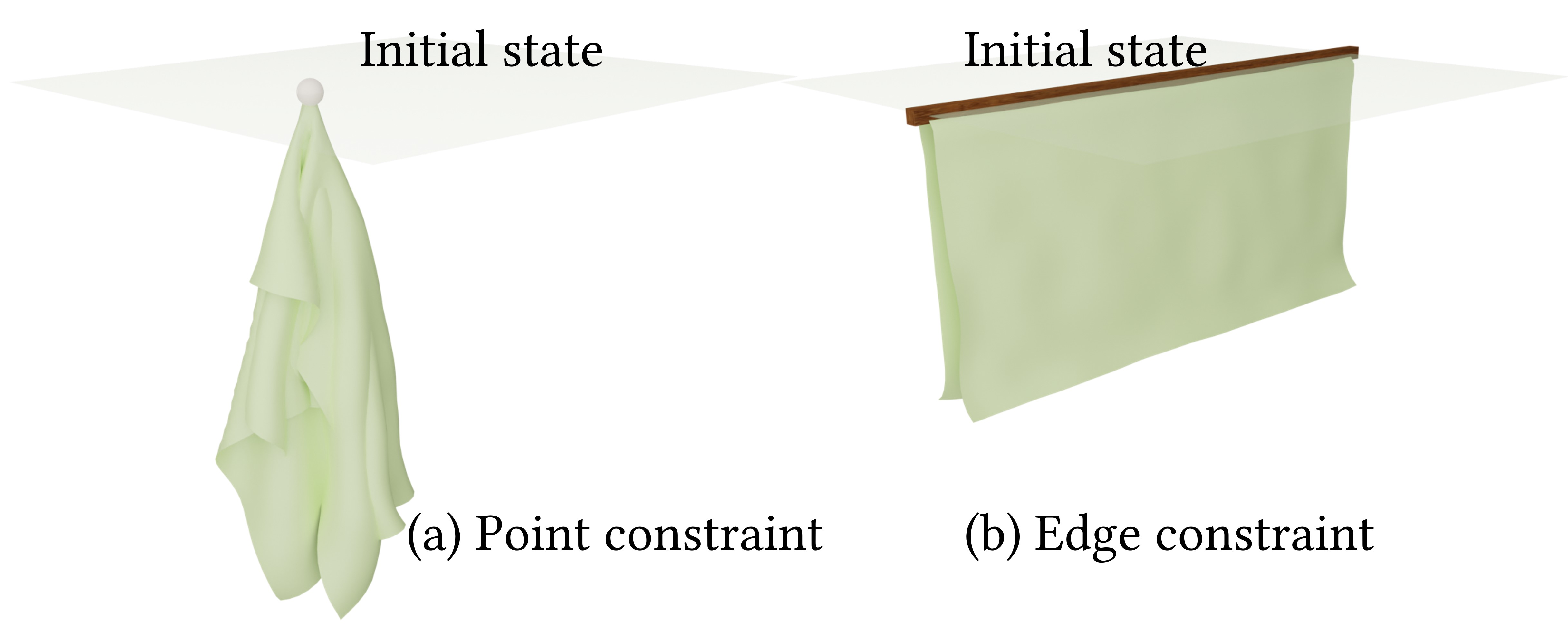}
	\caption
	{
        Napkin simulation upon convergence under gravity with \emph{non-boundary} constraints.
	}
	\label{fig:nonboundary_constraint}
\end{wrapfigure}

\section{Simulation Details} 
\label{sec:simulation_setup} 
We first describe the extension of NeuralClothSim for visualising trajectory to equilibria.  
Then, similar to the previous section, we provide boundary and loading conditions for all experiments here. 
\newpage
\subsection{Cloth Trajectory Visualisation}
\label{ssec:appendix_temporal_interpolation}
We next visualise the %
transition 
from the reference to the equilibrium state: We extend the NDF \eqref{eq:deformation_field} to $\mathbf{u} (\boldsymbol{\xi}, t;\Theta)$ modelling time-dependent deformations, $ \forall t \in [0, T]$, with $T=1$.
For a smooth and physically-plausible interpolation from the initial state $\bar{\mathbf{x}}(\boldsymbol{\xi})$ to the deformed $\mathbf{x}(\boldsymbol{\xi}, T)$,
we impose initial conditions and a temporal regulariser.

\emph{Initial Cloth Configuration.}
If we optimise the time-dependent NDF only with the potential energy loss \eqref{eq:main_loss}, the model finds the converged %
equilibrium states of the underlying cloth model $\forall t \in [0, T]$. 
To start from the initial undeformed cloth state,
initial conditions leading to zero displacement and velocity need to be explicitly incorporated. 
Hence, we use the function $\mathcal{I}(t):= t^2$ as an additional multiplying factor in \eqref{eq:deformation_field}
leading to $\mathbf{u}(\boldsymbol{\xi}, 0) = \mathbf{0}, \;\text{and} \;\dot{\mathbf{u}}(\boldsymbol{\xi}, 0) = \mathbf{0}$.

\emph{Temporal Smoothing.}
Without any temporal prior, the transition from the reference to the equilibrium state will be too swift and not smooth. 
Therefore, we use an additional regularisation loss $\mathcal{L}_t(\Theta) := \frac{|\Omega|}{N_\Omega N_t}\sum_{i=1}^{N_\Omega} \sum_{j=1}^{N_t} \frac{1}{2} \rho |\dot{\mathbf{u}}(\boldsymbol{\xi}_i, t_j;\Theta)|^2 $ constraining the cloth velocity $\dot{\mathbf{u}}$.
Specifically, the smooth optimised trajectory $\mathbf{u}^* (\boldsymbol{\xi}, t;\Theta)$ is obtained with the final loss 
$ \mathcal{L} + \mathcal{L}_t$, where, similar to $\mathcal{L}_t(\Theta)$,  
the physics loss $\mathcal{L}(\Theta)$ is now evaluated over the entire parametric-temporal domain.

Additionally, for some examples, such as sleeve compression/torsion, we drive changes in the deformation trajectory by imposing time-varying Dirichlet boundary conditions. 
We note that the time-stepping is performed purely for visualisation, and we do not model the simulation dynamics that would require
taking into account inertial and damping effects. 
For linear isotropic material, we set $\rho = \qty{0.144}{\kg\per\metre\squared}$ and for the non-linear orthotropic material from \citet{clyde2017modeling}.

\subsection{Napkin} 
The force in this experiment is defined as $\mathbf{f} = [0, -9.8 \rho, 0]^\top$ and the boundary conditions read $\partial \Omega = \{(0,0), (0,L)\}$. 
The result of a napkin droop with a fixed corner is shown 
in Fig.~\ref{fig:teaser}-(main matter). 
NDF in this experiment is parametrised as follows: 
\begin{align}
\begin{split}
    & \mathbf{u}(\boldsymbol\xi, t; \Theta)  = \mathcal{F}_\Theta(\boldsymbol\xi, t) \mathcal{I}(t) \mathcal{B_\mathrm{top\_left\_corner}}(\boldsymbol\xi), \\
     \text{ s.t. }  & \mathcal{B_\mathrm{top\_left\_corner}}(\xi^1, \xi^2) := 1- e ^{-((\xi^1 )^2 + (\xi^2- L)^2)/\sigma}. 
\end{split}
\end{align}

The experimental result of a napkin droop with moving corners is shown in Fig.~\ref{fig:overview}-(main matter). 
The boundary condition read $ \partial \Omega = \{(0,L), (L,L)\}$. 
The NDF parametrisation in this scenario is as follows: 
\begin{align}
\begin{split}
    & \mathbf{u} = \mathcal{F}_\Theta \mathcal{I} \mathcal{B_\mathrm{top\_left}} \mathcal{B_\mathrm{top\_right}} + 
    (1 - \mathcal{B_\mathrm{top\_left}}) \mathcal{B_\mathrm{motion}} - (1 - \mathcal{B_\mathrm{top\_right}}) \mathcal{B_\mathrm{motion}}, \\
 & \text{ s.t. } \mathcal{B_\mathrm{top\_left}}(\xi^1,\xi^2) := 1 - e ^{-((\xi^1)^2 + (\xi^2-L)^2)/\sigma},  \\
 & \mathcal{B_\mathrm{top\_right}}(\xi^1,\xi^2) := 1 - e ^{-((\xi^1-L)^2) + (\xi^2 - L)^2)/\sigma}, \text{ and } \\
 & \mathcal{B_\mathrm{motion}}(t) := [0.2 t, 0, 0]^T. 
\end{split}
\end{align}

The experimental result for napkin droop with fixed edges is shown in Fig.~\ref{fig:non_linear_strain}. 
The boundary conditions are defined as $ \partial \Omega = \{(\xi^1,0), (0, \xi^2)\}, \forall (\xi^1, \xi^2) \in [0,L]^2$ and the NDF parameterisation reads 
\begin{align}
\begin{split}
    & \mathbf{u} = \mathcal{F}_\Theta \mathcal{I} \mathcal{B_\mathrm{left\_edge}} \mathcal{B_\mathrm{right\_edge}} \\
 \text{ s.t. }
 & \mathcal{B_\mathrm{left\_edge}}(\xi^1) := 1 - e ^{-(\xi^1 )^2 /\sigma}, \text{ and } \\ 
 & \mathcal{B_\mathrm{right\_edge}}(\xi^2 ):= 1 - e ^{-(\xi^2 )^2 /\sigma}. 
 \end{split} 
\end{align}

\subsection{Sleeve}\label{ssec:sleeve} 
In the experiment with a sleeve, no external force is exerted: $\mathbf{f} = [0, 0, 0]^T$. %
The boundary region is defined by: $\partial \Omega = \{(\xi^1,0), (\xi^1,L)\}, \forall  \xi^1 \in [0,2\pi)$. 
The NDF is parametrised as follows: 
\begin{align}
\begin{split}
    & \mathbf{u} = \mathcal{F}_\Theta \mathcal{I} \mathcal{B_\mathrm{bottom\_rim}} \mathcal{B_\mathrm{top\_rim}} 
     + (1 - \mathcal{B_\mathrm{bottom\_rim}}) \mathcal{B_\mathrm{motion}} - (1 - \mathcal{B_\mathrm{top\_rim}}) \mathcal{B_\mathrm{motion}}, \\
\text{ s.t. }
 & \mathcal{B_\mathrm{bottom\_rim}}(\xi^2) := 1 - e ^{-(\xi^2)^2/\sigma}, \\
 & \mathcal{B_\mathrm{top\_rim}}(\xi^2) := 1 - e ^{-(\xi^2 - L)^2/\sigma}, \text{ and } \\
 & \mathcal{B_\mathrm{motion}}(t) := [0, 0.1 t, 0]^\top. 
\end{split}
\end{align}

\begin{figure}
\centering
	\includegraphics[width=\linewidth]{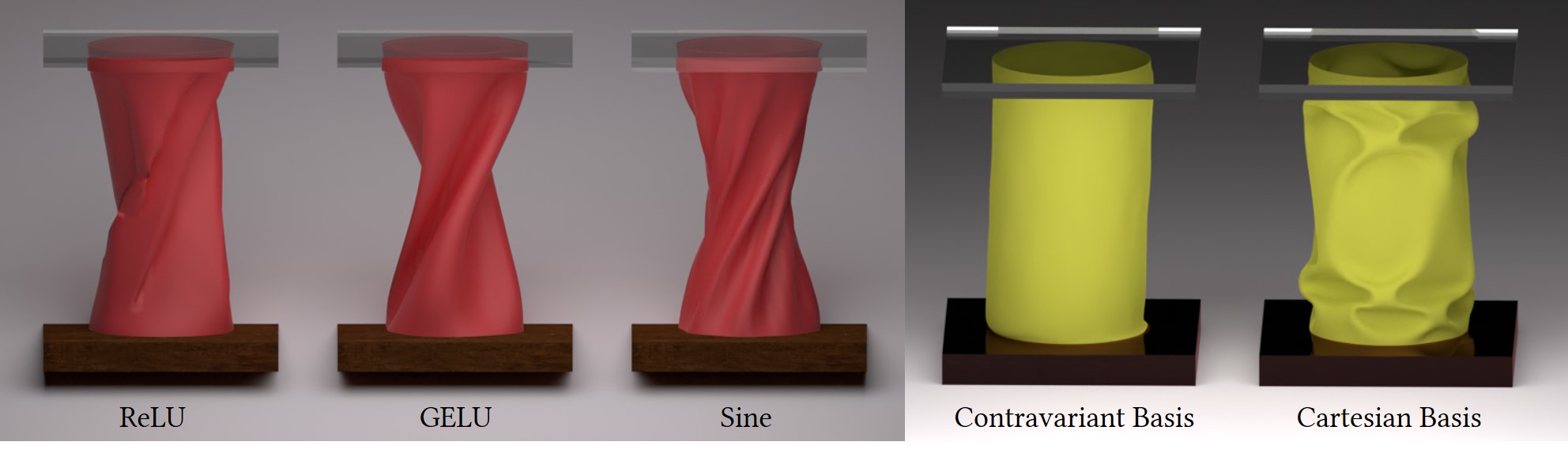}
	\caption
	{
        \textbf{Activations (left).} Results of our method with different activation functions 
        (ReLU, GELU and Siren). 
        \textbf{Contravariant \textit{vs} Cartesian basis (right).} 
        Prediction of NDF output in the Cartesian coordinate system is well conditioned compared to the local contravariant coordinate system. 
	}
	\label{fig:ablation_activation}
\end{figure}

Sleeve twist is achieved by introducing rotation displacement
$\theta  = \frac{3\pi}{4}$. 
The NDF in this scenario is parametrised as follows: 
\begin{align}
\begin{split}
    & \mathbf{u} = \mathcal{F}_\Theta \mathcal{I} (1-\mathcal{B_\mathrm{bottom\_rim}})(1-\mathcal{B_\mathrm{top\_rim}})
     - \mathcal{B_\mathrm{bottom\_rim}} \mathcal{B_\mathrm{bottom\_motion}} + \mathcal{B_\mathrm{top\_rim}} \mathcal{B_\mathrm{top\_motion}}, \\
 \text{ s.t. }
 & \mathcal{B_\mathrm{bottom\_rim}}(\xi^2) := e ^{-(\xi^2)^2/\sigma}, \\
 & \mathcal{B_\mathrm{top\_rim}}(\xi^2) := e ^{-(\xi^2 - L)^2/\sigma}, \\
& \mathcal{B_\mathrm{bottom\_motion}}(\xi^1, t) := \begin{bmatrix}
           R(\cos(\xi^1-\theta t)-\cos \xi^1) \\
           0 \\
          R(\sin(\xi^1-\theta t)-\sin \xi^1)
         \end{bmatrix}, \text{ and } \\
     & \mathcal{B_\mathrm{top\_motion}}(\xi^1, t) := \begin{bmatrix}
           R(\cos(\xi^1+\theta t)-\cos \xi^1) \\
           0 \\
          R(\sin(\xi^1+\theta t)-\sin \xi^1)
         \end{bmatrix}. 
    \end{split} 
  \end{align} 
We demonstrate sleeve torsion in 
Fig.~\ref{fig:ablation_activation}-(left) and buckling in \ref{fig:ablation_activation}-(right). 
\subsection{Skirt}
See Fig.~\ref{fig:resolution_inconsistency} for the experimental results with skirt. 
The reference skirt geometry is defined as: 
\begin{align}
\begin{split}
    \mathbf{\bar x}(\boldsymbol{\xi}) &= [r \cos\xi^1, \xi^2, r \sin\xi^1]^T, \; \forall  \xi^1 \in [0,2\pi); \xi^2 \in [0,L], \\
     \text{ s.t. } r(\xi^2) &:= \frac{(R_\mathrm{top} - R_\mathrm{bottom})  \xi^2}{L} + R_\mathrm{bottom}. 
    \label{eq:skirt}
\end{split}
\end{align}
The skirt deform in this experiment under gravity, \textit{i.e.,} 
$\mathbf{f} = [0, -9.8 \rho, 0]^T$; 
the boundary region is given by $\partial \Omega = \{(\xi^1,L)\}, \forall  \xi^1 \in [0,2\pi)$. 
NDF is parametrised as follows: 
\begin{align}
\begin{split}
    & \mathbf{u} = \mathcal{F}_\Theta \mathcal{I} (1-\mathcal{B_\mathrm{top\_rim}}), \\
 \text{ s.t. } & \mathcal{B_\mathrm{top\_rim}}(\xi^2) := e ^{-(\xi^2 - L)^2/\sigma}.
\end{split} 
\end{align} 
The conditions for skirt twisting (angular displacement) are similar to those of the sleeve twist (applied at the top rim) in Sec.~\ref{ssec:sleeve}. %

\begin{figure}
	\includegraphics[width=\linewidth]{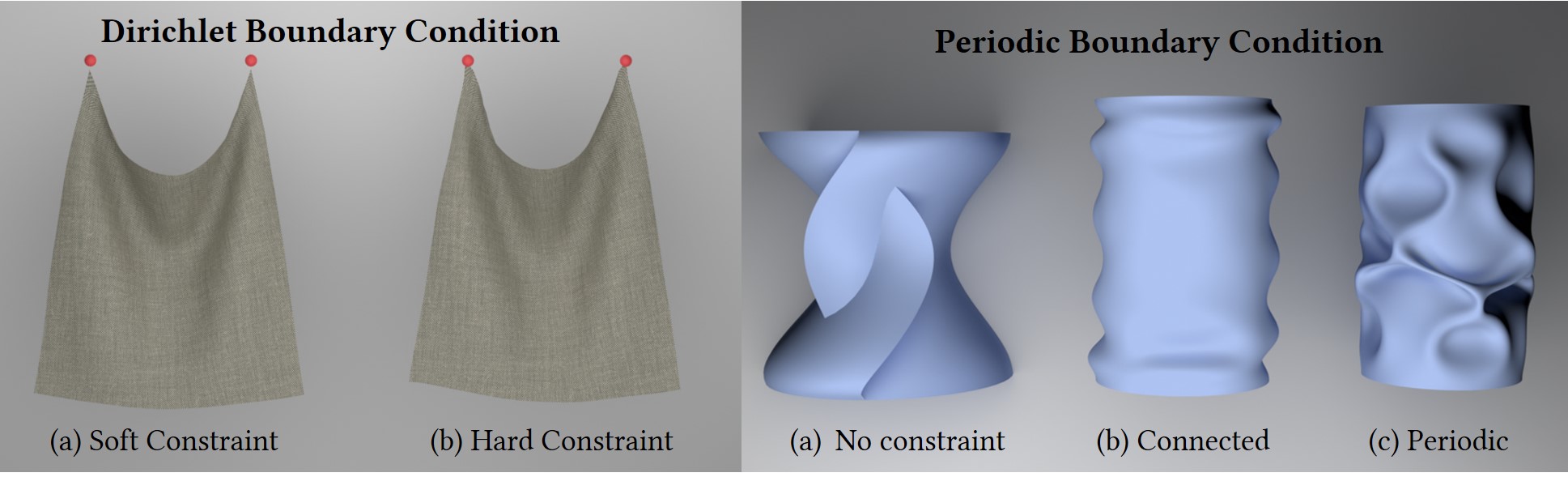}
	\caption
	{
        Ablation study for boundary conditions, with Dirichlet (top) and periodic (bottom) boundary conditions. 
	\label{fig:boundary_ablation}  
        }
\end{figure}
 
\begin{wrapfigure}[12]{R}{0.45\textwidth} %
\centering
        \vspace{-26pt}
	\includegraphics[width=\linewidth]{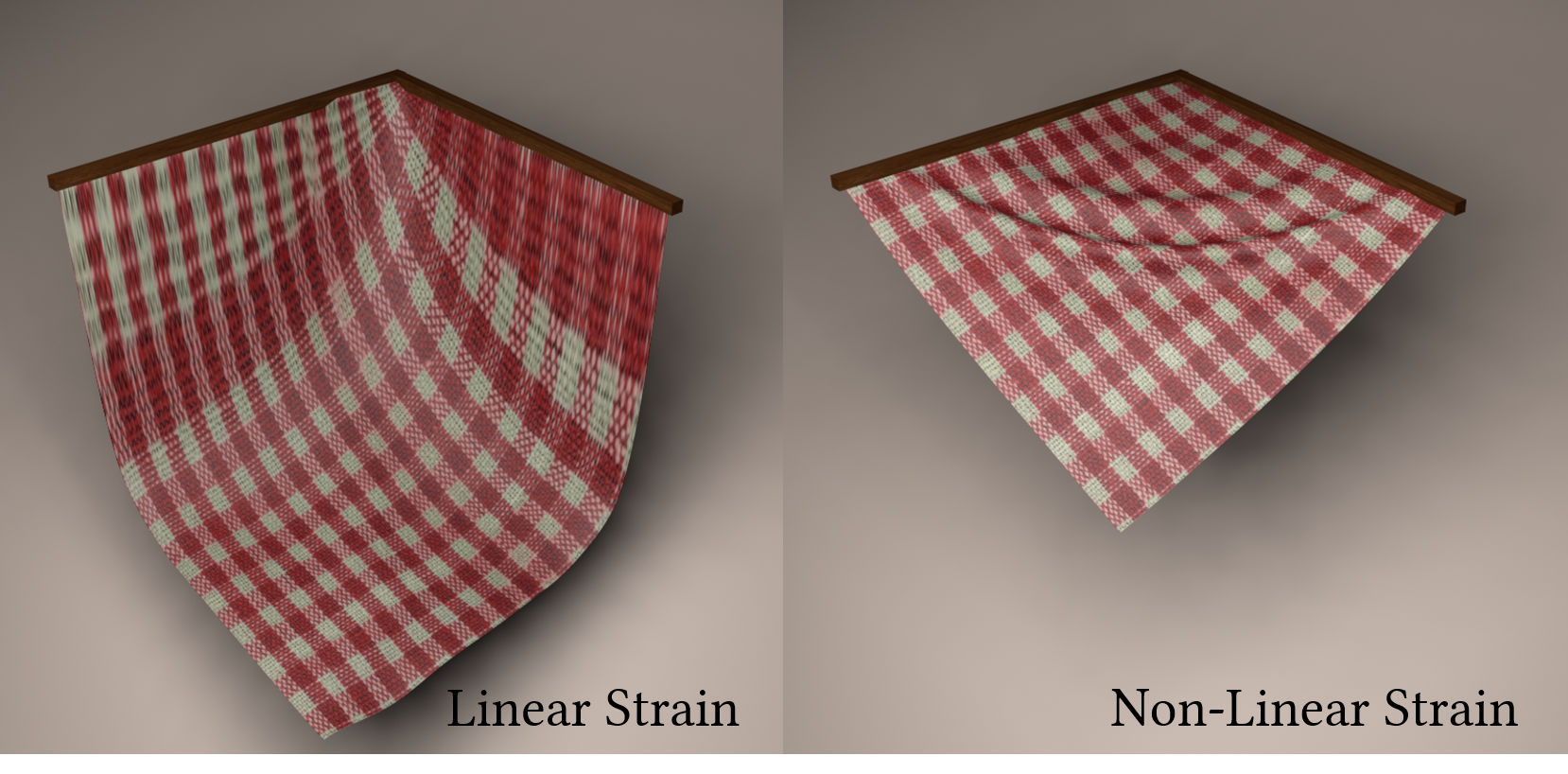}
	\caption
	{
	\textbf{Linear \textit{vs} non-linear strain.} 
        We demonstrate napkin drooping under a downward force. 
        Kirchhoff-Love strain is inherently highly non-linear. 
	}
	\label{fig:non_linear_strain}
        \vspace{-6mm}
\end{wrapfigure}
 
\section{Ablations}
\label{sec:appendix_ablations}
\subsection{Activation Function}

Experimental results for a sleeve twist with different  
activation functions in the NDF network are shown in Fig.~\ref{fig:ablation_activation}. 
While ReLU lacks support for higher-order derivatives leading to artefacts, 
a network with GELU activation can only represent low-frequency deformations.
Our usage of sine activation~\cite{sitzmann2020implicit} overcomes these limitations and successfully represents fine folds.

\newpage
\subsection{NDF Coordinate System} 
In the Kirchhoff-Love formulation, strain energy computation is performed in the local contravariant (or covariant) basis. %
This leaves us with an obvious choice of predicting covariant components of the NDF in a locally varying contravariant basis (\cref{fig:background_supplemental}-(b)).
Hence,
(a) we predict NDF in a contravariant basis and use it directly in strain calculation (ablated version), and (b) we predict NDF in a global basis and transform its components to a local basis before strain calculation. 
The second case leverages the knowledge of local basis (which is not guessed) and leads to better convergence (Fig.~\ref{fig:ablation_activation}).

\subsection{Non-linearity of Strains}
In the small-strain regime, linearised kinematics is often employed. %
However, accurate simulation of cloth quasistatics requires modelling of both rigid motion and non-linear deformation. 
Kirchhoff-Love membrane and bending strains are non-linear functions of the displacement field and 
non-linear strain calculation 
is decisive for obtaining 
realistic results. 
Thus, we evaluate the linear approximation of Kirchhoff-Love strain by omitting the non-linear terms in \eqref{eq:strain}-(main matter). 
In Fig.~\ref{fig:non_linear_strain}, we show that a linear approach leads to significant inaccuracies in modelling cloth bending under gravity.
\subsection{Boundary Constraints}
We perform ablations on 
the Dirichlet and periodic boundary conditions. 
We try a soft constraint variation, in which we impose the boundary condition as a loss term in addition to the Kirchhoff-Love energy. This requires empirically determining the optimal loss weight, takes much longer to train and does not guarantee satisfying boundary constraints, as shown in Fig.~\ref{fig:boundary_ablation}-(left). 
Our approach with hard constraints avoids all these problems. 
In the second example, we simulate the compression of a cylindrical sleeve as described in \cref{ssec:sleeve}. 
As seen in Fig.~\ref{fig:boundary_ablation}-(right), at $ \xi^1 = \pi$, cylinder (a:) is disconnected if no constraint is specified; (b:) is connected with $ \xi^1 \mapsto \cos \xi^1$; (c:) fully models continuity and differentiability forming folds with 
$\xi^1 \mapsto \{\cos \xi^1, \sin \xi^1\}$.

\section{Applications}
\label{sec:applications_suppl} 

\subsection{Material-conditioned NDFs} 

For simplicity, we choose the linear elastic materials, \ie
$\boldsymbol{\Phi} := \{\rho, h, E, \nu \}$.
Conditioning on $\boldsymbol{\Phi}$ allows us to adjust at test time mass density $\rho$, cloth thickness $h$, as well as the linear isotropic elastic properties of the material, \textit{i.e.,} Young's modulus $E$ and the Poisson's ratio $\nu$. 
The updated NDF---which is now a function of material as well---reads: 
    $\mathbf{u} (\boldsymbol{\xi}, t, \boldsymbol{\Phi};\Theta) = \mathcal{F}_\Theta (\boldsymbol{\xi}, t, \boldsymbol{\Phi}) \mathcal{I}(t) \mathcal{B}(\boldsymbol{\xi}),$
where $\boldsymbol{\Phi} \in [\boldsymbol{\Phi}_{\mathrm{min}}, \boldsymbol{\Phi}_{\mathrm{max}}]$ is the continuous range of material parameters. 
At each training iteration, we uniformly (at random) re-sample $\boldsymbol{\Phi}$ to explore the entire material domain. 
At test time, novel simulation can be generated with a single forward pass for any material $\boldsymbol{\Phi}$ in the valid material range. 
Unlike latent space conditioning in other fields and problems, the material space conditioning in NeuralClothSim has a direct physical (semantic) interpretation. 

As an example, we train an NDF conditioned on cloth thickness
and 
$\boldsymbol{\Phi} \in \{\rho, E, \nu \} \times [h_{\mathrm{min}}, h_{\mathrm{max}}]$ with $h_{\mathrm{min}} = \qty{0.0005}{\metre}$ and $h_{\mathrm{max}} = \qty{0.0025}{\metre}$. 
We visualise the simulated result %
for $h = \{0.0005, 0.0015, 0.0025\}$, in Fig.~\ref{fig:teaser}-(bottom right). 

\subsection{NDF Editing} 
\begin{wrapfigure}[22]{R}{0.5\textwidth}
\vspace{-5mm}
\centering
        \includegraphics[width=1.0\linewidth]{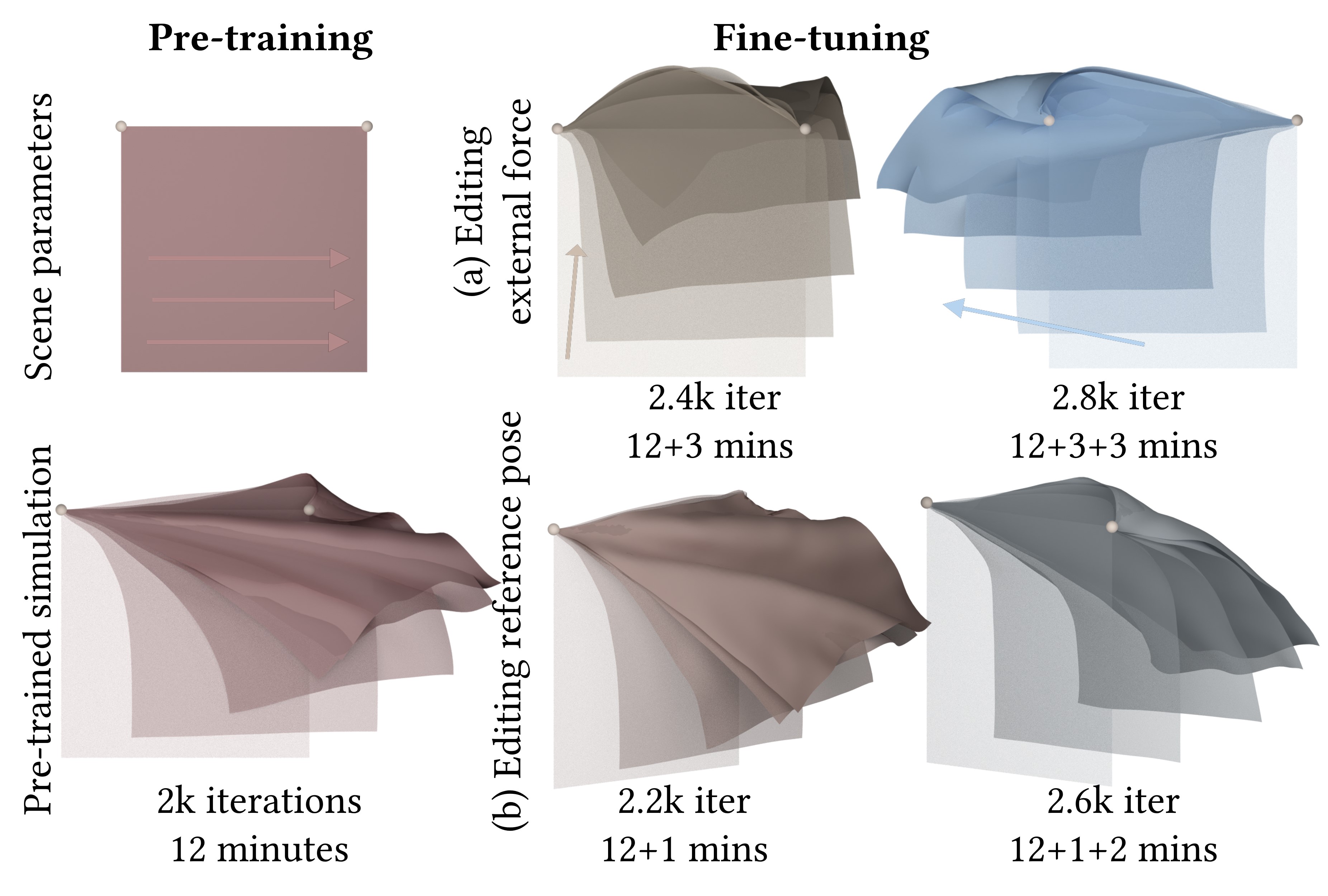}
	\caption
	{
	\textbf{Simulation editing with \ourtitle.}
        We show an example of a simulation pre-trained with a fixed reference state and external force.
        Once converged, we fine-tune the NDF with  smoothly varying external force (top) or the pose of the reference geometry (bottom) in each iteration. 
        Fine-tuning a pre-trained NDF with updated design parameters is faster and offers querying of physically-plausible intermediate simulations.
        }%
	\label{fig:simulation_editing}
     
\end{wrapfigure}

In movie and game production, a 3D artist’s workflow includes updating design parameters, which requires multiple repeated simulations from scratch. 
Such scene parameters include reference state geometry, external forces, and material properties. 
Material parameters typically constitute a low-dimensional space, so we propose to condition the NDF on material properties. 
However, other inputs such as shape and pose of reference state, as well as external force, are high-dimensional.
Instead of learning simulations over the entire scene space, %
we offer simulation editing the following way: the user can interrupt the training of NDF at any point, change the scene parameters and continue training for successive improvement. %
On the other hand, editing can also be done after full convergence (aka pre-training) and then fine-tuned with gradually modified design parameters. 
Editing an NDF provides multiple advantages over training a new NDF from scratch: It is computationally and memory efficient and allows access to interpolated simulations. 
\par
In the following, we demonstrate editing of the following scene parameters: (a) external force, and  (b) reference state geometry.
The key idea is to use the modified scene parameters in the loss function and update the NDF weights with gradient-based optimisation.
Specifically, given a cloth geometry $\mathbf{\bar{x}} $, external forces  $\mathbf{f}$,
we train an NDF to obtain a simulation $\mathbf{u}^*$ parameterised by network weights $\Theta^*$, as described in the main method.
As an editing objective, we would like to arrive at a novel simulation corresponding to external force $\mathbf{f}^I$ and/or reference geometry $\mathbf{\bar{x}}^I$ with $I \in \mathbb{N} $ training iterations. Here, $I$ is much smaller than the iterations needed for the convergence of the original simulation.
We can then fine-tune the pre-trained NDF over iterations $i \in \{0,...,I\} $ by minimising the loss function, $ \mathcal{L}(\Theta; \mathbf{f}^i, \mathbf{\bar{x}}^i) $ to obtain  edited and interpolated simulations $\mathbf{u}^i, \Theta^i$. 
Here, we assume a smooth transition of external force or the reference shape from the initial to the edited value, which can be obtained, for example, by linear interpolation, \textit{i.e.,} $\mathbf{f}^i = \operatorname{lerp}(\mathbf{f}, \mathbf{f}^I, \frac{i}{I})$.%

We next demonstrate results for dynamic editing of the pre-trained simulations. 
We conduct two experiments, \textit{i.e.,} editing external forces and editing the 6DoF pose of the reference geometry; the results are visualised in Fig.~\ref{fig:simulation_editing}. 
First, a short simulation of a napkin is pre-trained as an NDF with a fixed reference state and external force, which takes ${\approx}12$ minutes. 
In the first example (top row), we gradually vary the direction and magnitude of the external force by linearly interpolating between 
the original and the final forces. 
This leads to the motion of cloth towards the instantaneous force direction. 
In the second example (bottom row), we smoothly vary the reference poses and the corresponding position of the handles, 
generating novel edited simulations. 
Editing reference pose leads to the motion of the cloth towards a fixed force direction but originates from varying initial poses. 
Note how the change in the input scene parameters propagates to the entire 
simulation. 
Notably, fine-tuning is much faster and takes ${\approx}2$ minutes, leading to a time-saving of ${\approx}83\%$.
We show two intermediate simulations in Fig.~\ref{fig:simulation_editing}, and other edited simulations corresponding to each iteration can be queried as well. 
As the simulation is parameterised by the weights of a neural network (instead of meshes), our proposed way of simulation editing is memory-efficient. 

\begin{figure*}%
        \includegraphics[width=1.0\linewidth]{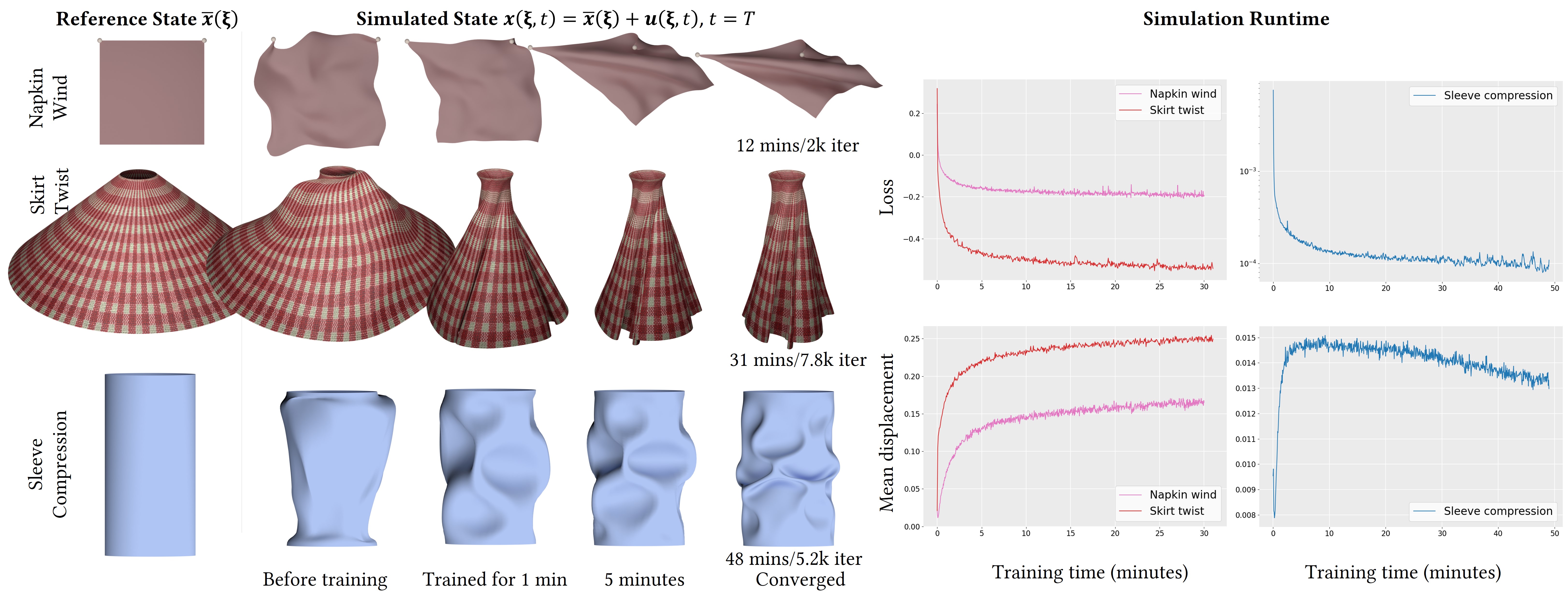}
	\caption
	{
	\textbf{Runtime analysis of NeuralClothSim.} 
        On the left, we visualise the evolution of the last frame ($T=1$) over the training iterations. On the right, the plot shows NDF convergence as a function of training time leading to refined simulations. %
	}
	\label{fig:runtime_ours_visual_plot}
\end{figure*}

\begin{wrapfigure}[13]{R}{0.5\textwidth}
 \centering
        \vspace{-14pt}
        \includegraphics[width=\linewidth]{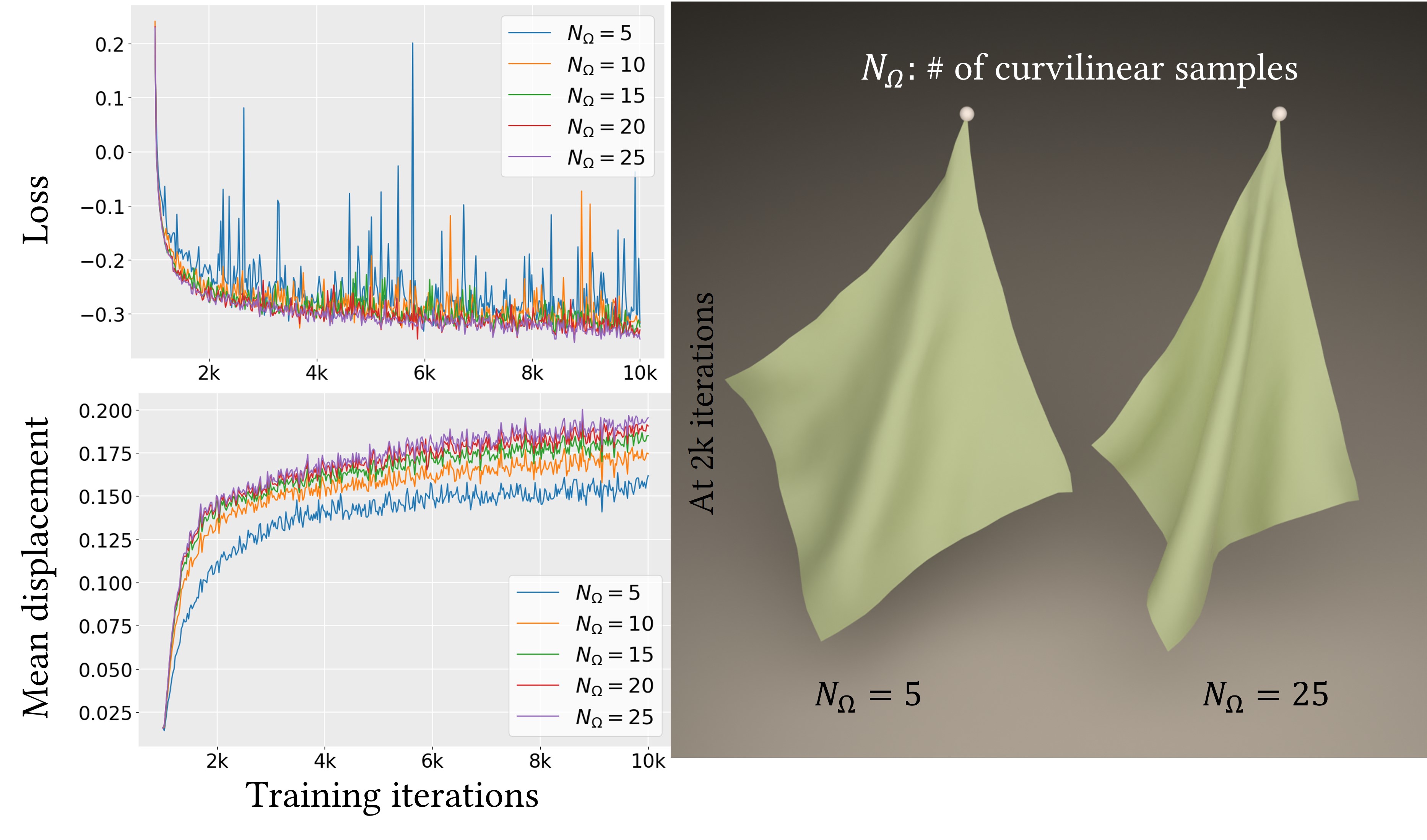}
	\caption
	{
	   \textbf{Analysis of the sampling strategies.} 
        We show the influence of the number of training points on the performance of our method.
	}
	\label{fig:robustness_sampling_thickness}
\end{wrapfigure}

\section{Performance}
\label{sec:performance}
The performance of a cloth simulator---such as computation time---is a crucial aspect of its usability. 
This work focuses on the fundamental challenges of developing an implicit neural quasistatic simulator with new characteristics. 
Our method does \textit{not} outperform the classical simulators in all aspects as they are well-engineered and highly optimised. 
Next, we provide a detailed analysis of NeuralClothSim's performance. 
\subsection{Runtime}
NeuralClothSim encodes the cloth equilibrium state as an NDF, and, consequently, 
the bulk of computation time lies in the NDF training (\textit{i.e.,} optimisation of the network weights). 
At inference, extracting the simulated states from NDF as meshes or point clouds requires a single forward pass and is, therefore, fast. 
In Fig.~\ref{fig:runtime_ours_visual_plot}, we provide a runtime analysis of three representative simulations as a function of training time. 
On the left, we visualise the evolution of the last frame (\ie, equilibria state), showing the refinement of the simulated state with increased training time.
Before training, the simulation state is the sum of the reference state and random noisy output from NDF. 
Within a few minutes of training, NDF generates a reasonable simulated state, which then converges within 30 minutes to one hour; see our supplementary video for the evolution of simulation states over training iterations. 
On the top-right of Fig.~\ref{fig:runtime_ours_visual_plot}, we plot loss values as a function of training time, which shows that our training is stable. 
As NeuralClothSim is an instance of a physics-informed neural network with a physics loss only (but no data term), the loss is not expected to converge to zero. 
We monitor the mean NDF over all sampled spatio-temporal points (Fig.~\ref{fig:runtime_ours_visual_plot}-(bottom right)) as an additional cue on the simulation refinement. 
Along with the loss, saturation in mean NDF can be used as a stopping criterion. %
Note that all our experiments are carried out on a single NVIDIA Quadro RTX 8000 GPU.

Similar to classical methods~\cite{choi2005stable}, simulation with our approach is not unique, as bifurcation due to buckling can lead to solutions with different folds and wrinkles. 
Among them, the selection of the simulation outcome depends on the NDF convergence.  %
Specifically, the randomness in training samples and weight initialisation introduces desirable optimisation path variations. 
In all cases, we observe NDF training to be numerically stable.

\begin{wrapfigure}[15]{R}{0.5\textwidth} %
\centering
        \vspace{-20pt}
	\includegraphics[width=\linewidth]{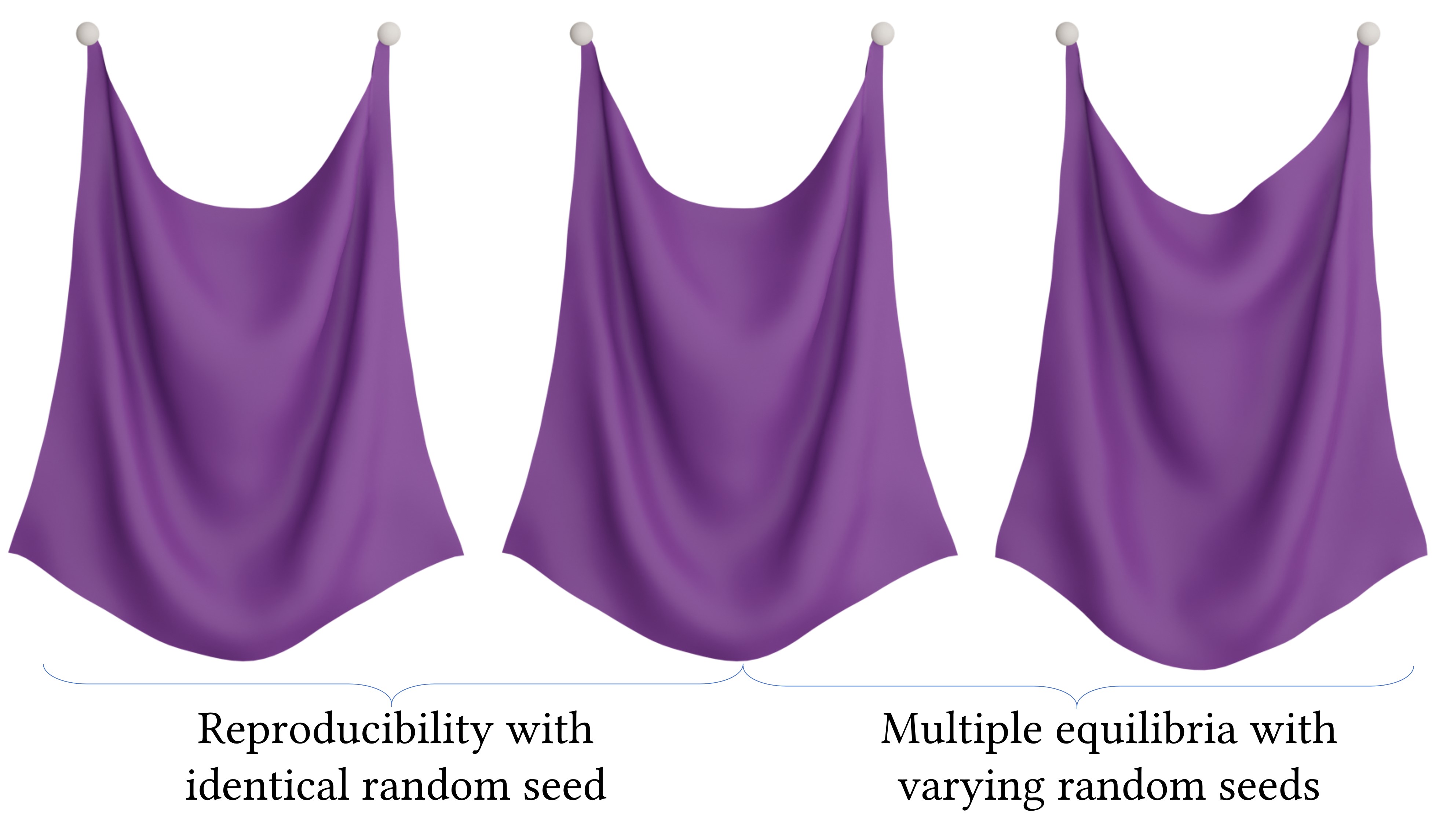}
	\caption
	{
        NDF weight initialisation allows us to control the simulation outcome. 
        We can generate multiple valid equilibrium solutions or reproduce a simulation. 
	}
	\label{fig:sensitivity_to_net_weights_initialisation}
\end{wrapfigure}

\subsection{Sampling Strategy}  %
Next, we study the influence of the number of training points on the performance of our method. 
Input samples to NDF include curvilinear $N_\Omega$ and temporal stratified $N_t$ coordinates (for trajectory visualisation) over which the loss is computed at each training iteration. 
For a napkin of size $\Omega=[0,1]^2$, we simulate for $t \in [0,1]$ by training NDF for 10k iterations with number of sampling points $N_\Omega \in \{5, 10, 15, 20, 25\}$.
Computation times for all experiments are comparable (and slightly higher for the higher number of samples) as they share the GPU memory and are processed in parallel. 
Fig.~\ref{fig:robustness_sampling_thickness} shows the qualitative and quantitative performance. 
We observe that higher $N_\Omega$ leads to faster learning, as seen in the qualitative result in the top row and the mean displacement plot in the bottom row. 
Furthermore, it leads to stable optimisation, as seen in the loss plot. 
Future work could explore advanced sampling techniques for improved performance.

\begin{figure}
 \centering
        \includegraphics[width=0.8\linewidth]{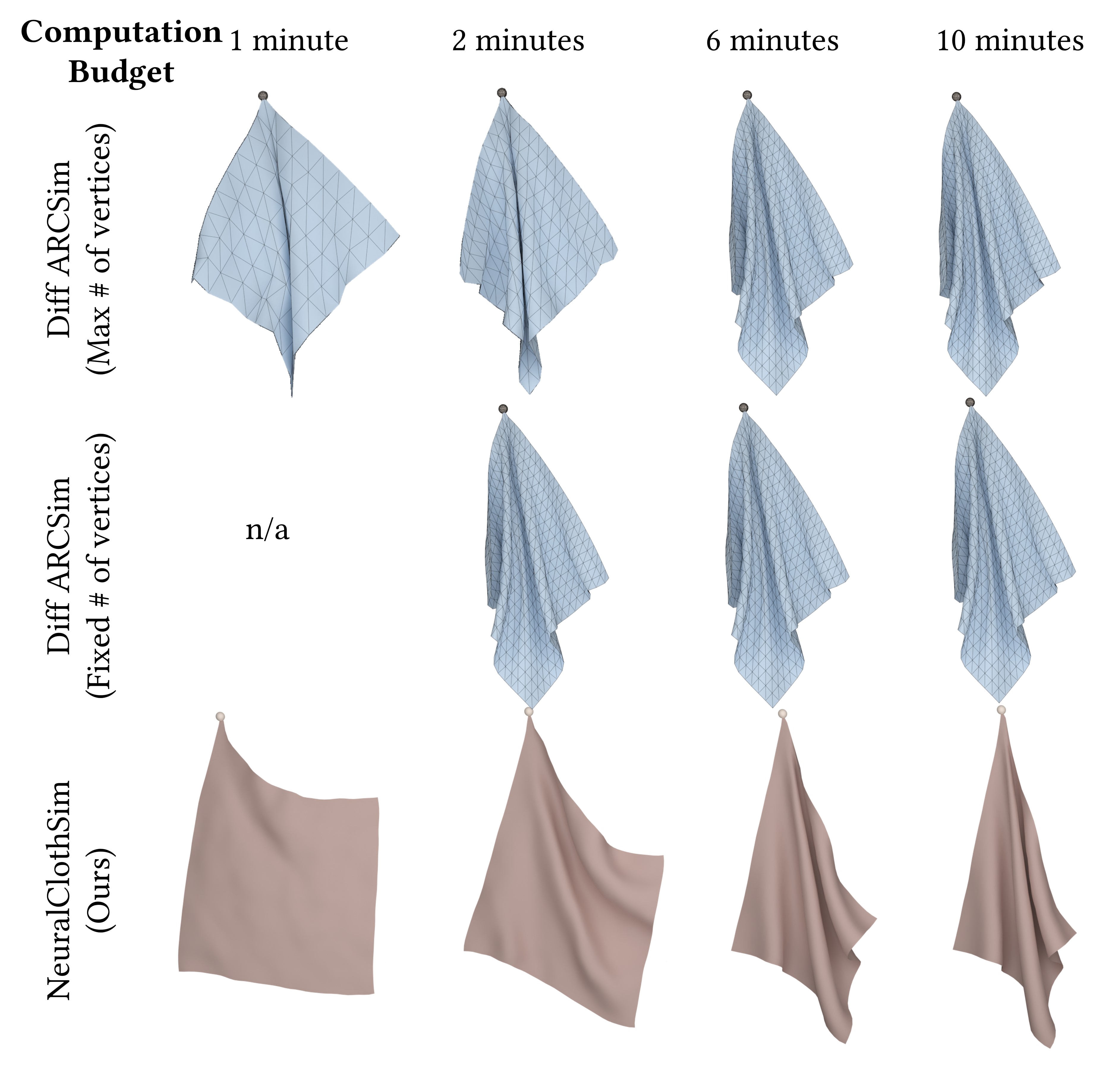}
	\caption
	{
	\textbf{Runtime comparison of DiffARCSim~\cite{Liang2019} and our approach.}
        Like most classical simulators, DiffARCSim integrates forward in time, solving for a 3D deformation field at each time step, in contrast to our approach which optimises for the 4D spatio-temporal NDF. 
        With decreasing computational budget, DiffARCSim produces converged simulated states of the cloth at low resolutions or only early frames at high resolutions. 
        On the other hand, NeuralClothSim offers partially converged simulations at arbitrary resolutions as the computational budget decreases. 
	}  
	\label{fig:runtime_comparison}
\end{figure}

\begin{figure}
        \includegraphics[width=\linewidth]{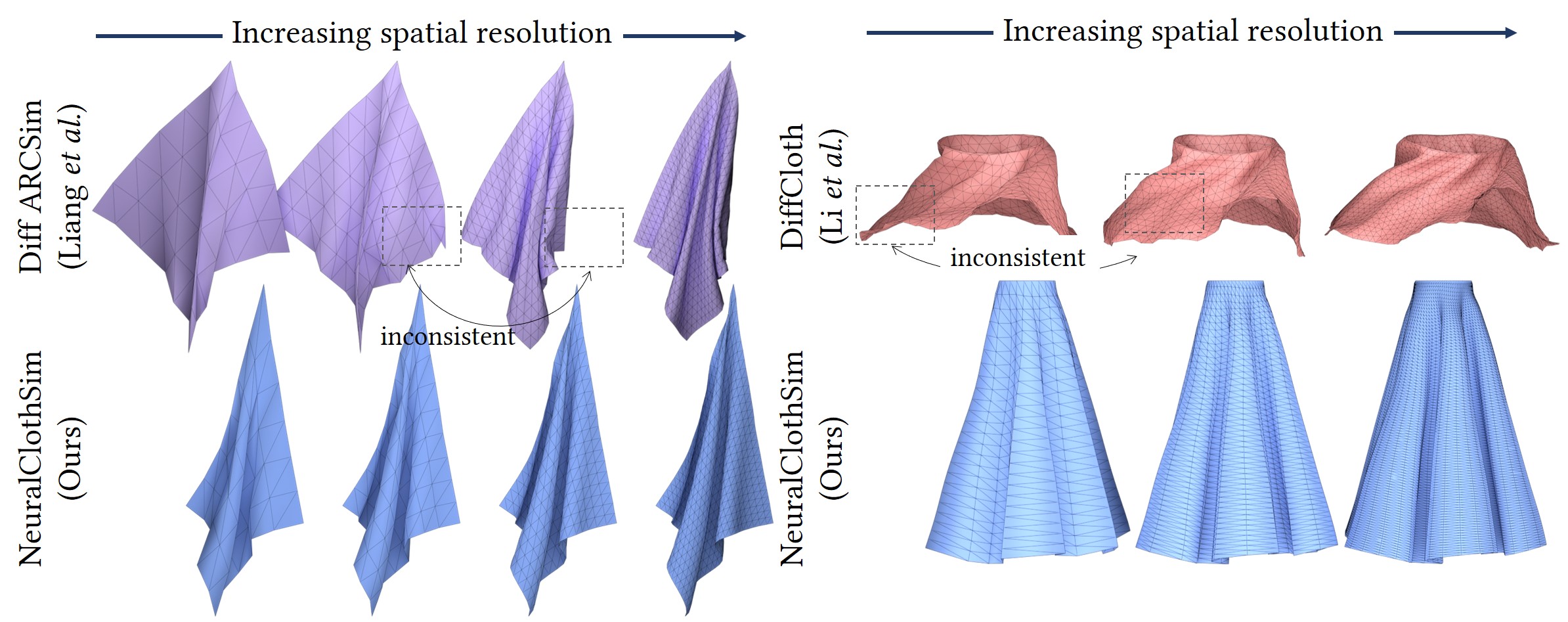}
	\caption
	{
	\textbf{Spatial and temporal surface consistency of state-of-the-art differentiable simulators and our approach.}
        Classical simulators such as ARCSim~\cite{Liang2019} and DiffCloth~\cite{li2022diffcloth} reproduce simulation outcomes when re-running at the same resolution.
        However, changing spatio-temporal resolution requires multiple runs and generates possibly different folds or wrinkles instead of refining (or previewing) the geometry. 
        Since we learn a continuous neural parameterised model, a converged (or partially converged) NDF provides consistent simulation when queried at different spatio-temporal inputs.
        Note that NeuralClothSim does not provide consistent refinement as a function of computation time (no speed \textit{vs} fidelity trade-off), but rather consistent simulation with respect to the spatio-temporal sampling (at a given computational budget). 
        }
	\label{fig:resolution_inconsistency}
\end{figure}

\subsection{Simulation Reproducibility}
Next, we investigate whether NeuralClothSim simulations are deterministic. 
Cloth simulation does not have a single ground truth; rather, it can have multiple equilibria solutions under the same input parameters (template, material, and boundary conditions). 
While FEM-based cloth simulators are designed to be deterministic, in practice, there are several factors---such as numerical precision and parallel computing---that can lead to slight variations in the simulation results between runs. 
We note that a mesh-based simulator running the same simulation scenario on different machines generates non-identical results (but reproducible ones on the same machine). 
Interestingly, we can replicate such behaviour by employing the  
sensitivity of our method to the initialisation of the neural network weights.
We conducted two experiments leading to the following observations: 1) We can obtain reproducible results if we set the random seed leading to the same network initialisation (\cref{fig:sensitivity_to_net_weights_initialisation}-(left)), and 2) We observe non-identical results if we do not set the random seed (\cref{fig:sensitivity_to_net_weights_initialisation}-(right)). 
\begin{wrapfigure}[16]{R}{0.5\textwidth}
\centering
        \vspace{-18pt}
	\includegraphics[width=\linewidth]{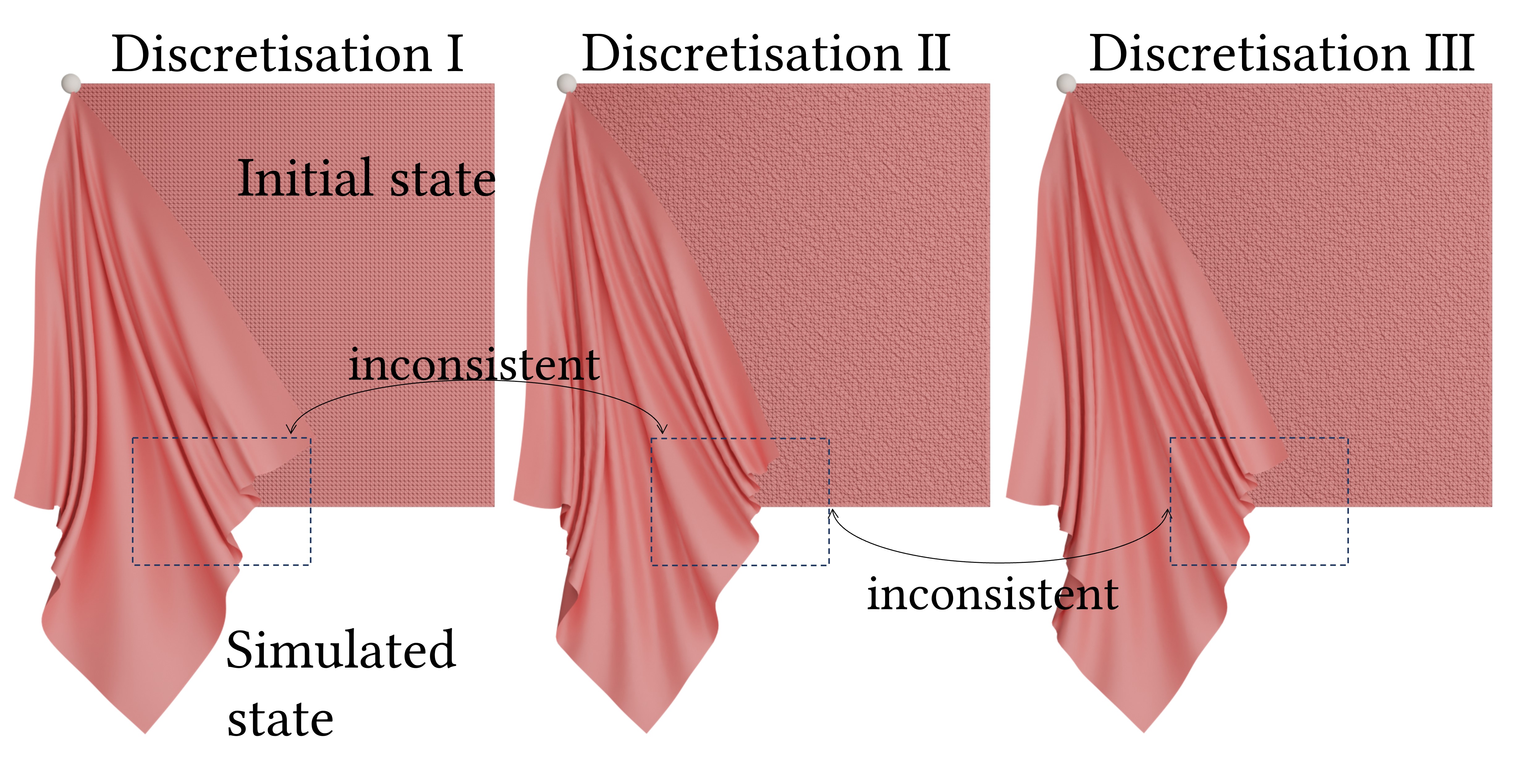}
	\caption
	{
        Visualisation of the inconsistencies observed in the results by FEM-based ARCSim~\cite{narain2012adaptive}, even at high resolutions.
        Our method leads to consistent results for much coarser discretisations (\cref{fig:initial_discrete_inconsistency_suppl}-main).
	}
	\label{fig:arcsim_inconsistency}
\end{wrapfigure}

\section{Additional Comparisons}
\begin{table}
	\begin{center}
		\caption
    	{
            \textbf{Conceptual comparison of our NeuralClothSim to previous state-of-the-art cloth simulators.} 
            Our approach %
            enables highly desired properties such as surface continuity, and consistent simulations (folds/wrinkles) at different discretisations of the initial mesh, material conditioning and simulation editing for updated parameters. 
            }
          \centering
            \begin{tabular}{ lcccc }
                \toprule
			                                    & Continuous                  &  Consistency      &  Sim. Editing   & Mat. Interpolation          \\
                \midrule
                \citet{narain2012adaptive}  		    & \xmark                        &  \xmark	    & \xmark  &   \xmark   \\
			\citet{Liang2019}  		    & \xmark                 	        & \xmark	    & \xmark  &   \xmark   \\
	        \citet{li2022diffcloth}  		& \xmark                 	        & \xmark	    &   \xmark &    \xmark   \\	
                \citet{zhang2022progressive}  		    & \xmark              	            & \cmark	  & \cmark &    \xmark    \\
			\textbf{Ours}  			            & \cmark                            & \cmark        & \cmark     &   \cmark \\
                \bottomrule
		\end{tabular}
        \label{tab:concecptual_comparison}
	\end{center}
\end{table}

NeuralClothSim is the first step towards neural implicit cloth quasistatics. 
Although less mature compared to FEM-based simulators, it offers several desired characteristics; See Table~\ref{tab:concecptual_comparison} for a comparison between existing cloth simulators and our approach.
\subsection{Runtime}

We compare the runtime of our method to those of the FEM-based simulator DiffARCSim~\cite{Liang2019, narain2012adaptive}. 
Since our approach does not support collisions, we turn off collision handling in DiffARCSim due to the computational overheads for a fair comparison. 
We simulate a napkin sequence, and our quasistatic result and the dynamic simulated state (after 1 $s$) from DiffARCSim are visualised in Fig.~\ref{fig:runtime_comparison}. 
For the same computation budget (runtime), we show the best simulated states for both methods. 
Therefore, we present two sets of results for DiffARCSim, \textit{i.e.,} simulated states for the given computational budget 1) with maximum mesh resolution (Fig.~\ref{fig:runtime_comparison}-(top row)) and 2) with fixed mesh resolution (Fig.~\ref{fig:runtime_comparison}-(middle row)). 
We notice that both methods refine the simulated states with increased runtime. 
With a decreasing computational budget, DiffARCSim produces converged simulated states of the cloth at low resolutions or only early frames at high resolutions.
On the other hand, NeuralClothSim offers partially converged simulations at arbitrary resolutions as the runtime decreases.

\subsection{Multi-Resolution Consistency}
Next, we show the comparison of NeuralClothSim to DiffARCSim, and DiffCloth~\cite{li2022diffcloth} in terms of the multi-resolution simulation consistency. 
We simulate 1) a napkin with a fixed corner under gravity, with our approach and ARCsim (Fig.~\ref{fig:resolution_inconsistency}, top two rows) and 
2) a twisting and twirling motion of the skirt with our approach and DiffCloth (Fig.~\ref{fig:resolution_inconsistency}, two bottom rows). 
The compared simulators operate on meshes of pre-defined resolution (as provided initially). 
Hence, they need to run from scratch for different mesh resolutions, and the simulation outcome are not guaranteed to be the same across these runs under different discretisations. 
Thus, increasing (or decreasing) spatial resolution can result in different folds or wrinkles instead of refining simulations at coarser resolutions. 
Unlike DiffARCSim and DiffCloth, our method provides consistent simulation at arbitrary resolutions.
The same 3D points remain unaltered in meshes extracted from NDF at different resolutions. 
We emphasise that we do not claim consistent refinement as a function of runtime but rather a consistent equilibrium state with respect to spatial sampling (at a given computational budget). 
This means that both converged or partially converged NDF provide consistent quasistatics when queried at different spatial inputs.

\begin{wrapfigure}[24]{R}{0.45\textwidth}
\centering
        \vspace{-16pt}
        \includegraphics[width=0.4\textwidth]{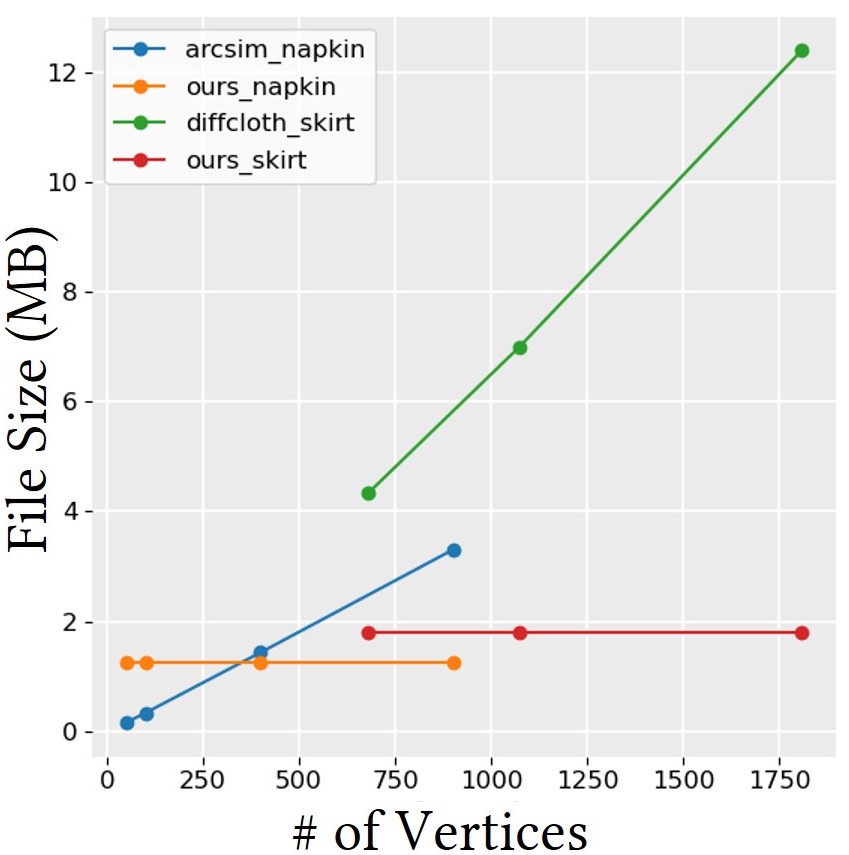}
	\caption
	{
	    \textbf{Memory efficiency}. 
        We plot the memory requirements for simulations generated by ours and DiffARCSim~\cite{Liang2019}, and DiffCloth~\cite{li2022diffcloth}. %
        The simulations are chosen to be of similar complexity and are visualised in Fig.~\ref{fig:resolution_inconsistency}.
        The constant memory requirement of our approach is due to the compressing property of the MLP weights that encode the simulations. %
        }
	\label{fig:memory_resolution}
\end{wrapfigure}

Our comparison deviates from the literature, as the primary reason for using different spatio-temporal resolutions is to adjust runtime and memory usage. 
For example, the recent method of~\cite{zhang2022progressive} produces artefact-free  %
previewing geometries (at various approximation levels) by biasing their solutions with shell forces and energies evaluated on the finest-level model. 
This approach offers a trade-off between runtime \textit{vs} resolution while maintaining simulation consistency.
In contrast, with NeuralClothSim, simulation is consistent at arbitrary resolutions at 
any moment during the NDF training, 
which, we believe, is still beneficial for many downstream tasks. 
Of course, 
ARCSim and DiffCloth also support very high resolutions, which 
eventually enables 
browsing the simulations at different mesh resolutions (while maintaining mesh consistency across the levels); 
however, at the cost of high memory consumption. 
Moreover, in their case, methods for inverse problems that estimate the simulation parameters from simulated states cannot use adaptive, \eg, coarse-to-fine and importance sampling. 
In contrast, our continuous formulation offers clear advantages in this regard.

\begin{figure}
\centering
	\includegraphics[width=1\linewidth]{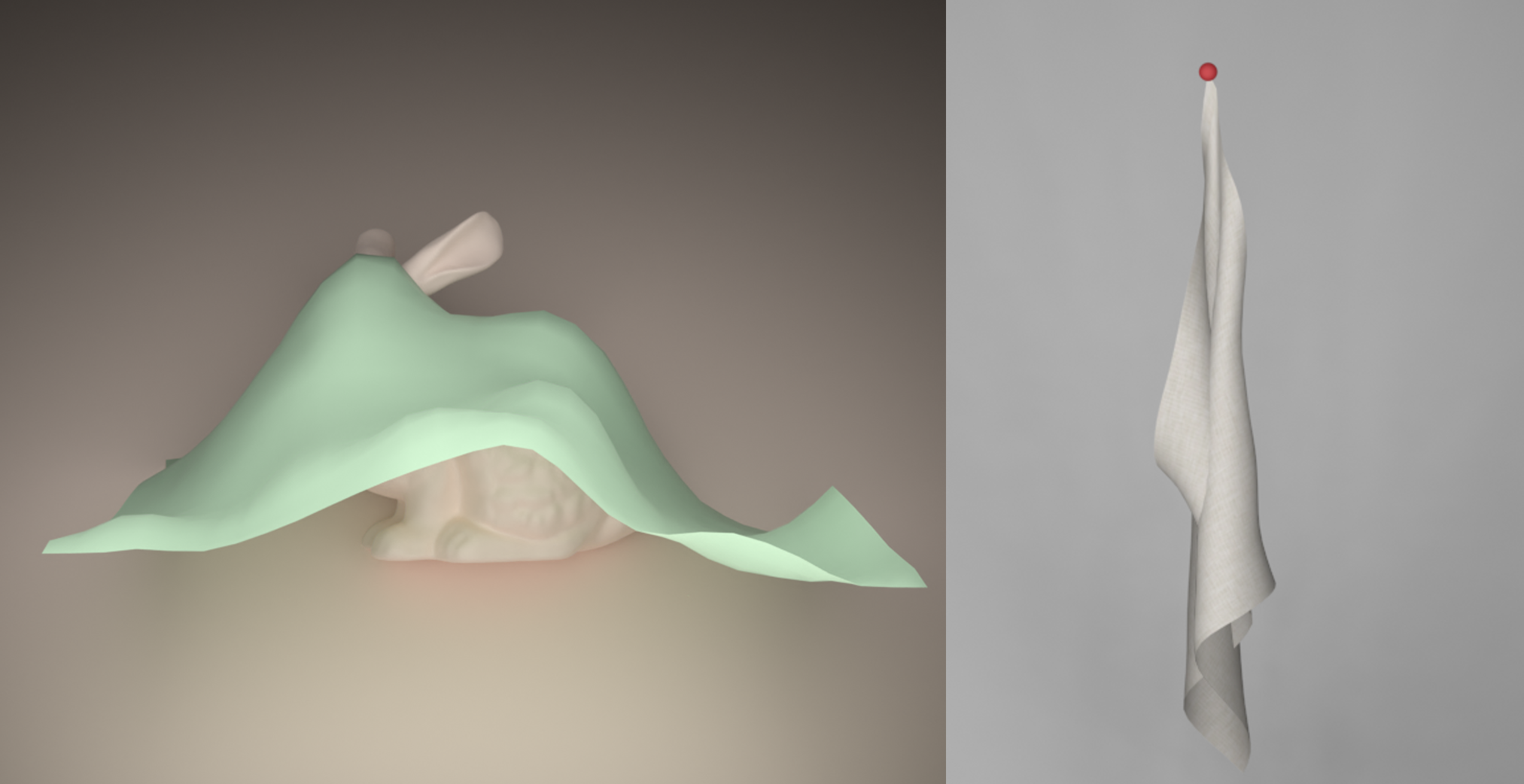}
	\caption
	{
        \textbf{Limitations.}
        Our approach does not handle collisions, contacts and frictions at the moment, since the focus of this work is on the fundamental challenges of developing a neural cloth simulator. These examples show inaccuracies due to the simplifications made in one possible extension \eqref{eq:collisions}. 
	}
	\label{fig:limitation}
\end{figure}
 
\section{Collision} 
\label{ssec:futher_discussion} 
In a preliminary experiment, we model collisions with external objects %
following earlier neural methods~\cite{santesteban2022snug, bertiche2022neural}, \textit{i.e.,} 
we define an additional loss term $\mathcal{L}_\mathrm{collision}(\Theta)$ that penalises collisions, leading to: 
\begin{align}
\label{eq:collisions}
\begin{split}
    \mathcal{L} &= \mathcal{L}_\mathrm{physics}(\Theta) +  \lambda \mathcal{L}_\mathrm{collision}(\Theta),\;\text{with} \\
    \mathcal{L}_\mathrm{collision}(\Theta) &=  \frac{|\Omega|}{N_\Omega N_t}\sum_{i=1}^{N_\Omega} \sum_{j=1}^{N_t}  \max(\epsilon - \operatorname{SDF}(\mathbf{x}(\boldsymbol{\xi}_i, t_j;\Theta)), 0), 
\end{split}
\end{align}
where $\operatorname{SDF}(\mathbf{x})$ is the signed distance to the object, 
$\epsilon$ is a small safety margin between cloth and object to ensure robustness,
$ \lambda$ is the weight for the collision term, and $\mathcal{L}_\mathrm{physics}(\Theta)$ is our main thin-shell loss in Eq.~\eqref{eq:main_loss}-(main matter). 
We set $\lambda=1000$, $\epsilon=0.001$, and use a pre-trained SDF network encoding signed distance function.
Specifically, we employ the method of~\citet{sitzmann2020implicit} to fit an SDF network on an oriented point cloud, where an Eikonal regularisation is used in addition to the SDF and normal loss.
Fig.~\ref{fig:limitation}-(left) visualises a simulation result for a piece of cloth falling on the Stanford bunny;
see our supplementary video for the full simulation. 
We observe that the cloth coarsely respects the object contours, although constraints in Eq.~\eqref{eq:collisions} are soft and do not guarantee %
physically realistic deformations. 
Difficulties in training PINN with multiple loss terms were previously reported in the literature~\cite{hao2022physics} and 
future research is necessary to further investigate collision handling in the context of NDFs. 
\par

\section{Extended Discussion and Limitations} \label{sec:limitations}

This article addresses the fundamental challenges of cloth simulation with NDFs. 
All in all, we find the proposed design and the obtained experimental results very encouraging and see multiple avenues for future research. 
Our current quasistatic approach is the first step towards implicit neural simulation. It would be a promising direction to add dynamic effects such as inertia and damping in this setting. 
Moreover, our simulator does not handle 
contacts, friction and collisions which will be necessary for many potential applications beyond those demonstrated in this article. 
This is, however, a standalone research question 
in the new context. 

Several 
limitations of NeuralClothSim 
originate from NDF modelling as a single MLP:

First, MLP weights have a global effect on the simulation,  whereas the movement of mesh vertices affects only the local neighbourhood. %
While this global nature offers continuity and differentiability, 
we believe exploring alternative network parameterisations that bring the best of both representations could bring improvements in future. 
Second, our results are currently empirical: While we observe expected results in all our experiments, there are no convergence guarantees or upper bounds on accuracy. 
Finally, periodic boundary conditions aid mainly with simple geometries; the extension to more complex garments needs further exploration. 
Future work could also explore 
modelling different types of human clothing 
with the help of the proposed implicit neural framework.

Summa summarum, NeuralClothSim is the first step towards neural implicit cloth simulation, which we believe can become a powerful addition to the class of cloth simulators. 
Inverse problems in vision and graphics could also benefit from its consistency (\eg, multi-resolution data generation), and adaptivity. %

\end{document}